%% file: LAT_Stack.tex
\documentclass[reprint,amsmath,amssymb,aps,nofootinbib,noeprint]{revtex4-2}

\usepackage{epsfig}
\usepackage{verbatim}
\usepackage{epstopdf}
\usepackage{graphicx}
\usepackage{aas_macros}
\usepackage{epsf}
\usepackage{amsmath}
\usepackage{tikz}
\usepackage{enumitem}
\usepackage{adjustbox}
\usepackage{threeparttable}
\usepackage{ragged2e}
\usepackage{MnSymbol}
\usepackage{mathrsfs}
\usepackage{booktabs}
\usepackage{xcolor}

\usepackage{tabularray}
\DefTblrTemplate{caption-sep}{default}{.\enskip}
\usepackage{tabularx}
\usepackage{multirow}
\usepackage{placeins}

\usepackage{rotating}

\definecolor{darkgreen}{rgb}{0.0,0.5,0.0}
\usepackage[colorlinks,citecolor=darkgreen]{hyperref}

\newlength{\MapGeV}
\newcommand{\MapLabelX}{85.1}
\newcommand{\MapLabelY}{78.0}
\newcommand{\MapLabelFont}{\footnotesize}
\definecolor{MapBlue}{rgb}{0,0.3,1}\definecolor{MapRed}{rgb}{0.85,0,0}\definecolor{MapBlack}{rgb}{0,0,0}
\newcommand{\MapTwoLabel}{\rotatebox{90}{\scriptsize $Z_l=2$}}\newcommand{\MapTwoX}{92.6}\newcommand{\MapTwoY}{81.5}
\newcommand{\MapRedLabel}{{\scriptsize $72\GeV$ triad}}\newcommand{\MapRedX}{74}\newcommand{\MapRedY}{71}
\newcommand{\MapTriLabel}{\rotatebox{90}{\scriptsize triads}}\newcommand{\MapTriX}{93.2}\newcommand{\MapTriY}{80.5}
\newcommand{\MapExtraLabels}{\node[anchor=east, font=\MapLabelFont, inner sep=0pt, text=MapBlue, xshift={(\MapTwoX-102.15)*\MapGeV}, yshift={(\MapTwoY-101.11)*\MapGeV}] at (img.north east) {\MapTwoLabel};\node[anchor=west, font=\MapLabelFont, inner sep=0pt, text=MapRed, xshift={(\MapRedX-102.15)*\MapGeV}, yshift={(\MapRedY-101.11)*\MapGeV}] at (img.north east) {\MapRedLabel};\node[anchor=west, font=\MapLabelFont, inner sep=0pt, text=MapBlack, xshift={(\MapTriX-102.15)*\MapGeV}, yshift={(\MapTriY-101.11)*\MapGeV}] at (img.north east) {\MapTriLabel};}
\newcommand{\MapPanel}[5]{\begin{tikzpicture}\node[anchor=south west, inner sep=0pt] (img) at (0,0) {\includegraphics[width=#1\textwidth]{#2}};\node[align=center, font=\MapLabelFont, inner sep=0pt, xshift={(\MapLabelX+(#3)-102.15)*\MapGeV}, yshift={(\MapLabelY+(#4)-101.11)*\MapGeV}] at (img.north east) {#5};\MapExtraLabels\end{tikzpicture}}
\newcommand{\RowLabelFont}{\large}
\newlength{\RowLabelDX}
\newlength{\RowLabelDY}
\newcommand{\MapPanelRow}[6]{\begin{tikzpicture}\node[anchor=south west, inner sep=0pt] (img) at (0,0) {\includegraphics[width=#1\textwidth]{#2}};\node[align=center, font=\MapLabelFont, inner sep=0pt, xshift={(\MapLabelX+(#3)-102.15)*\MapGeV}, yshift={(\MapLabelY+(#4)-101.11)*\MapGeV}] at (img.north east) {#5};\node[overlay, anchor=north west, inner sep=0pt, font=\RowLabelFont, text=blue, xshift=\RowLabelDX, yshift=\RowLabelDY] at (img.north west) {#6};\MapExtraLabels\end{tikzpicture}}

\usepackage[leftmargin=0pt,rightmargin=0pt,skipabove=2pt,skipbelow=2pt,innerleftmargin=5pt,innerrightmargin=5pt,innertopmargin=5pt,innerbottommargin=5pt,linewidth=0pt,innerlinewidth=0pt,middlelinewidth=0pt,outerlinewidth=0pt,roundcorner=0pt]{mdframed}

\input{Definitions.tex}

\let\DgrdotOrig\dgrdot\renewcommand{\dgrdot}{\smash{\DgrdotOrig}}
\let\MindotOrig\mindot\renewcommand{\mindot}{\smash{\MindotOrig}}

\newcommand{\Myfr}{{k}}
\newcommand{\Myc}{{\mathsf{c}}}
\newcommand{\MyC}{{\mathsf{C}}}
\newcommand{\Myh}{{\mathsf{h}}}
\newcommand{\MyH}{{\mathsf{H}}}
\newcommand{\Myn}{{\mathsf{n}}}
\newcommand{\Myf}{{\mathsf{f}}}
\newcommand{\MyROI}{{\textrm{ROI}}}
\newcommand{\SAll}{{\texttt{S}}}
\newcommand{\SE}{{\texttt{E}}}
\newcommand{\SW}{{\texttt{W}}}
\newcommand{\SO}{{\texttt{O}}}
\newcommand{\SCW}{\ensuremath{\mathtt{C}_{\mathtt{W}}}}
\newcommand{\SMW}{\ensuremath{\mathtt{M}_{\mathtt{W}}}}
\newcommand{\SCT}{\ensuremath{\mathtt{S}_{3}}}
\newcommand{\SCTb}{\ensuremath{\mathtt{S}_{3\mathtt{b}}}}
\newcommand{\SJ}{{\texttt{J}}}

\setcitestyle{super}

\makeatletter
\def\@hangfrom@section#1#2#3{\@hangfrom{#1#2}#3}
\def\@hangfroms@section#1#2{#1#2}
\makeatother

\newcommand{\TblNote}[1]{\par\smallskip{\footnotesize\justifying\noindent #1\par}}

\begin{document}

\title{Rest-frame annihilation lines of $\sim70$ GeV WIMPs in 23,000 stacked galaxy clusters}

\author{Uri Keshet}
\address{
    Physics Department, Ben-Gurion University of the Negev, POB 653, Be'er-Sheva 84105, Israel
}
\email{keshet.uri@gmail.com}

\begin{abstract}
\begin{mdframed}[backgroundcolor=black!5]
Weakly interacting massive particle (WIMP) annihilation features were reported as a 43 GeV line in three \emph{Fermi}-LAT galaxy clusters ($\mathtt{S}_3$), as triad $\{\gamma\gamma, \gamma Z,\gamma h\}$ lines from $\sim70$ GeV WIMPs in LAT data cross-correlated ($\mathtt{C}_{\mathtt{W}}$) with western eROSITA X-rays, and as a nonad (combining three triads) of two cross-annihilating WIMPs in $\mathtt{C}_{\mathtt{W}}$.
A blind scan of LAT 16.3-year data around 23,343 high-latitude clusters in their rest frames finds three triads, at 70 GeV ($2.8\sigma$ trial-corrected, henceforth), $72$ GeV ($1.3\sigma$; strengthening to $2.7\sigma$ if mass-fixed by $\mathtt{S}_3$, $4.0\sigma$ combined with $\mathtt{C}_{\mathtt{W}}$ masked around the clusters, $5.7\sigma$ if also combined with $\mathtt{S}_3$), and 74--75 GeV: a triplet not only rare ($3.5\sigma$) but also forming a $\{70,74\}$ GeV nonad ($3.4\sigma$; strengthening to $4.0\sigma$, $4.6\sigma$, and $6.1\sigma$, respectively).
The spectral features, robust in and out of the $\mathtt{C}_{\mathtt{W}}$ region and $\mathtt{S}_3$ clusters, appear in all Galactic quadrants and only in the cluster rest frame (vanishing upon redshift rescaling or permutation),
fade away from cluster centers, are absent from their surrounding stacked fields,
and like $\mathtt{C}_{\mathtt{W}}$ indicate intrinsic few $10^{-19}$ cm$^3$ s$^{-1}$ cross sections (uncorrected)
across samples. All features are largely explained by cross-annihilating $69.7^{+0.4}_{-1.0}$ and $74.5^{+1.1}_{-0.6}$ GeV WIMPs.
\end{mdframed}
\end{abstract}

\maketitle

\section{Introduction}
\label{sec:Intro}

About five sixths of the matter in the Universe is non-baryonic, or dark, as inferred from the dynamics of stars, galaxies, and cluster members \cite{Kapteyn1922, Oort1932, Zwicky33, Babcock1939, RubinFord70}, from gravitational lensing \cite{TysonEtAl90, CloweEtAl06}
and X-ray-emitting intracluster gas \cite{FabricantEtAl80,KeshetEtAl21}, and from the growth of large-scale structure (LSS) and the cosmic microwave background (CMB) \cite{BlumenthalEtAl84, Peebles1982, SmootEtAl92}.
Weakly interacting massive particles (WIMPs) \cite{LeeWeinberg77} remain a leading dark matter (DM) candidate \cite{BertoneHooper18, ArbeyMahmoudi21}, pursued by direct detection \cite{AalbersEtAl23LZ, AprileEtAl23XENONnT}, by collider searches \cite{AadEtAl21ATLAS, SirunyanEtAl21CMS}, and indirectly through WIMP annihilation products \cite{Ackermann2015_FermiLAT, AbramowskiEtAl12HESS, AartsenEtAl17IceCube, AguilarEtAl16AMS}.

The leading annihilation channel may be velocity dependent, for example because $s$-wave annihilation is suppressed or forbidden \cite{GriestSeckel91, KumarMarfatia13, BerlinEtAl14, DAgnoloRuderman15}, as motivated by CMB and dwarf-galaxy constraints \cite{AghanimEtAl20Planck, Ackermann2015_FermiLAT}.
For $p$-wave annihilation, the cross section scales as the squared relative velocity, and the annihilation rate is proportional to an effective cross section \cite{DiamantiEtAl14, BoddyEtAl18}
\begin{equation}\label{eq:EffCS}
\overline{\sigma v}_{\text{eff}} = \overline{\sigma v}_p \,\beta^2 \,B \coma
\end{equation}
where $\overline{\sigma v}_p$ is the intrinsic cross section, $\beta^2$ is the line-of-sight velocity (normalized to the speed $c$ of light) dispersion squared (six times smaller than its relative-velocity counterpart), and $B$ is a substructure boost factor.

The deep potential wells of galaxy clusters, with $\beta^2\sim10^{-5}$, impose stronger $p$-wave limits\cite{AckermannEtAl15Clusters, AbramowskiEtAl12HESS} than $\beta^2\sim10^{-9}$ dwarf spheroidals (some \v{C}erenkov limits\cite{AleksicEtAl10MAGIC, VERITAS12_Coma} are less constraining).
Barring the Galactic center ($\beta^2\sim10^{-7}$), whose limits are more model dependent \cite{KarwinEtAl17, JohnsonEtAl19pWave}, clusters yield the strongest limits on $\overline{\sigma v}_p$: by Eq.~\eqref{eq:EffCS}, the best cluster limits on continuum emission \cite{AckermannEtAl15Clusters, AbramowskiEtAl12HESS} correspond to $\overline{\sigma v}_p\sim10^{-18}\cm^3\se^{-1}$, down to $\sim10^{-19}\cm^3\se^{-1}$ for $B>1$ models\cite{AckermannEtAl15Clusters}.
Such large intrinsic cross sections are natural in models with enhanced late-time annihilation, such as Sommerfeld or Breit--Wigner enhancement \cite{HisanoEtAl04, ArkaniHamedEtAl09, IbeEtAl09}, or with a relic abundance not set by thermal freeze-out \cite{KamionkowskiTurner90, GelminiGondolo06, MoroiRandall20, KaplanEtAl09}.

Monochromatic $\chi\chi\to\gamma\gamma$, $\gamma Z$, and $\gamma h$ lines form a triad with energies fixed by the mean mass $\overline{m}_\chi$ of the two annihilating WIMPs (possibly of different species), providing a distinct fingerprint \cite{BergstromSnellman88, JacksonEtAl10}.
Searches for such lines in the \emph{Fermi} Large Area Telescope (LAT) and DAMPE data toward the Galactic center\cite{AckermannEtAl13, AckermannEtAl15Lines, AlemannoEtAl22Lines} and nearby galaxy clusters\cite{AndersonEtAl16} placed upper limits on the signal.
A deeper study\cite{Keshet26Corr}, applying matched filters to the cross-correlation of 16.3 years of LAT data with the eROSITA X-ray map of the western Galactic hemisphere (henceforth \SCW; Appendix~\ref{app:Notation} summarizes all notations), found $\overline{m}_\chi c^2\simeq71.9$ and $67.5\GeV$ triads (before redshift corrections discussed below), each at a local $Z$-score $Z_l\simeq3.0$ and $1.8\sigma$ and $1.7\sigma$ globally
(trial corrected, henceforth whenever using $\sigma$ notation).

Interestingly, the $\gamma Z$ channel of the $\sim72\GeV$ triad coincides with a $43.2\GeV$ line (denoted $L_{43}$)
found \cite{FanEtAl26} by stacking the Virgo, Fornax, and Ophiuchus clusters (denoted \SCT) at $4.3\sigma$ ($2.7\sigma$ under different morphology and trial assumptions\cite{Profumo26}, denoted \SCTb).
At the mass implied by the $43.2\GeV$ line, the triad reaches $3.0\sigma$; Fisher\cite{Fisher25}-combining these observations supports $\overline{m}_\chi c^2\simeq72.1\GeV$ WIMPs at $5.1\sigma$ ($3.8\sigma$) using \SCT\ (\SCTb).
The triad appears in different Galactic sectors, strengthens under cosmological (but not negative) redshift $z$ broadening, peaks at $\Delta z\simeq 0.05$ consistent with the expectation from LSS, and is absent from the ambient (much deeper, X-ray-uncorrelated) spectrum.
The $43\GeV$ line does not carry the triad: its two companion channels reach $2.4\sigma$ by themselves.

A nonad, largely combining the leading $71.9$ and $67.5\GeV$ triads (with an additional, intermediate, weaker triad), was blindly found in \SCW\ at $2.5\sigma$, rising to $2.9\sigma$ with one mass fixed by $L_{43}$, and to $5.0\sigma$ ($3.8\sigma$) when combined with \SCT\ (\SCTb).
This leading \SCW\ nonad, consisting of the nine redshifted lines of two different self- and cross-annihilating WIMP species,
gives $\{m_1,m_2\} c^2\simeq\{66.4,71.9\}\GeV$.
The higher mass, corresponding to $72.2$--$72.6\GeV$ when redshift-corrected (for a non-dominant $\gamma\gamma$ line),
coincides with the $L_{43}$-implied mass.
Alternatively, the leading triads, if emitted at the anticipated LSS peak $z\simeq0.03$, correspond for dominant $\gamma\gamma$ channels to $\{m_1,m_2\}c^2\simeq\{69.5,74.1\}\GeV$ WIMPs, here with the cross-annihilation, $\sim71.8\GeV$ triad coincident with the $L_{43}$ mass.

A cross-correlation analysis like \SCW\ cannot distinguish between different emitting LSS, and it blends their unknown redshifts.
Stacking cataloged clusters instead allows shifting the spectrum of each cluster to its own rest frame before co-addition.
A cluster line then adds coherently, whereas features fixed in the observer frame, such as foregrounds and instrumental artifacts, are smeared by the spread of cluster redshifts.
This coherence renders the stacked rest-frame spectrum a demanding test of the reported triads, largely immune to systematics.

Co-adding LAT photons around many galaxy clusters, without weights or emission templates, averages down the foreground and background (henceforth field) emission and exposes faint cluster signals, such as the Compton emission of virial shocks\cite{ReissKeshet18}, the hadronic emission of cluster cores\cite{Keshet25PaperII}, and the peripheral quenching and inward emergence of excess discrete sources \cite{KeshetPartIII26}.
Here we stack the centers of clusters from three large catalogs at high spectral resolution in the rest frame, searching for sharp spectral features such as DM annihilation lines, with enough photons to probe the intrinsic $p$-wave cross sections of lines below the previous continuum limits.

After outlining the methodology in \S\ref{sec:Methodology}, both blind and prior-based stacking results are presented in \S\ref{sec:Results},
and their properties are quantified and shown to be consistent with cluster WIMP signals in \S\ref{sec:Tests}.
A joint analysis combining stacking and a version of \SCW\ masked around the stacked clusters is presented in \S\ref{sec:Joint}. The overall results are modeled by two cross-annihilating WIMP species in \S\ref{sec:Model} and summarized in \S\ref{sec:Summary}.

\section{Methodology}
\label{sec:Methodology}

We reduce a clean, high spectral-resolution cut of LAT photons and comprehensively mask all cataloged \gama-ray sources (see \S\ref{subsec:FermiLAT}).
These photons are stacked around cataloged cluster centers, after subtracting the local field of each cluster and shifting its spectrum to its rest frame (\S\ref{subsec:Stacking}).
The same stacking without the shift, \ie in the observer frame, serves as a control.
The cluster sample consists of disjoint sets drawn from three catalogs: MCXC\cite{PiffarettiEtAl11}, eROSITA\cite{BulbulEtAl24}, and DESI\cite{WenHan24}.
Fairly low redshift ($z<0.5$), high quality (based on contamination probability or richness), high-latitude clusters are selected and grouped into samples by catalog and sky region (\S\ref{subsec:Samples}).

The stacked spectrum, with errors estimated from the field-expected counts, is detrended against smooth residuals in order to isolate narrow features (\S\ref{subsec:Detrending}).
The detrended spectrum is then analyzed with the same native-resolution spectral apertures (matched filters; \S\ref{subsec:MatchedFilters}) used in the \SCW\ analysis to detect single lines and kinematically constrained (triad and nonad) line complexes.
Besides blind scans, we also consider spectral features a priori fixed by the $L_{43}$ line or the \SCW\ triad harboring it, including companion lines and associated nonads (\S\ref{subsec:FixedTests}).

Each statistic is calibrated by Poisson Monte Carlo simulations that match the pipeline, including binning, photon statistics, spectral detrending, and the handling of photons shared by clusters that overlap on the sky.
Stacks at random cluster longitudes and with permuted cluster redshifts serve as empirical checks (\S\ref{subsec:Significance}).

The brightness of each line is converted into an annihilation cross section by assigning a cusped or cored DM profile to each cluster based on its cataloged parameters (\S\ref{subsec:CrossSections}), and corrected, where so labeled, for the part of the signal lost in the pipeline, mostly in  detrending (\S\ref{subsec:Losses}).

The results are later tested (in \S\ref{sec:Results}, \S\ref{sec:Tests}, and Appendix~\ref{app:Dz}) by
rebinning, permuting the cluster redshifts (which should erase rest-frame features), scaling or broadening the redshift correction, and contrasting independent cluster subsamples: separate catalogs, inside and outside the \SCW\ region, separate hemispheres, redshift ranges, and bins of predicted signal, X-ray-weighted signal (Appendix~\ref{app:Samples}), or mass, all of which should share the same cross section for a true WIMP signal.

Controls (\S\ref{sec:Tests}, Appendix~\ref{app:Nulls}) shift cluster locations, stack the field alone, examine quieter regions (at higher, $|b|>30\dgr$ latitudes and farther from any cluster in the sample), remove individual clusters and sky regions (including Virgo, Fornax, and the Magellanic Clouds), and vary the analysis choices.
Mock signals injected into simulated backgrounds quantify the biases and losses of the pipeline (Appendix~\ref{app:Mock}).

\subsection{Fermi-LAT data, binning, and source mask}
\label{subsec:FermiLAT}

Archival Pass 8 (P8R3; Ref.~\citenum{BruelEtAl18Pass8R3}) LAT data are used from mission weeks 9 through 859 ($\sim16.3$ years), in the highest-purity \texttt{ULTRACLEANVETO} class (\texttt{evclass=1024}), reduced with the Fermi Science Tools (version \texttt{2.2.0}) with the \texttt{P8R3\_ULTRACLEANVETO\_V3} exposure.
Earth-limb photons are removed by a $90\dgr$ zenith-angle cut \cite{AbdoEtAl09EarthLimb}, and good time intervals follow the standard \texttt{(DATA\_QUAL > 0 \&\& LAT\_CONFIG==1)} filter.
Photons are binned on an order-10 HEALPix grid \cite{GorskiEtAl05} of pixel solid angle $\delta\Omega\simeq1.0\times10^{-6}\sr$ and $\sim3\mindot4$ pixel separation, enclosing $\sim10$ pixels within each cluster's region of interest (ROI), defined below as the radius $0\dgrdot1$ disk around its center.

The energy range $0.1$--$1000\GeV$ is divided into 256 logarithmic energy bins (native bins), each of fractional width $s_0\simeq3.7\%$, which
oversamples the $\sim7\%$ LAT $68\%$ energy containment in the $4$--$100\GeV$ range \cite{AckermannEtAl15Lines}.
The \SCW\ analysis examined a similar, as well as a $\sim$doubly refined, energy resolution, finding that both resolve the same features, which somewhat strengthen at the refined resolution\cite{Keshet26Corr}.
Spectral features are scanned in the 3--$100\GeV$ range of the stacked spectrum (single lines from $4\GeV$, and the channels of triads and nonads down to $3\GeV$; see \S\ref{subsec:Significance}). At lower energies, the point spread function (PSF) is too broad for the small ROIs, and at higher energies, photons are scarce and increasingly attenuated by the extragalactic background light at the cluster redshifts \cite{FranceschiniEtAl08, DominguezEtAl11}.

All pixels within $\theta_{\min}=0\dgrdot5$ (the $68\%$ PSF containment at $\sim2\GeV$) of a source in the LAT 4FGL-DR4 catalog \cite{AbdollahiEtAl20, FGL4DR4} are masked. The mask removes sources that could otherwise imprint line-like features on the stacked spectrum, as reported in the spectra of individual active galactic nuclei \cite{ZhouEtAl21, KangEtAl26a, Kang26b, KangEtAl26c}.
Sensitivity tests to the binning,
$\theta_{\min}$,
and all other analysis parameters introduced above and below indicate that our conclusions do not hinge on any particular choice (\S\ref{subsec:RobustTests}, Appendix~\ref{app:Nulls}).

\subsection{Field subtraction and rest-frame stacking}
\label{subsec:Stacking}

\begin{table*}[htb!]
    \centering
    \begin{talltblr}[
        caption = {Stacking analysis parameters and cuts (catalogs and samples of \S\ref{subsec:Samples}).\label{tab:StackParameters}
},
    ]{        width = \textwidth,
        colspec = {|l | c | c | c | c | c | c | c | c |},
        colsep = 4pt, rowsep=1pt,
        column{1} = {rightsep=8pt},
        hline{1,Z} = {0.08em},
        hline{2,5,11,13,15} = {0.05em},
        cell{11}{3} = {c=7}{c},
        cell{12}{3} = {c=7}{c},
    }
        Parameter & Symbol & MCXC & $\Delta$eROSITA & $\Delta$DESI & \SAll & \SE & \SW & \SO \\
        Maximal redshift & $z_{\max}$ & 0.5 & 0.5 & 0.5 & --- & --- & --- & --- \\
        Quality & --- & --- & $\texttt{PCONT}<0.5$ & $\lambda_{500}>40$ & --- & --- & --- & --- \\
        Latitude cut & $|b|_{\min}$ & $10\dgr$ & $15\dgr$ & $10\dgr$ or\TblrNote{\dag} $15\dgr$ & --- & --- & --- & --- \\
        Number of clusters & $N_c$ & 1628 & 7833 & 13882 & 23343 & 10274 & 13069 & 14992 \\
        Fully masked ROIs & --- & 13.7\% & 8.6\% & 8.3\% & 8.8\% & 9.4\% & 8.3\% & 9.2\% \\
        Masked ROI pixels & --- & 16.6\% & 11.5\% & 11.3\% & 11.8\% & 12.5\% & 11.1\% & 12.3\% \\
        Median redshift & med$(z)$ & 0.131 & 0.238 & 0.360 & 0.303 & 0.340 & 0.279 & 0.311 \\
        Median mass, catalog\TblrNote{\S} & med$(M_{500}/10^{14}M_{\odot})$ & 1.7 & 1.7 & 2.3 & 2.2 & 2.3 & 2.1 & 2.2 \\
        Median radius, from $M_{500}$\TblrNote{\P} & med$(R_{500}/\textrm{Mpc})$ & 0.82 & 0.79 & 0.84 & 0.83 & 0.85 & 0.82 & 0.84 \\
        ROI angle & $\theta_{\MyROI}$ & $0\dgrdot1$ & & & & & & \\
        Field outer angle & $\theta_{f}$ & $1\dgrdot0$ & & & & & & \\
        Kinematic $\mathcal{J}$-factor \eqref{eq:Jc}\TblrNote{\P} & $\mathcal{J}/(10^{5}M_\odot^2\,\textrm{kpc}^{-5})$ & 2.25 & 3.18 & 2.82 & 8.25 & 3.09 & 5.16 & 4.97 \\
        Effective $\mathcal{J}$-factor \eqref{eq:Jtilde}\TblrNote{\ddag\P} & $\tilde{\mathcal{J}}/(M_\odot^2\,\textrm{kpc}^{-5})$ & 69.7 & 21.1 & 16.1 & 21.4 & 21.1 & 21.6 & 22.7 \\
        Retained fraction, $\gamma\gamma$ line & $\mathcal{F}_{\gamma\gamma}$ & 0.60 & 0.60 & 0.61 & 0.62 & 0.60 & 0.61 & 0.61 \\
        Retained fraction, $\gamma Z$ line & $\mathcal{F}_{\gamma Z}$ & 0.60 & 0.60 & 0.60 & 0.61 & 0.60 & 0.61 & 0.60 \\
        Retained fraction, $\gamma h$ line & $\mathcal{F}_{\gamma h}$ & 0.79 & 0.78 & 0.78 & 0.77 & 0.79 & 0.77 & 0.78 \\
    \end{talltblr}
    \TblNote{$^\dagger$Only one cluster at $|b|<15\dgr$.\\
    $^\ddagger$Converts the stacked brightness to a cross section (Eq.~\eqref{eq:SigmaStack}); computed with the $\psi_\Myc$ of the NFW profile, for the $\gamma\gamma$ line (slightly energy dependent) of the fixed-mass triad (\S\ref{subsec:FixedTests}). \\
    $^\S$Over the clusters with a valid catalog mass (all but 949 $\Delta$eROSITA clusters).\\
    $^\P$Using estimated masses for the 949 $\Delta$eROSITA clusters without a valid catalog mass (\S\ref{subsec:CrossSections}).
    }
\end{table*}

Our stacking procedure follows Refs.~\citenum{ReissKeshet18, Keshet25PaperII, KeshetPartIII26}, but focuses here on the very centers of clusters, with a much finer spectral binning, and after the photons in each ROI are corrected for the redshift of the cluster in order to keep rest-frame spectral features aligned.

For each channel $\Myfr$ of photon energy $\epsilon$, cluster $\Myc$, and non-masked HEALPix pixel $\Myh$, let
\begin{equation}\label{eq:DeltaN}
  \Delta \Myn_{\Myfr}(\Myc,\Myh) = \Myn_{\Myfr}(\Myh)-\Myf_{\Myfr}(\Myc,\Myh)
\end{equation}
be the excess in the number $\Myn$ of detected photons with respect to the expected number $\Myf$ of field photons.
The observer-frame (redshifted) excess brightness of a cluster's ROI, per unit $\ln\epsilon$ (i.e., $\epsilon\,I_\epsilon$), becomes
\begin{equation}\label{eq:Ic}
  \Delta I_{\Myfr}(\Myc)= \frac{\epsilon_{\Myfr}^2 \sum_{\Myh\in \MyH} \Delta \Myn_{\Myfr}(\Myc,\Myh)}
  {\delta\epsilon_{\Myfr}\,\delta\Omega \sum_{\Myh\in \MyH} \mathcal{E}_{\Myfr}(\Myh) } \coma
\end{equation}
with $\epsilon_{\Myfr}$ the mean energy of channel $\Myfr$ (the geometric mean of its edges) and $\delta\epsilon_{\Myfr}$ its width, $\MyH(\Myc)$ the set of non-masked HEALPix pixels in the ROI of cluster $\Myc$, and $\mathcal{E}_{\Myfr}(\Myh)$ the exposure (effective area times livetime, in $\cm^2\se$) of pixel $\Myh$ at the energy of channel $\Myfr$ (averaged between its two edges).

The ROI of each cluster is identically defined, as a disk about the center of the cluster with radius $\theta_{\MyROI}\simeq0\dgrdot1$: the $68\%$ LAT containment radius at $30$--$100\GeV$, providing a good signal-to-field photon ratio.
The field is defined for simplicity based on the mean photon count
over the non-masked pixels $\Myh'$ in an annulus extending from $\theta_{\MyROI}$ to $\theta_f=1\dgrdot0$, such that $\Myf_{\Myfr}(\Myc,\Myh)=\mathcal{E}_{\Myfr}(\Myh)\sum_{\Myh'}\Myn_{\Myfr}(\Myh')/\sum_{\Myh'}\mathcal{E}_{\Myfr}(\Myh')$.
Modest changes to $\theta_f$ or correcting for field curvature using two concentric annuli changes nothing significant on the small relevant scales.

The field excludes the ROIs of all stacked clusters in the combined three-catalog sample, as do the fields of the random-longitude stacks (\S\ref{subsec:Significance}), so each cluster has the same inferred spectrum in every subsample considered.
This choice is mainly for convenience: the ROIs of the other catalogs cover $\sim3\%$ of an MCXC field and $\lesssim1\%$ of an eROSITA or DESI field, and leaving them in the fields would change the single-catalog results by $|\Delta Z_l|<0.05$.
Photons in pixels shared by the ROIs of several clusters are counted in each ROI, so their covariance is included in the errors and simulations (\S\ref{subsec:Detrending}).

Henceforth, consider $\Myfr$ in the rest frame and $\Myfr'$ in the observer frame.
Summing photons over a sample $\MyC$ of $N_c$ clusters gives the mean excess brightness in the observer frame,
\begin{equation}
\label{eq:ExcessIj}
\Delta I_{\Myfr'}(\MyC)
= \frac{\epsilon_{\Myfr'}^2\sum_{\Myc\in\MyC} \sum_{\Myh\in \MyH} \Delta \Myn_{\Myfr'}(\Myc,\Myh)}{\delta\epsilon_{\Myfr'}\,\delta\Omega \sum_{\Myc\in\MyC} \sum_{\Myh\in \MyH} \mathcal{E}_{\Myfr'}(\Myh) } \coma
\end{equation}
binned by observer-frame channel $\Myfr'$.
This standard stacking smears any narrow emission feature over the spread of cluster redshifts, and is used as a control.
Aligning the cluster spectra in their rest frames instead adds sharp features coherently,
so we bin the same photons by their emitted energy, in rest-frame channels $\Myfr$, giving the aligned excess brightness
\begin{equation}
\label{eq:deltaI}
\Delta\tilde{I}_{\Myfr}(\MyC)
= \frac{\sum_{\Myc\in\MyC} (\epsilon_{\Myfr}^2/\delta\epsilon_{\Myfr}) \sum_{\Myh\in \MyH} \Delta \Myn_{\Myfr'}(\Myc,\Myh)}{\delta\Omega \sum_{\Myc\in\MyC} \sum_{\Myh\in \MyH} \mathcal{E}_{\Myfr'}(\Myh) } \fin
\end{equation}

In these equations, $\Myfr'(\Myc)$ is the observer-frame channel containing the redshifted energy $(1+z_{\Myc})^{-1}\epsilon_{\Myfr}$, given the catalog estimate $z_{\Myc}$ of the specific cluster's redshift.
The aligned excess brightness $\smash{\Delta\tilde{I}_\Myfr}$ is thus the exposure-weighted mean of the observed cluster brightness \eqref{eq:Ic}, stacked at the emitted photon energies and enhanced by the respective factors $(\epsilon_\Myfr^2/\delta\epsilon_\Myfr)/(\epsilon_{\Myfr'}^2/\delta\epsilon_{\Myfr'})=\xi_\Myc\equiv\epsilon_\Myfr/\epsilon_{\Myfr'}\simeq1+z_\Myc$.
This factor $\xi_\Myc$ enhancement ensures that all photons in a rest-frame channel $\Myfr$ receive the same weight, $\epsilon_\Myfr^2/\delta\epsilon_\Myfr$, so the clusters contribute in proportion to their excess counts, as they do in each observer-frame channel of Eq.~\eqref{eq:ExcessIj}.
Apart from these factors, $\smash{\Delta\tilde{I}_\Myfr}$ remains in the observer frame, because fully converting it to the emitted brightness would boost each cluster by an additional factor $(1+z_\Myc)^3$, amplifying observer-frame residuals and injecting an uncontrolled variance.

As the channels are logarithmically spaced, the spectrum of each cluster is in practice shifted by the integer number of channels nearest to $\ln(1+z_\Myc)/\ln(1+s_0)$.
The rest-frame energies of all clusters are thus aligned to within $\pm s_0/2$.
This residual misalignment, uniform within $\pm1.8\%$ ($1.1\%$ rms), broadens a line by only $\sim1\%$, in quadrature with the $\sim7\%$ LAT energy resolution, and is included in the mock tests (Appendix~\ref{app:Mock}).

More generally, replacing $(1+z_\Myc)^{-1}$ by $(1+z_\Myc)^{-\alpha}$ in the definition of $\Myfr'(\Myc)$, i.e., shifting each spectrum by $\alpha\ln(1+z_\Myc)$ in $\ln\epsilon$, interpolates between the observer frame \eqref{eq:ExcessIj} for $\alpha=0$, and (assuming accurate catalog redshifts) the cluster frame \eqref{eq:deltaI} for $\alpha=1$.
A line emitted by the clusters is sharpest at $\alpha=1$, whereas an instrumental or foreground feature is sharpest at $\alpha=0$.
For a wide range of redshifts, true WIMP annihilation lines would emerge only as $\alpha$ approaches unity, and diminish for larger $\alpha$: a test that most artifacts and systematics would fail.

The field, being mostly local (Galactic emission, Solar System foregrounds, instrumental noise), is subtracted in the observer frame, cluster by cluster, before the redshift alignment.
Without this subtraction, detrending would have to absorb the spectral structure of the field, which is brighter than the cluster excess.

Finally, both stacked spectra are detrended (\S\ref{subsec:Detrending}) to remove any smooth residual continuum. Denoting detrending by $\delta$, the resulting observer-frame ($\alpha=0$) and cluster-frame ($\alpha=1$) spectra are respectively $\delta I_0\equiv\delta(\Delta I_{\Myfr'})$ and $\delta I\equiv\delta(\Delta\smash{\tilde{I}}_\Myfr)$.
While these two quantities differ in channel labels, observer-frame $\Myfr'$ vs. rest-frame $\Myfr$, and the observed brightness \eqref{eq:Ic} enters $\Delta\smash{\tilde{I}}$ enhanced by one factor of $\xi_\Myc$,
both stacked spectra combine the same photon counts and exposures, weighting all photons within a bin equally.

\subsection{Cluster catalogs, cuts, and samples}
\label{subsec:Samples}

As the MCXC, eROSITA, and DESI catalogs overlap, we remove from eROSITA every cluster within $0\dgrdot02$ of an MCXC cluster, and from DESI every cluster within $0\dgrdot02$ of an MCXC or eROSITA cluster.
A more conservative exclusion of neighboring clusters is used in \S\ref{subsec:Scaling}.
The resulting disjoint catalogs, denoted MCXC, $\Delta$eROSITA, and $\Delta$DESI, are analyzed separately and later co-added.

Clusters are restricted to $z<0.5$, avoiding excessive redshift corrections.
Rest-frame spectral features above $\sim15\GeV$ are then observed above $10\GeV$, where the LAT angular resolution varies slowly with energy, thus mitigating the energy-dependent bias due to the fixed ROI.
The numerous ($1.6\times10^6$) DESI clusters are filtered to the subset with spectroscopic redshifts, keeping the stacking manageable.

We select high-quality clusters, with contamination probability $\texttt{PCONT}<0.5$ in $\Delta$eROSITA and richness $\lambda_{500}>40$ in $\Delta$DESI.
Clusters are selected with latitudes $|b|>|b|_{\min}$, with $|b|_{\min}=15\dgr$ limiting Galactic-plane contamination.
For the smaller MCXC, this cut is relaxed to $|b|_{\min}=10\dgr$, taking advantage of its generally nearer and better studied clusters; this choice slightly raises the relative contribution of MCXC to the signal (see \S\ref{subsec:CrossSections}).

In addition to the full sample (denoted \SAll) and a sample containing each catalog separately, we also consider all clusters in the eastern (\SE) and western (\SW) Galactic hemispheres, and all clusters whose ROIs lie outside (\SO) the \SCW\ sky region, the latter spanning longitudes $\sim180\dgr$--$310\dgr$ at $|b|\ge30\dgr$.
Thus, \SE\ and \SO\ share no ROI photons with \SCW,
so their results can be combined with those of \SCW.
The samples are summarized in Table~\ref{tab:StackParameters} and their composition is given in Appendix~\ref{app:Samples}.

The \SCW\ analysis, if masked (and then denoted \SMW) within $0\dgrdot1$ of any stacked cluster, does not use any of the ROI photons in any of our samples.
Our joint analysis (denoted \SJ; see \S\ref{sec:Joint}) pairs \SAll\ with \SMW, so the corresponding two $Z_l$ values inferred at each energy or mass can be co-added.

The eROSITA catalog covers mainly the western Galactic hemisphere, so \SE\ consists of MCXC, $\Delta$DESI clusters, and only two $\Delta$eROSITA clusters near the boundary of the eROSITA sky.
Virgo and Fornax, where $L_{43}$ was reported, contribute negligibly: removing them, together with the two $z<0.01$ $\Delta$eROSITA clusters within their $\theta_{500}$,
changes the $Z_l$ of the main triads we report (in \S\ref{sec:Results}) by $\le0.02$.
Ophiuchus lies below the latitude cut.

Here, subscript $500$ refers (henceforth) to the radius $R_{500}$, enclosing a mean density $500$ times the critical density of the Universe at the cluster redshift $z_\Myc$.
The angle $\theta_{500}$ subtended by $R_{500}$ is taken from the catalogs when provided, or from the estimated $R_{500}$ otherwise.
We adopt a flat Planck cosmology \cite{AghanimEtAl20Planck}, of Hubble constant $H_0=67.4\,\mathrm{km\,s^{-1}\,Mpc^{-1}}$,
and density parameters $\Omega_b=0.049$ of baryons, $\Omega_c=0.265$ of cold DM, and $\Omega_m=\Omega_c+\Omega_b$ of all matter.

\subsection{Null errors and spectral detrending}
\label{subsec:Detrending}

The error $\sigma_\Myfr$ of channel $\Myfr$ of the aligned stacked spectrum is the standard deviation of $\Delta\tilde{I}_\Myfr$ under the null hypothesis that every ROI detects only field photons.
The photon counts of different pixels are then independent Poisson variables, of mean $\Myf_{\Myfr'}(\Myc,\Myh)$ in the ROI of cluster $\Myc$, and Eqs.~\eqref{eq:DeltaN} and \eqref{eq:deltaI} give
\begin{eqnarray}\label{eq:Var}
\sigma_\Myfr^2 & = & \left[\frac{\epsilon_\Myfr^2/\delta\epsilon_\Myfr}{\delta\Omega\sum_{\Myc\in\MyC}\sum_{\Myh\in\MyH}\mathcal{E}_{\Myfr'}(\Myh)}\right]^2\\
& & \times\sum_{\Myc\in\MyC}\left[\sum_{\Myh\in\MyH}\nu_\Myh\,\Myf_{\Myfr'}(\Myc,\Myh)+\frac{\Myf_{\Myfr'}^2(\Myc,\MyH)}{\Myn_{\Myfr'}(\MyH_\Myf)}\right] \nonumber \fin
\end{eqnarray}
Here, $\Myf_{\Myfr'}(\Myc,\MyH)\equiv\sum_{\Myh\in\MyH}\Myf_{\Myfr'}(\Myc,\Myh)$ is the field-expected photon count in the ROI of cluster $\Myc$, $\Myn_{\Myfr'}(\MyH_\Myf)\equiv\sum_{\Myh'\in\MyH_\Myf}\Myn_{\Myfr'}(\Myh')$ is the count detected in its field annulus $\MyH_\Myf(\Myc)$, and $\nu_\Myh$ is the number of clusters in $\MyC$ whose ROIs contain pixel $\Myh$ and whose spectra are shifted by the same number of channels as that of $\Myc$.

The first term in the last brackets is the Poisson variance of the field photons expected in the ROI.
A photon in a pixel shared by a number $\nu_\Myh$ of such ROIs is counted $\nu_\Myh$ times in channel $\Myfr$. Summed over these $\nu_\Myh$ clusters, the term $\nu_\Myh\Myf_{\Myfr'}(\Myc,\Myh)$ thus gives $\nu_\Myh^2$ rather than $\nu_\Myh$ times its variance.
The second term is the variance of the field estimate,
smaller than the first by the $\sim1\%$ ratio of ROI-to-field exposures.
The errors $\sigma_{\Myfr'}$ of the observer-frame stacked spectrum \eqref{eq:ExcessIj} follow from Eq.~\eqref{eq:Var} but with $\epsilon_{\Myfr'}$ and $\delta\epsilon_{\Myfr'}$ in the prefactor, and with $\nu_\Myh$ counting all the ROIs in $\MyC$ that contain $\Myh$.
As $\sigma_\Myfr$ depends only on the fields, exposures, and ROI overlaps, it is independent of the photons detected in the ROIs.

The detrending $\delta$ removes the smooth continuum of a stacked spectrum in two steps, using all channels with field photons (well beyond the $3$--$100\GeV$ analysis range).
First, the stacked field spectrum is fitted as the exponential of a cubic polynomial in $\ln\epsilon$.
A Gaussian process regression (GPR) \cite{RasmussenWilliams06} smooths the channel-by-channel ratio of $\Delta\smash{\tilde{I}}_\Myfr$ to this fit.
The outcome, multiplied by the fit, estimates the continuum of $\Delta\smash{\tilde{I}}_\Myfr$, which is subtracted to reveal narrow spectral features.

The GPR (see Appendix \ref{app:Detrending}) fits no functional form: its two hyperparameters define only its prior covariance, and only one of them --- the correlation length --- is fixed a priori. We choose it sufficiently large ($0.27$ dex in energy: $\sim9$ times the energy half-width of a line dispersed by the LAT) such that narrow lines pass largely intact, whereas broader features are removed with the continuum.
The second hyperparameter --- the prior covariance amplitude --- is set to the value that maximizes the marginal likelihood of the data \cite{RasmussenWilliams06}.

In the second step, a linear function of $\ln\epsilon$, fitted by least squares with weights $\sigma_\Myfr^{-2}$ to the $3$--$100\GeV$ residuals, is subtracted from all channels, giving $\delta I_\Myfr$.
This subtraction sets the weighted mean and slope of $\delta I_\Myfr$ over the analysis range to zero, which the GPR, fitted over the much wider range, need not do.
The observer-frame $\delta I_0$ follows likewise from $\Delta I_{\Myfr'}$, with its own field fit and errors.
See Appendix~\ref{app:Detrending} for details.

Detrending absorbs part of any real line, an effect we conservatively neglect throughout the nominal analysis (except for loss-corrected cross sections; \S\ref{subsec:Losses}).
A more careful approach is channel-masked detrending, which gives zero weight, in both steps above, to the tested lines.
We adopt this approach in \S\ref{subsec:SumSig}, where the continuum is interpolated after excluding $[0.95,1.05]$ (nominally; about $1.5$ standard deviations of the IRF core) times each line energy of every statistic (see Appendix~\ref{app:Nulls}).

For fixed hyperparameters and field, both steps are linear in $\Delta\smash{\tilde{I}}$: the GPR estimate is then a weighted linear combination of the data,
with weights independent of the ROI photons.
Then $\delta I_\Myfr=\sum_j\mathcal{D}_{\Myfr j}\Delta\tilde{I}_j$, with a fixed detrending matrix $\mathcal{D}$ for each sample.

The null calibrations of \S\ref{subsec:Significance} pass every realization through the same detrending procedure: the Poisson Monte Carlo applies the $\mathcal{D}$ of the data, and the random-longitude stacks and redshift permutations rebuild it for each realization (Appendix~\ref{app:Detrending}).
These calibrations also include the covariances that Eq.~\eqref{eq:Var} omits. Monte Carlo includes all of them except the covariance between overlapping field estimates,
as it does not draw the noise of the field estimates ($\lesssim2\%$ of the cluster-frame null variance; Appendix~\ref{app:Detrending}).
The detrended spectra, divided by $\sigma_\Myfr$, have standard deviations of $0.9$--$1.1$ over $3$--$100\GeV$ in all samples of Table~\ref{tab:StackParameters}.

\subsection{Line, triad, and nonad matched filters}
\label{subsec:MatchedFilters}

Spectral features are scanned using the same matched filters used in the \SCW\ analysis.
These apertures are based on the energy dispersion of the LAT \texttt{P8R3\_ULTRACLEANVETO\_V2} instrument response functions (IRFs; applicable also to \texttt{V3}), summed over event types and averaged over incidence angle with effective-area weights.
For details, see Ref. \citenum{Keshet26Corr}.

The matched filter $\mathbb{M}_i$ of a spectral line $i$ centered at energy $\epsilon_0$ is defined by its weights $w_\Myfr$.
These weights are the averaged IRF of the emission line in channel $\Myfr$, normalized to unit integral over $\ln\epsilon$.
Applied to the detrended stacked spectrum $\delta I=\delta(\Delta\tilde{I})$ of \S\ref{subsec:Stacking}, the filter gives the integrated line brightness $\mathcal{I}_i=\mathbb{M}_i[\delta I]$ and its local $Z$-score,
\begin{equation}\label{eq:MatchedFilter}
\mathcal{I}(\epsilon_0) = \frac{\sum_\Myfr w_\Myfr y_\Myfr}{\sum_\Myfr w_\Myfr^2} \quad \mbox{and} \quad Z_l(\epsilon_0) = \frac{\sum_\Myfr w_\Myfr y_\Myfr}{\sqrt{\sum_\Myfr w_\Myfr^2 \sigma_\Myfr^2 }} \fin
\end{equation}
Here, $\delta I_\Myfr = y_\Myfr\pm\sigma_\Myfr$ is the detrended spectrum (and likewise $\delta I_0$, in channels $\Myfr'$), with the errors $\sigma_\Myfr$ of Eq.~\eqref{eq:Var}, i.e., of the stacked spectrum before detrending, as in the \SCW\ analysis.

Detrending, which fits a continuum to the noisy spectrum, absorbs part of the noise and correlates the channels.
Therefore, in the absence of a signal, $Z_l$ does not follow a unit Gaussian.
The Monte Carlo realizations, random-longitude stacks, and redshift permutations of \S\ref{subsec:Significance}
are detrended and filtered as the data, so their $Z_l$ distribution includes the absorbed noise and the channel correlations.

In the cluster-frame spectrum $\delta I$, the filter of a line of rest-frame energy $\epsilon_0$ uses the IRF at $\epsilon_0$.
The LAT, however, records the photons of this line from cluster $\Myc$ at the lower energy $\epsilon_0/(1+z_\Myc)$, and disperses them with its IRF at that energy.
As the fractional energy resolution of the LAT\cite{AckermannEtAl15Lines} is nearly constant at $4$--$100\GeV$,
and the shift to the $z<0.5$ cluster frame is modest and preserves fractional widths, the stacked line has nearly the same shape as the IRF at $\epsilon_0$.
The mock signals of Appendix~\ref{app:Mock}, dispersed at their observed energies, include this small residual mismatch.

Two WIMPs (of either one or two species) of mean mass-energy $E_\chi\equiv\overline{m}_\chi c^2$, annihilating into a photon and a mass $m_\varphi$ neutral boson $\varphi\in\{\gamma,Z,h\}$, produce a line at
\begin{equation}\label{eq:kin}
\epsilon \simeq E_\chi - \frac{m_\varphi^2c^4}{4E_\chi} \coma
\end{equation}
as dictated by energy-momentum conservation.
The $\gamma\gamma$, $\gamma Z$, and $\gamma h$ lines of a triad are thus kinematically fixed by $\overline{m}_\chi$ alone, given the $m_Zc^2\simeq91.19\GeV$ and $m_hc^2\simeq125.2\GeV$ masses of the $Z$ and Higgs bosons \cite{PDG2024}.
Such triads, from either self- or cross-annihilating WIMPs, fall within our $3$--$100\GeV$ window for $\overline{m}_\chi c^2=64.1$--$100\GeV$; at the low mass end, the $\gamma h$ line approaches our predetermined $3\GeV$ detectability limit.
The three lines of a triad thus span a wide energy range, \eg $17.7$--$72.1\GeV$ for $\overline{m}_\chi c^2=72.1\GeV$, so that a triad filter acts as a natural veto against artifacts localized in energy.

We scan for triads of any $\overline{m}_\chi$ by sliding across the spectrum a matched filter composed of three IRF copies rigidly placed by Eq.~\eqref{eq:kin};
the $\gamma Z$ copy is convolved with the relativistic Breit--Wigner mass distribution of the $Z$ boson \cite{PDG2024}, of width $\Gamma_Z\simeq2.5\GeV$.
The distribution is sampled at $Z$ masses within $\pm5\Gamma_Z$ of $m_Z c^2$, in steps of $\Gamma_Z/2$, each giving an IRF copy on the native grid at the $\gamma Z$ energy of Eq.~\eqref{eq:kin}; a step shifts this energy by less than one native bin.
The local $Z_{l,i}$ of the $n$ lines of a line complex combine to
\begin{equation}\label{eq:CoAddedZ}
Z_l = n^{-1/2}\sum_{i=1}^{n} Z_{l,i} \coma
\end{equation}
with $n=3$ for a triad.
The $Z_l$ values of the triad channels are denoted $\{Z_{\gamma\gamma}$, $Z_{\gamma Z}$, $Z_{\gamma h}\}$.
Where the symbol of a spectral feature (such as $L_{43}$, or its associated triad $T_{72}$ or companions $C_{72}$, defined below) is quoted as a value, it denotes the $Z_l$ of that feature.

Different WIMP species that can both self- and cross-annihilate within the $64\lesssim \overline{m}_\chi c^2/\mbox{GeV}\lesssim100$ window would produce three triads, of masses $m_1$, $m_2$, and their
\begin{equation}\label{eq:nonad}
  \overline{m}_\chi =\frac{m_1+m_2}{2}
\end{equation}
mean, the latter with $\gamma Z$ and $\gamma h$ energies slightly above the arithmetic means of their $m_1$ and $m_2$ counterparts.
The nine lines of these three triads, kinematically fixed by $m_1$ and $m_2$, are referred to as a nonad; several triads do not imply a nonad, but an observed nonad would strongly indicate cross-annihilating WIMPs.

Co-adding the nine lines by Eq.~\eqref{eq:CoAddedZ} with $n=9$, without correcting for their overlaps, would overestimate $Z_l$: the $\gamma\gamma$ (and, to a lesser degree, the $\gamma Z$) lines of adjacent triads overlap within the IRF width, so their $Z_{l,i}$ share photons.
Instead, as in the \SCW\ analysis, our nonad filter applies the matched filter of Eq.~\eqref{eq:MatchedFilter} to a single template, the sum of the nine unit-integral IRF copies: the simplest choice that normalizes the overlaps correctly.
While this template weights the nine lines equally, it does not require them to be equally bright; unequal line brightness levels only reduce the sensitivity of the filter.

The nonad filter thus counts each photon once. In contrast, a part of a nonad purposely filtered alone (a line, a triad, or a pair of triads) also collects the photons of the overlapping lines of the other triads, so its brightness, cross section, and $Z_l$ include photons that it shares with the rest of the nonad.
We therefore quantify the contribution of such a part by its conditional significance $Z_c$ beyond its overlap with the rest of the nonad, computed from the null covariance of the overlapping components (Appendix~\ref{app:JointStats}). Examples are the $Z_c$ of the intermediate triad beyond the two outer triads, and of the outer triads beyond the intermediate one.

Local $Z_l$ maxima are identified on logarithmic grids of $0.2\%$ steps in the line energy $\epsilon_0$ for lines, $0.1\%$ in $\overline{m}_\chi$ for triads, and $0.5\%$ in each mass for nonads. The outer ($m_1$ and $m_2$) triads of a nonad are kept apart by a window $f_t=0.02$, i.e., $m_2/m_1>(1+f_t)/(1-f_t)\simeq1.04$.
The $\Delta\textrm{TS}=1$ bounds of these maxima ($\epsilon_0-\Delta\epsilon_-$ and $\epsilon_0+\Delta\epsilon_+$ for lines), where the test statistic $\textrm{TS}=Z_l^2$ drops by unity, reflect the likelihood geometry rather than Gaussian uncertainties. The dispersion of blindly recovered mock signals (below) gauges the actual uncertainty. For a nonad, the bounds on each mass are profiled, i.e., at each value of one mass, the other mass is varied to maximize $Z_l$.

Multiple features are extracted iteratively, each with an exclusion window $[1-f,1+f]\epsilon_0$ that may not overlap with the windows of stronger, \ie earlier extracted, features. We nominally pick $f=0.1$ for lines, $f=0.015$ for the $\gamma\gamma$ channel of triads, and
$f=0.05$ for nonads, around each of their two masses $m_1$ and $m_2$, i.e., the $\gamma\gamma$ energies of their outer triads (neither window may overlap either window of a previous nonad).
The intermediate triad of a nonad has no window of its own, to allow alternative nonad readings that reuse it as an outer triad.
Close pairs of lines are also probed with $f=0.03$,
an exclusion window spanning roughly the Gaussian core of the IRF without its tails.

The matched filters are tested, and their biases and losses quantified, by injecting mock signals into simulated backgrounds (Appendix~\ref{app:Mock}).
Blind scans recover the energy of an injected line (the mass of an injected triad) with a negligible, $\lesssim0.3\%$ bias and a robust dispersion of $1.5$--$2.6\%$ ($0.4$--$0.9\%$), for signals injected with $3\le Z_l\le5$. This dispersion is $1.483$ times the median absolute deviation (MAD).

\subsection{Fixed-mass tests implied by \SCT\ or \SCW}
\label{subsec:FixedTests}

Besides blind line, triad, and nonad scans, we also test spectral features a priori fixed by the previously reported $\sim43\GeV$ line\cite{FanEtAl26} or $\sim72\GeV$ triad\cite{Keshet26Corr}.
A fixed $\overline{m}_\chi c^2\simeq72.05\GeV$ (henceforth rounded to $72.1\GeV$) mean WIMP mass, and its fixed-mass triad (denoted $T_{72}$), follow approximately from \SCW\ or from adopting $43.2\GeV$ as the emitted energy of a $\gamma Z$ channel (henceforth the $\gamma Z$ reading of $L_{43}$).
The $\pm0.5\GeV$ uncertainty in $L_{43}$ energy ($43.7\pm0.6\GeV$ in a cluster sample\cite{FanEtAl26} larger than \SCT) corresponds to $\pm0.4\GeV$ in WIMP mass.
This mass is consistent with the $71.9\GeV$ triad of \SCW, approximately redshift-corrected\cite{Keshet26Corr} to rest-frame $72.2$--$72.6\GeV$
(see Appendix~\ref{app:eROSITA}).
As \SCT\ and \SCW\ independently imply nearly the same WIMP mass, our fixed-mass tests can use either analysis as prior information.
Testing spectral features with $\overline{m}_\chi$ fixed a priori then avoids its scan and related
multiple-trial (look-elsewhere) penalty.

In addition to the fixed-mass $L_{43}$ and $T_{72}$, we also record the companions (denoted $C_{72}$) of $L_{43}$ in $T_{72}$, namely its $\gamma\gamma$ and $\gamma h$ lines. These lines are sufficiently separated to share no photons with each other or with $L_{43}$, so they are co-added by Eq.~\eqref{eq:CoAddedZ} with $n=2$.
We also record, for completeness, the $Z_l$ of the less likely\cite{Keshet26Corr} $87.8\GeV$ triad implied if $L_{43}$ is read as a $\gamma h$ line (henceforth the $\gamma h$ reading of $L_{43}$).
For tests that include \SCW\ data (\SCW, \SMW, and \SJ), the mass is fixed by $L_{43}$.

Similarly tested are nonads harboring $T_{72}$ as their lower, upper, or intermediate (cross-annihilation) triad, henceforth the low, high, and intermediate $T_{72}$ readings.
The high reading is supported by the modest redshift correction of the \SCW-leading, $\{66.4, 71.9\}\GeV$ nonad.
The intermediate reading is supported both by a second, $\{67.4, 75.6\}\GeV$ nonad $Z_l$ maximum in \SCW, and by a possible $\gamma\gamma$-dominated redshift correction of the two leading \SCW\ triads,
and is thus chosen as our nominal reading; its nonad is denoted $N_{72}$.
Note that for comparable abundances and cross sections, cross-annihilation would give the brightest triad of a nonad, as it involves twice as many pairs as self-annihilation.
We also take the strongest of the three readings of the nonad, denoted $N_{72}^{\max}$, as a statistic.

For each nonad reading, Eq.~\eqref{eq:nonad} confines its two masses to a line in the $\{m_1,m_2\}$ plane, \eg
$(m_1+m_2)c^2=144.1\GeV$ for the intermediate reading.
We scan the free mass along each line in $0.1\%$ steps,
and take the strongest interior $Z_l$ maximum as the statistic.
Edge maxima and nonads with any line outside $(3.5,95)\GeV$, too close to the edges of the analysis range (\S\ref{subsec:Significance}), are discarded.
These selections are made alike in the data and in the nominal null realizations (but see Appendix~\ref{app:Nulls}).

We also test, with no scan, the blind $\{66.4, 71.9\}\GeV$ nonad of \SCW, with both masses nominally fixed at $\{68.4, 74.1\}\GeV$ by correcting for the anticipated\cite{Keshet26Corr} $z=0.03$ peak redshift; different redshift corrections are discussed in \S\ref{subsec:FixedMass} and Appendix~\ref{app:JointStats}.

\subsection{Null distributions and global significance}
\label{subsec:Significance}

We calibrate each statistic of stacking by comparing it with its null distribution, obtained by applying the full pipeline, including detrending and the scan, to many realizations without a cluster signal. Its $p$-value ($p$, henceforth)
in this distribution is quoted as a one-sided Gaussian significance, referred to as the calibrated significance of the statistic.
Each statistic, including the strongest maximum of a scan, is computed alike in the data and in every realization. The calibrated significance of a scan thus includes the correction for multiple trials (the look-elsewhere effect), whereas the fixed-mass tests of $L_{43}$, $T_{72}$, and $C_{72}$ (see \S\ref{subsec:FixedTests}) involve no scan, so every calibrated significance is global.

The calibrated significance is denoted $Z_g$, with a superscript for the checks and approximations, according to the null distribution used:
\begin{enumerate}[leftmargin=12pt, itemsep=1pt]
\item Poisson Monte Carlo simulations --- $Z_g$, nominal; $\sim10^4$--$10^6$ realizations --- keeping the LAT binning, exposure, and photon statistics of each real sample, drawing Poisson ROI counts of field-expected mean (plus any smooth excess removed by detrending), with the photons of each shared pixel drawn once and added to every ROI that contains it.
\item Random cluster longitudes
--- $Z_g^{\smash[t]{(r)}}$, a check; $\sim10^4$ realizations for \SAll; a few hundred for the other samples --- every ROI is moved to a random longitude at its own Galactic latitude, away from all stacked clusters, while preserving the cluster redshift, thus building control clusters from the data.
\item Redshift permutations --- $Z_g^{\smash[t]{(z)}}$, a check; $\sim10^3$--$10^5$ realizations --- redshifts are randomly permuted among the clusters of a sample, destroying the rest-frame alignment while keeping all photons and positions.
\item Gaussian noise --- $Z_g^{\smash[t]{(n)}}$, an approximation --- realizations of Gaussian noise with the measured errors of the native bins, scanned and selected as the data.
\item Analytic \smash{\v{S}id\'{a}k} estimate --- $Z_g^{\smash[t]{(a)}}$, an approximation --- computed from the local $p$-value and the effective number of trials of the scan.
\end{enumerate}

The smooth excess restored in item 1 includes the part of any real line that detrending absorbs, thus raising the null, so this nominal calibration is conservative.
The channel-masked detrending of \S\ref{subsec:SumSig} follows item 1 but restores the data detrended with all tested lines masked. Each masked statistic is then rescaled, in data and null realizations alike, to the null standard deviation of its nominal counterpart; this leaves its own calibration unchanged but keeps its nominal weight in \SJ\ (relative to \SMW) and in the scans (relative to other masses).

Items 2 and 3, built directly from the data, include
structure
that Monte Carlo does not model. Items 4 and 5, quoted for comparison, assume a unit Gaussian null for $Z_l$ (Appendix~\ref{app:Nulls}).
Edge maxima, at grid points with a neighbor outside the scanned range, are dropped from the data.
Also dropped are features with any line below $3.5\GeV$ or above $95\GeV$, too close to the edges of the $3$--$100\GeV$ window for the filter and the detrending to be reliable.

Even for a fixed-mass test, the calibrated significance differs from $Z_l$. Equation~\eqref{eq:MatchedFilter} normalizes $Z_l$ by the errors before detrending, but detrending, which fits the continuum to the noisy spectrum, also absorbs part of the noise that the filter collects. In the null realizations, $Z_l$ thus scatters less than a unit Gaussian
(standard deviation $\sim0.84$ for the $T_{72}$ of \SAll\ in Monte Carlo, and $0.81$ in the random-longitude stacks built from the data), so a given $Z_l$ corresponds to a higher calibrated significance.

The joint analysis \SJ\ co-adds the $Z_l$ values of \SAll\ and of \SMW,
two practically independent datasets, which share no ROI photons, with different null distributions. Each of its joint null realizations, referred to as a null pair, therefore pairs a Monte Carlo realization of stacking
with a shuffled-contrast \SMW\ sky (the \SMW\ sky with its X-ray contrast
shuffled among the unmasked pixels,
added to the smooth continuum of the \SMW\ data,
then passed through
the full \SCW\ pipeline, detrending included), or, as a check giving $Z_g^{\smash[t]{(n)}}$, with a Gaussian realization of \SMW; see Appendix~\ref{app:JointStats}.

Independent tests, such as a fixed-mass statistic of stacking and the $L_{43}$ of \SCT\ or \SCTb, are combined through their $p$-values by Fisher's method.
Every numerically calibrated significance derived here
rests on at least ten (and typically $>50$) null realizations that surpass the observed statistic.

\subsection{Inferring $\overline{\sigma v}_p$ from stacked clusters}
\label{subsec:CrossSections}

Next, the stacked line brightness is related to the annihilation cross section of Eq.~\eqref{eq:EffCS}.
Note that the intrinsic cross section $\overline{\sigma v}_p$ of this equation refers to the $\gamma\gamma$ channel, of photon multiplicity $Y=2$. For a different channel $i$, of multiplicity $Y_i$, one should replace $\overline{\sigma v}_p$ by $(Y_i/2)\overline{\sigma v}_{p,i}$.
Similarly, our $\overline{\sigma v}_p$, defined through the line-of-sight velocity dispersion, would become six times smaller if defined through the 3D relative velocity \cite{Keshet26Corr}.

The annihilation luminosity of a single WIMP species $\chi$, of mass $m_\chi$, mass density $\rho_\chi$, and density parameter $\Omega_\chi$, in a cluster $\Myc$ at luminosity distance $d_L(z)$ is proportional to
\begin{equation}\label{eq:Lchi}
  \int\limits_\Myc\rho_\chi^2\beta^2\,dV \equiv \left(\frac{\Omega_\chi}{\Omega_m}\right)^2 B \int\limits^{\infty} \rho_0^2\beta^2\,dV \equiv \left(\frac{\Omega_\chi}{\Omega_m}\right)^2 B \mathcal{J}_\Myc d_L^2\, \mathcal{S} \coma
\end{equation}
where $\rho_0$ is a smooth mass-density model, the conventional factor
\begin{equation}\label{eq:Jc}
  \mathcal{J}_\Myc\equiv \frac{G M_{500}^3}{32\,c^2 d_L^{2} R_{500}^{4}} \coma
\end{equation}
$G$ is Newton's constant, and the dimensionless shape parameter is $\mathcal{S}\sim 10$ for typical smooth $\rho_0^2\beta^2$ profiles.
The sum $\mathcal{J}\equiv\sum_\Myc \mathcal{J}_\Myc$ over a sample gauges its contribution to the total line flux.

Our nominal smooth reference model is the cusped NFW\cite{NFW97} profile of DM halos, $\rho_0\propto (1+c_{500} r/R_{500})^{-2}/r$, where $\beta^2$ vanishes toward the center under dynamical equilibrium. Solving the isotropic Jeans equation for $\beta^2$, Eq.~\eqref{eq:Lchi} yields $\mathcal{S}_{\text{NFW}}=[4/(3\pi)]\left[36 \zeta(3)-53+\pi ^2\right] (c_{500}+1)^3 c_{500}^4 /\left[(c_{500}+1) \ln (c_{500}+1)-c_{500}\right]^3\simeq 19$ for a concentration parameter $c_{500}\simeq 3$ (with $\mathcal{S}_{\text{NFW}}\simeq12$--$29$ for $c_{500}=2$--$4$), where $\zeta(3) \simeq 1.20$ is Ap\'{e}ry's constant.

We also consider a quasi-isothermal cored profile $\rho_0(r) \propto [1 + (g_c r/R_{500})^2]^{-1}$, with a uniform $\beta^2=GM_{500}/(2c^2R_{500})$.
Here, $\mathcal{S}_{\textrm{core}} \equiv g_c^3/(g_c-\arctan g_c)^2 \simeq 14$ for the core parameter $g_c \simeq 10$ (with $\sim0.2$ dex scatter) of massive clusters inferred from X-ray hydrostatic mass profiles \cite{EttoriEtAl19}, and $\mathcal{S}_{\textrm{core}} \simeq 28$ for $g_c \simeq 25$ (with $\sim0.2$ dex scatter) inferred from strong lensing \cite{RichardEtAl10, CaminhaEtAl19, LimousinEtAl22}.
The former (adopted below) cored profiles yield an annihilation signal slightly weaker than a cusped profile, whose $\beta^2$ vanishes toward the center \cite{BoddyEtAl18}.

We may now write the observed energy flux from cluster $\Myc$ at emitted photon energy $\epsilon_i$ as
\begin{equation}\label{eq:LineFlux}
    F_{i,\Myc} = \frac{Y_i \epsilon_i}{8\pi} \frac{\overline{\sigma v}_{p,i}}{\eta_\chi} \left(\frac{\Omega_\chi}{m_\chi \Omega_m}\right)^2 B\,\mathcal{S}\,\mathcal{J}_\Myc
\end{equation}
in channel $i$ from DM species $\chi$, where the symmetry factor $\eta_\chi$ is 1 for a self-conjugate (Majorana) WIMP or $2$ for a particle-antiparticle (Dirac) pair.
For any reference profile $\mathcal{S}$, the stacked flux determines $\overline{\sigma v}_{p,i}/D$, where the dimensionless $D\equiv(\Omega_c/\Omega_\chi)^2\eta_\chi/B$ collects structure deviations and substructure corrections in $B$, as well as the unknown dark-sector parameters, and is normalized as $D=1$ for a single Majorana species distributed as the reference model ($B=1$).

As the aligned excess brightness \eqref{eq:deltaI} is an exposure-weighted mean over the unmasked ROI pixels, a channel of integrated brightness $\mathcal{I}_i$ implies
\begin{equation}\label{eq:SigmaStack}
\frac{\overline{\sigma v}_{p,i}}{D} = \frac{8\pi}{Y_i\epsilon_i\mathcal{S}}\left(\frac{m_\chi \Omega_m}{\Omega_c}\right)^2 \frac{\pi\theta_\MyROI^2\mathcal{I}_i}{\tilde{\mathcal{J}}} \coma
\end{equation}
where we defined an effective mean $\mathcal{J}$-factor
\begin{equation}\label{eq:Jtilde}
  \tilde{\mathcal{J}}\equiv \frac{\sum_\Myc \xi_\Myc\psi_\Myc \mathcal{J}_\Myc \mathcal{E}_\Myc}{\sum_\Myc \mathcal{E}_\Myc}\coma
\end{equation}
and $\mathcal{E}_\Myc$ is the exposure of cluster $\Myc$ at the observed line energy, summed over its unmasked ROI pixels.
Cluster $\Myc$ thus contributes $\tilde{\mathcal{J}}_\Myc\equiv\xi_\Myc\psi_\Myc\mathcal{J}_\Myc\mathcal{E}_\Myc/\sum_{\Myc'}\mathcal{E}_{\Myc'}$ to $\tilde{\mathcal{J}}$, and a line signal $\propto\tilde{\mathcal{J}}_\Myc$ to the stacked brightness.
The factor $\xi_\Myc\simeq1+z_\Myc$ in $\tilde{\mathcal{J}}$ cancels the corresponding weight in Eq.~\eqref{eq:deltaI} propagated into $\mathcal{I}_i=\mathbb{M}_i[\delta I]$.

The $0\dgrdot1$ ROI collects only part of the annihilation signal of each cluster, especially at low energies, where the PSF is broad; in addition, the source mask removes some ROI and field pixels, changing the part of the signal sampled in each cluster.
These effects are accounted for by the factor
\begin{equation}
\psi_\Myc\equiv\frac{\mathcal{I}_\Myc^{\mbox{\tiny ROI}}-\mathcal{I}_\Myc^{\mbox{\tiny field}}}{F_{i,\Myc}/(\pi\theta_\MyROI^2)} \coma
\end{equation}
where $\mathcal{I}_\Myc^{\mbox{\tiny ROI}}$ and $\mathcal{I}_\Myc^{\mbox{\tiny field}}$ are the modeled line brightness of the cluster's ROI and of its field estimate, respectively, as observed through the PSF, averaged with exposure weights over the unmasked pixels.
Thus, $\psi_\Myc$ tends to the PSF containment of the ROI for compact emission and vanishes for emission much broader than the field annulus; a cluster whose ROI is fully masked has $\mathcal{E}_\Myc=0$, and no weight in Eq.~\eqref{eq:Jtilde}.

We approximate $\psi_\Myc$ accordingly, by projecting the modeled emission along the line of sight, convolving the projected emission with the PSF of the IRF at the observed line energy (summed over event types and averaged over incidence angle, like the energy dispersion; \S\ref{subsec:MatchedFilters}), and averaging the resulting brightness with exposure weights over the unmasked pixels of the ROI and of the field estimate.
As above, we adopt $c_{500}=3$ for the NFW profile and a core angle $\theta_{500}/10$ for the cored profile.
The less concentrated, $\propto\rho_0^2$ cored emission gives a smaller $\psi_\Myc$, which, with $\mathcal{S}_{\textrm{core}}$, raises the cross sections inferred from \SAll\ by a factor
of $\sim1.6$.
For the $\gamma\gamma$ line of the $72.1\GeV$ triad, the NFW profile gives $\psi_\Myc\simeq0.57$--$0.76$ for $90\%$ of the clusters with unmasked ROI pixels (median $0.64$), and the cored profile a median $6\%$ lower.
Table~\ref{tab:StackParameters} lists $\mathcal{J}$ and (for the $\gamma\gamma$ line at $72.1\GeV$) $\tilde{\mathcal{J}}$ in the different samples.

As alluded to in \S\ref{subsec:Samples}, the relaxed latitude cut of MCXC raises its share of $\mathcal{J}$ to $27\%$, comparable to the $39\%$ and $34\%$ of $\Delta$eROSITA and $\Delta$DESI.
However, as nearby clusters extend farther beyond the fixed ROI, the three catalogs contribute respectively $22\%$, $33\%$, and $45\%$ of the $\tilde{\mathcal{J}}$ of \SAll.
The ROIs of 2049 out of the 23,343 clusters ($8.8\%$) lie entirely in 4FGL-masked pixels, so these clusters contribute to $\mathcal{J}$, but not to the signal or to $\tilde{\mathcal{J}}$.

We adopt the $M_{500}$ of each catalog, as reported, with the implied $R_{500}$.
Appendix~\ref{app:MassCal} shows how the results change if the MCXC and $\Delta$DESI masses are placed on the eROSITA mass scale instead.
Estimating $\mathcal{J}_\Myc$, and thus $\tilde{\mathcal{J}}$, requires $M_{500}$, but 949 $\Delta$eROSITA clusters have no valid catalog mass, most of them lacking also $R_{500}$ and an X-ray luminosity $L_{500}$.
Of these, seven clusters are provided with $L_{500}$, so are assigned the $M_{500}$ of the luminosity--mass relation fitted to $\Delta$eROSITA (Appendix~\ref{app:Samples}).
For the remaining 942 clusters, we adopt the median valid $M_{500}$ of the 100 $\Delta$eROSITA clusters nearest in redshift.
These 949 clusters contribute only $\sim3\%$ of the $\tilde{\mathcal{J}}$ of \SAll, and their masses do not affect the stacked spectra.

\subsection{Pipeline losses}
\label{subsec:Losses}

The pipeline loses part of a real signal, mostly through detrending.
In the mock tests of Appendix~\ref{app:Mock}, which inject line and triad signals into simulated backgrounds, the measured brightness and $Z_l$ fall below their injected values even without noise, by $(13\pm4)\%$ for lines and by $(37\pm3)\%$ for triads (median and scatter over the injection energies). The triad loss is $\sim$the same whether the three channels are injected with equal $Z_l$ or with equal brightness.
The loss is larger for a triad because the linear fit of the second detrending step (\S\ref{subsec:Detrending}) spans its three lines and thus couples them, which adds a loss of $\sim20\%$ (Appendix~\ref{app:Mock}).
We never correct $Z_l$ for this loss,
so the quoted $Z_l$ understate the injected signal, whereas the global $Z_g$ values, calibrated by null realizations that pass through the same pipeline (\S\ref{subsec:Significance}), need no correction.

Only the cross sections are corrected for this loss, and only where so stated.
For each channel $i$ of $T_{72}$, the correction divides the measured cross section by the fraction $\mathcal{F}_i$ that the pipeline retains.
We estimate $\mathcal{F}_i$ by injecting into each cluster of each \S\ref{subsec:Samples} sample a triad at the fixed mass, with equal brightness in its three channels, as expected counts in an otherwise empty sky.
The line flux of each cluster, given by Eq.~\eqref{eq:LineFlux} and thus proportional to its $\mathcal{J}_\Myc$, is redshifted, spread over its ROI and field as for $\psi_\Myc$, and dispersed by the IRF at its observed energy.
The stacking, detrending, and matched filters of the sample then measure the brightness of each line of the injected triad, and Eq.~\eqref{eq:SigmaStack} converts it to a cross section.
The ratio of this cross section to the true one is $\mathcal{F}_i$.

The entire pipeline --- field subtraction, stacking, detrending ($\mathcal{D}$ fixed by the data), matched filters, and conversion to cross section --- is linear in photon counts, with weights fixed by data and catalogs, so $\mathcal{F}_i$ does not depend on injected signal strength, nor on any signal already in the data.
The pipeline adds no appreciable bias to the measured brightness; for example, with no signal injected we find on average $Z_l\simeq 0.03$ (dispersion $\sim0.84$) for $T_{72}$ in Monte Carlo realizations of \SAll\ and $Z_l\simeq-0.07$ in random-longitude stacks.
Linearity thus implies that our $\mathcal{F}_i$ values relate mean measured to true cross sections also in noisy data.
Refitting the GPR amplitude to each injected spectrum, instead of fixing $\mathcal{D}$ by the data, would change $\mathcal{F}_i$ by $\lesssim11\%$ (Monte Carlo medians for injected $T_{72}$ of $Z_l\simeq1.5$--$6$).

In \SAll, we find (see Table~\ref{tab:StackParameters}) $\mathcal{F}_i\simeq \{0.62, 0.61, 0.77\}$ for the $\{\gamma\gamma, \gamma Z, \gamma h\}$ lines, whereas detrending alone retains $\{0.60, 0.60, 0.76\}$ of their brightness (Appendix~\ref{app:Mock}).
The difference arises because each matched filter also collects part of the IRF tails of the other two lines of the triad, which raises the cross sections inferred before detrending by $\{3, 2, 0\}\%$.
Across all samples and subsamples considered, we find $\mathcal{F}_i$ spanning the narrow ranges $\{0.59$--$0.63$, $0.60$--$0.63$, $0.77$--$0.81\}$, so comparing cross sections as measured in different samples is meaningful.

In \SCW\ and \SMW, the loss factors follow those of the \SCW\ analysis\cite{Keshet26Corr} assuming a signal spread over the modeled distribution (Appendix~\ref{app:JointStats}), even when the line templates are placed for simplicity at the peak redshift $z=0.03$ of the model.

Loss-corrected cross sections, given by Eq.~\eqref{eq:SigmaStack} with $\mathcal{I}_i\to\mathcal{I}_i/\mathcal{F}_i$, are extrapolations, quoted only where labeled as such.
We focus in particular on the intrinsic $\gamma\gamma$ cross section
\begin{equation}\label{eq:Sigma0}
  \Sigma_0\equiv\overline{\sigma v}_{p,\gamma\gamma}/D \coma
\end{equation}
its measured value
\begin{equation}\label{eq:Sigma}
  \Sigma\equiv\frac{4\pi}{\epsilon_{\gamma\gamma}\mathcal{S}}\left(\frac{m_\chi \Omega_m}{\Omega_c}\right)^2 \frac{\pi\theta_\MyROI^2\mathcal{I}_{\gamma\gamma}}{\tilde{\mathcal{J}}} \coma
\end{equation}
and its loss-corrected $\Sigma_0\simeq \Sigma/\mathcal{F}_{\gamma\gamma}$ estimate.

\section{Stacking results}
\label{sec:Results}

This section presents the results of stacking.
We first present the rest-frame spectra (\S\ref{subsec:Spectra}) and their blind lines, triads (\S\ref{subsec:Blind}), and nonads (\S\ref{subsec:Nonads}).
Results are then shown for a priori fixed WIMP masses, for lines and triads (\S\ref{subsec:FixedMass}) and for nonads (\S\ref{subsec:FixedNonads}).
The results discussed here are based only on LAT data and on the cataloged positions and redshifts of the clusters.
Previous analyses (\SCT\ and \SCW) enter only after the blind study, as prior information on WIMP masses and line energies, and when giving combined significance values.
Figure panels involving \SCW\ or \SJ\ (which uses the eROSITA map through \SMW) are discussed in \S\ref{sec:Joint}.

\subsection{Rest-frame spectra}
\label{subsec:Spectra}

\begin{figure*}[t!]
    \centering
    \includegraphics[width=0.49\textwidth]{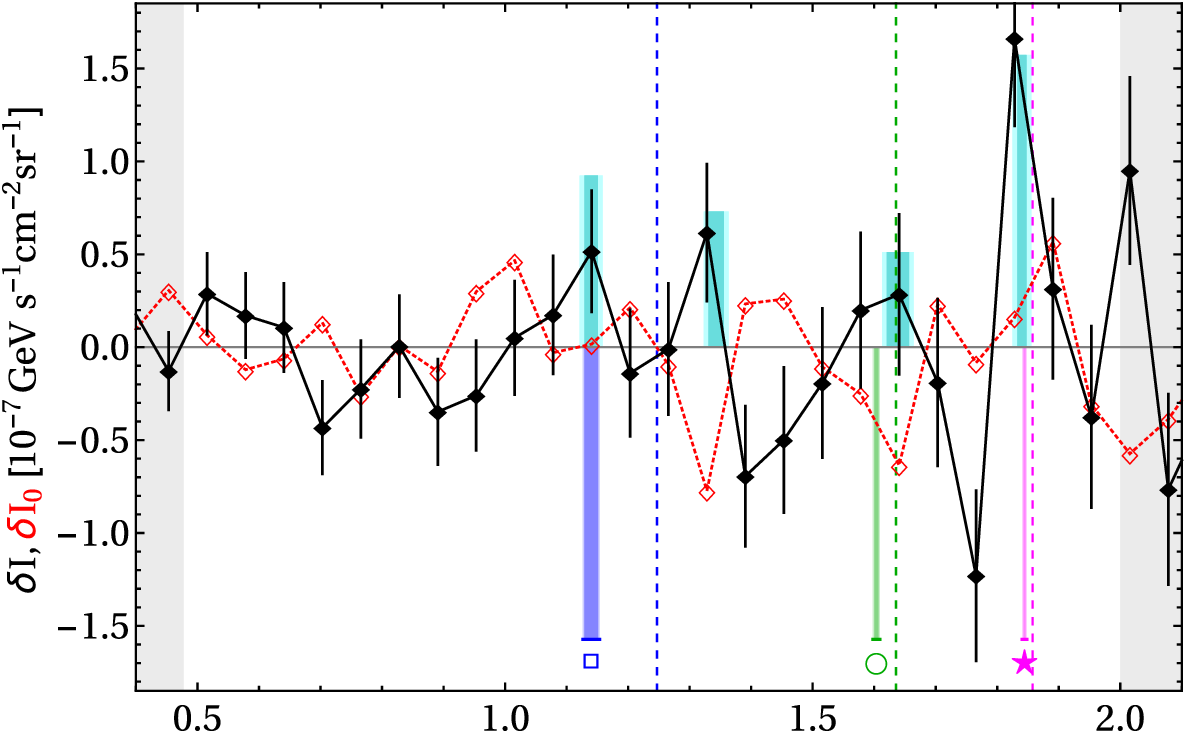}\hfill
    \includegraphics[width=0.49\textwidth]{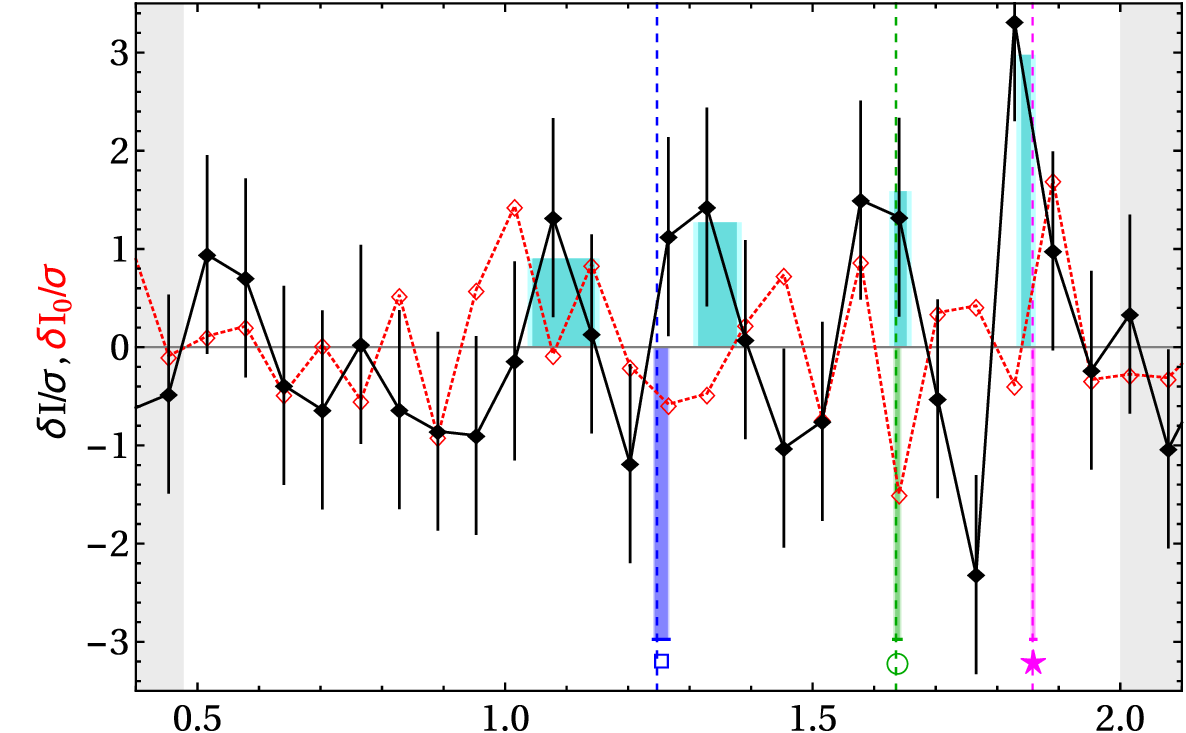}\\
    \includegraphics[width=0.49\textwidth]{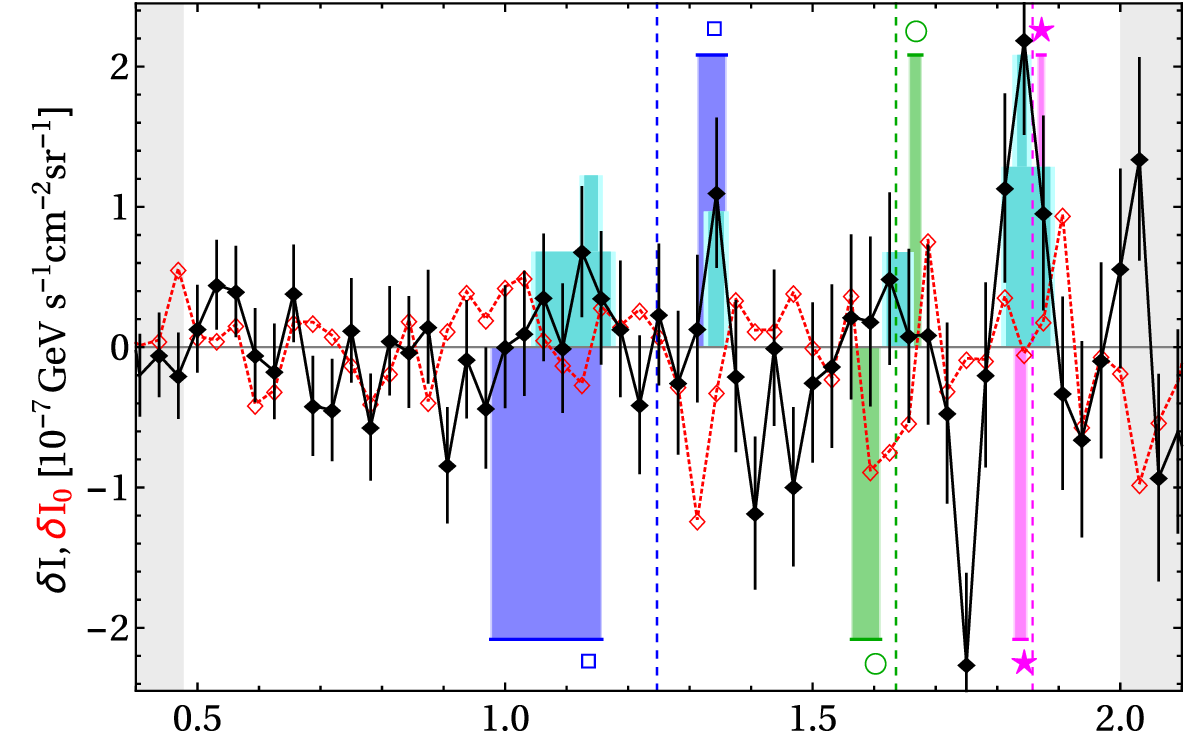}\hfill
    \includegraphics[width=0.49\textwidth]{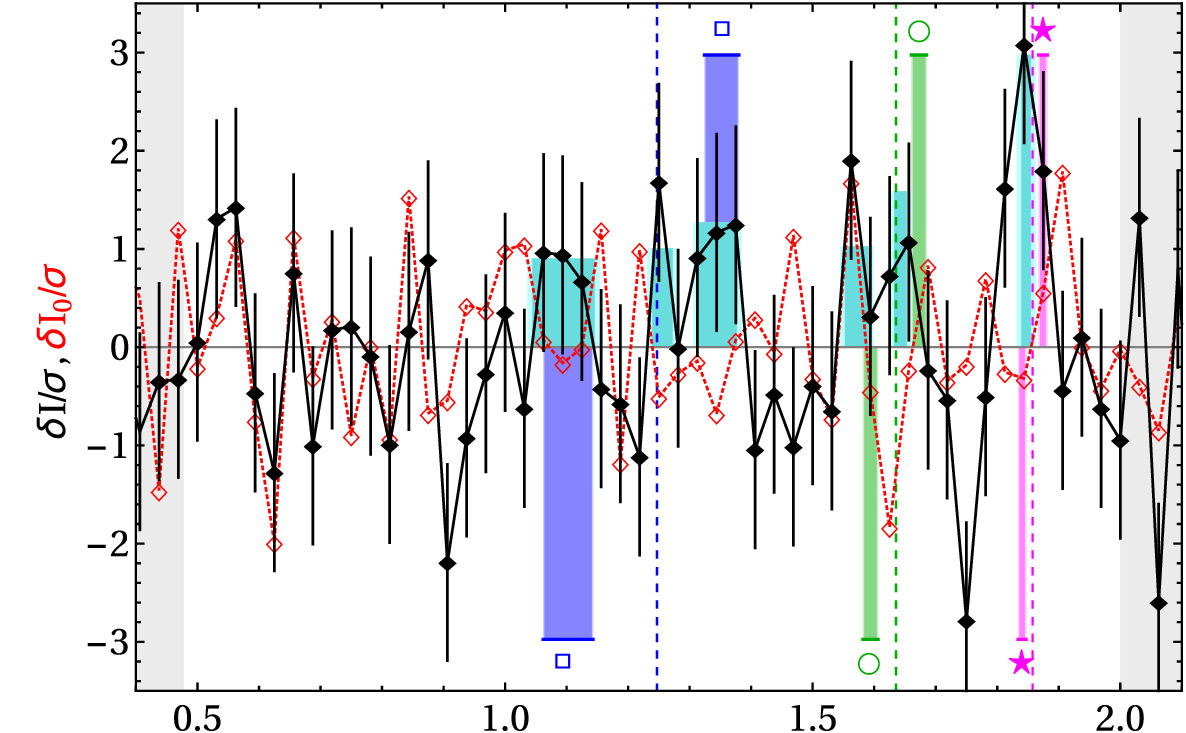}\\
    \includegraphics[width=0.49\textwidth]{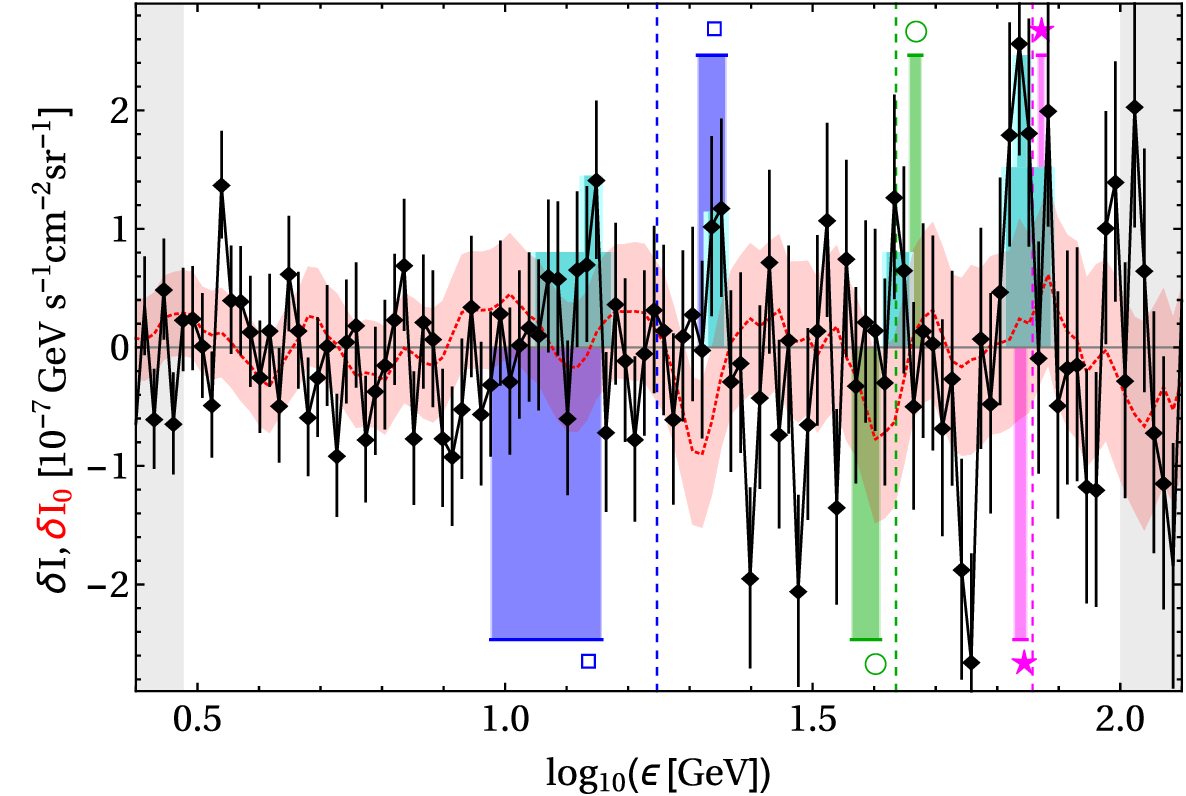}\hfill
    \includegraphics[width=0.49\textwidth]{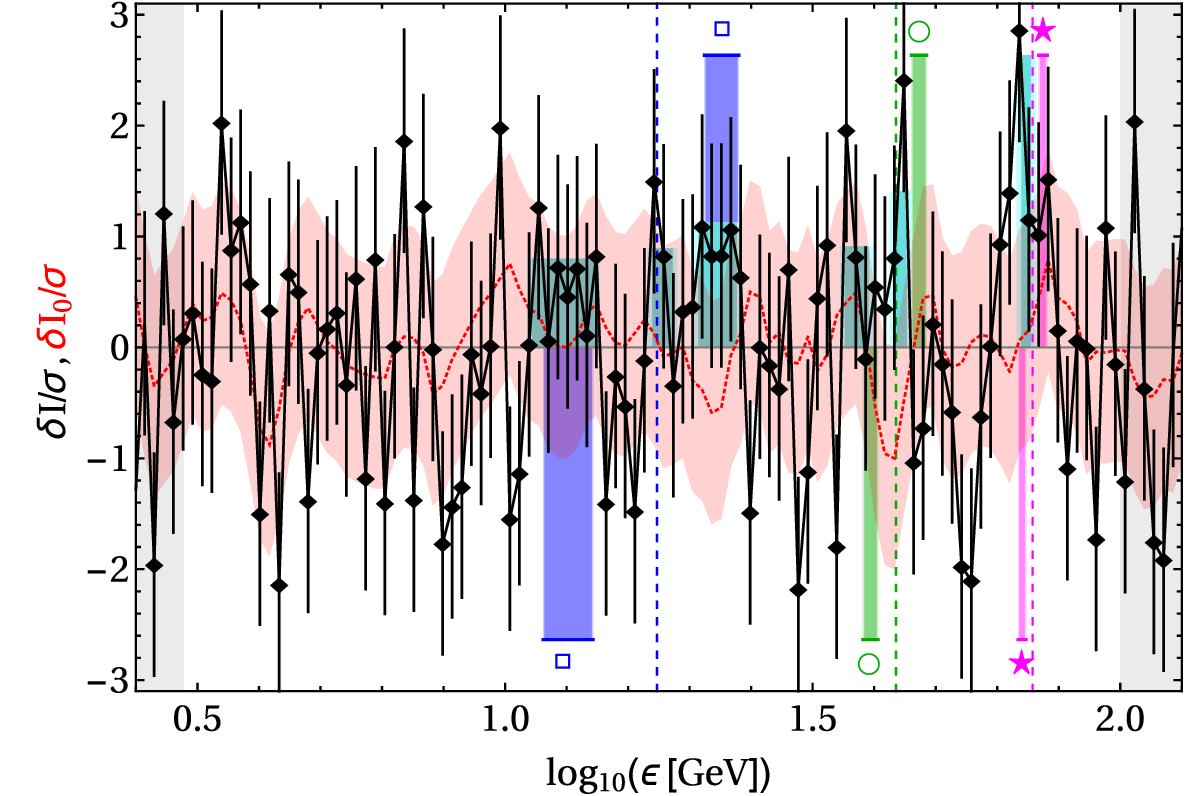}
    \caption{
        Stacked LAT spectrum of all clusters (\SAll; left column) and its combination with \SMW\ (\SJ; right column),
        before (empty red diamonds) and after (filled black diamonds) correcting for galaxy clusters
        (see text), rebinned at low (top row: $s\simeq 4s_0 \simeq14.8\%$; $f=0.1$ line exclusion), medium (middle row: $s \simeq 2s_0\simeq7.4\%$; $f=0.03$), or high (bottom row: native $s=s_0\simeq3.7\%$; $f=0.03$; red spectrum replaced by shaded $\pm1\sigma$ band over the running four-native-bin mean) energy resolution.
        Emission lines with $Z_l>1$ picked up by the matched filter within the $3$--$100\GeV$ scan range (unshaded) are shown
        (cyan bars of likelihood-bound widths
        and height $\propto Z_l$).
        The strongest blind nonad is shown as the $\gamma\gamma$ (magenta stars), $\gamma Z$ (green circles), and $\gamma h$ (blue squares) channels (fixed-height bars indicating mass bounds) of its low (negative bars) and high (positive bars) triads; the top row shows only the strongest blind triad. The a priori $T_{72}$ triad (dashed vertical lines) lies within $0.1\GeV$ of the intermediate (cross-annihilation) triad of each nonad.
        \label{fig:Spectra}
    }
\end{figure*}

Figure~\ref{fig:Spectra} presents the rebinned rest-frame spectrum stacked over \SAll, including all sample clusters (left column).
The right column, combining \SAll\ with \SMW\ into a joint (\SJ) dataset, is discussed in \S\ref{subsec:JointSpectra}.
Spectra of the disjoint hemisphere samples \SE\ and \SW\ are deferred to Appendix~\ref{app:SubSpectra}.

The spectra are shown at different resolutions, ranging from low (top row; fractional width $s\simeq15\%$, grouping four native bins), through the LAT $\sim68\%$ energy-containment (middle row; $s\simeq7.4\%$, grouping two native bins), to high (our native $s=s_0$ binning; bottom row).
Each spectrum is shown both before ($\delta I_0$; empty red diamonds) and after ($\delta I$; filled black diamonds) boosting to the cluster frame.
Note that the rebinned spectra are sensitive to the choice of bins.

The cluster-frame boost reveals a peak near $70\GeV$ and weaker features at lower energies, absent or diminished in the observer frame.
These features can be quantified without matched filters or a chosen energy, using the pull of each rebinned bin over the $3$--$100\GeV$ scan range: its detrended residual over its null error, $\delta I/\sigma$ in the cluster frame and $\delta I_0/\sigma$ in the observer frame, taking into account all binning phases (the bins of Fig.~\ref{fig:Spectra} shifted by one native bin at a time) and calibrating by the Poisson Monte Carlo
of each frame.

At low resolution, the largest cluster-frame pull in Fig.~\ref{fig:Spectra}, $\delta I/\sigma\simeq3.5$ near $67.3\GeV$, has $p\simeq0.005$ (averaging over the four binning phases gives $p\simeq0.004$), compared to only $1.8$ with $p\simeq0.6$ ($p\simeq 0.9$ averaged) in the observer frame.
At medium resolution, the largest cluster-frame pull, $\delta I/\sigma\simeq3.3$ near $69.8\GeV$, has $p\simeq0.04$ (averaging over the two binning phases gives $p\simeq0.01$), compared to only $1.7$ with $p\simeq0.9$ ($p\simeq0.96$ averaged) in the observer frame.
Combined, in five of the six low-to-medium resolution binnings, the largest cluster-frame pull ($3.2$--$3.5$) lies at $67$--$72\GeV$, whereas no observer-frame pull exceeds $1.9$ in any binning.

At low resolution, the number of cluster-frame bins with $\delta I/\sigma>3$ has $p\simeq0.03$ ($0.005$ phase-averaged) and the number with $\delta I/\sigma>2.5$ has $p\simeq0.1$ ($0.01$ averaged), whereas the observer frame shows no bin above $\delta I_0/\sigma=2$ in any binning phase.
At medium resolution, the cluster-frame numbers become $p\simeq0.08$ ($0.02$ averaged) for $\delta I/\sigma>3$ and $p\simeq0.3$ ($0.04$ averaged) for $\delta I/\sigma>2.5$, while the observer frame again shows no bin above $\delta I_0/\sigma=2$.
Combined, five of the six binnings exceed $\delta I/\sigma=3$ at $67$--$72\GeV$.

Unlike peaks, deficits appear in both frames, with similar phase-averaged significance.
At low resolution, the strongest $\delta I/\sigma\simeq-2.7$ deficit near $58\GeV$ has $p\simeq0.06$ (phase-averaged $p\simeq0.03$), compared to observer frame $p\simeq0.1$ ($p\simeq0.03$).
Medium resolution shows a stronger, $\delta I/\sigma\simeq-3.5$ negative pull near $56\GeV$ with $p\simeq0.006$ (phase averaging gives $p\simeq0.02$), compared to observer frame $p\simeq0.08$ ($p\simeq0.03$).
However, the pipeline, which loses part of any real signal mostly through detrending (see \S\ref{subsec:Losses}), lowers the surrounding continuum: a triad at the measured $\Sigma$
deepens the $56$--$58\GeV$ cluster-frame deficit, raising the above $p$ values to $0.01$ (medium resolution; $0.05$ phase-averaged) and $0.1$ (low resolution; $0.09$ phase-averaged).
Once the triad is accounted for, the phase-averaged cluster-frame deficits are thus no more significant than the observer-frame ones.

Native binning over-resolves any line structure, and as expected, neither frame then shows a significant excess ($p\gtrsim0.15$). Here, matched filters provide more appropriate diagnostics, as we next discuss.

\begin{table*}[t]
    \centering
    \caption{Strongest blind spectral features in the cluster-frame spectrum $\delta I$ of \SAll.\label{tab:FeaturesS}}
    \setlength{\tabcolsep}{4pt}
    \begin{tabular}{l c c c c c c}
        \toprule
        Feature & $\epsilon$ & $Z_l$ & $Z_g\,(Z_g^{\smash[t]{(a)}})\,[Z_g^{\smash[t]{(n)}}]$ & $\{\epsilon_{\gamma\gamma},\epsilon_{\gamma Z},\epsilon_{\gamma h}\}$ & $\{Z_{\gamma\gamma},Z_{\gamma Z},Z_{\gamma h}\}$ & $\{\overline{\sigma v}_{p,\gamma\gamma},\overline{\sigma v}_{p,\gamma Z},\overline{\sigma v}_{p,\gamma h}\}/D$ \\
         & (GeV) & & & (GeV) & & ($10^{-19}\cm^3\se^{-1}$) \\
        \midrule
        \multirow{4}{*}{\shortstack[l]{Lines\\($f=0.1$)}} & $69.3_{-1.2}^{+1.1}$ & $3.5$ & $2.1\,(2.6)\,[2.1]$ & & & \\
         & $13.9_{-0.4}^{+0.3}$ & $2.0$ & ---\,(0.6)\,[---] & & & \\
         & $22.1_{-0.7}^{+0.6}$ & $1.6$ & --- & & & \\
         & $43.5_{-1.7}^{+1.9}$ & $1.1$ & --- & & & \\
        \midrule
        \multirow{3}{*}{\shortstack[l]{Triads\\($f=0.015$)}} & $69.9_{-0.2}^{+0.2}$ & $3.3$ & $2.8$$\,(2.5)\,[2.2]$ & $\{69.9,\,40.1,\,13.8\}$ & $\{3.4,\,0.3,\,2.0\}$ & $\{2.0\pm0.6,\,0.7\pm2.1,\,10.2\pm5.0\}$ \\
         & $72.0_{-0.4}^{+0.4}$ & $2.2$ & $1.1$$\,(0.9)\,[0.8]$ & $\{72.0,\,43.1,\,17.6\}$ & $\{2.8,\,0.9,\,0.1\}$ & $\{1.7\pm0.6,\,1.8\pm2.0,\,0.5\pm4.1\}$ \\
         & $74.6_{-0.5}^{+0.4}$ & $2.2$ & $1.1$$\,(0.9)\,[0.8]$ & $\{74.6,\,46.7,\,22.0\}$ & $\{2.1,\,0.1,\,1.6\}$ & $\{1.4\pm0.6,\,0.2\pm2.0,\,5.7\pm3.6\}$ \\
        \midrule
        \multirow{2}{*}{\shortstack[l]{Nonad\\($f=0.05$)}} & $69.8_{-2.3}^{+0.3}$ & \multirow{2}{*}{$3.2$} & \multirow{2}{*}{$3.3\,(2.4)\,[2.1]$} & & & \\
         & $74.5_{-0.7}^{+0.5}$ & & & & & \\
        \bottomrule
    \end{tabular}
 \TblNote{Matched-filter features of \SAll, away from the scan edges: lines with $Z_l>1$ ($n_G=4.2\pm1.6$ such lines expected in Gaussian noise; $1.0\pm1.0$ with $Z_l>2$),
 triads with $Z_l>2$ ($n_G=0.5\pm1.0$;
 three or more in $<0.5\%$ of Monte Carlo realizations, while three $f=0.01$ triads at least as strong as these are calibrated as $3.5\sigma$),
 and the nonad with $Z_l>1$ ($n_G=0.7\pm0.7$; $n_G=0.2\pm0.4$ for $Z_l>2$).
 Here, $n_G$ is the mean $\pm$ standard deviation over $5\times10^4$ Gaussian spectra with the measured errors of the native bins, scanned and selected as the data. Energies and masses ($\overline{m}_\chi c^2$ for triads; $m_1c^2$ and $m_2c^2$ for the nonad, scanned with $f_t=0.02$) carry asymmetric $\Delta\textrm{TS}=1$ bounds, which are not Gaussian uncertainties; the channel energies $\{\epsilon_{\gamma\gamma},\epsilon_{\gamma Z},\epsilon_{\gamma h}\}$ of the triads are those of Eq.~\eqref{eq:kin}. Global scores: the nominal Monte Carlo $Z_g$, (the analytic \smash{\v{S}id\'{a}k} $Z_g^{\smash[t]{(a)}}$ in parentheses), and [$Z_g^{\smash[t]{(n)}}$ from Gaussian realizations in brackets; Appendix~\ref{app:Nulls}], shown when positive.
 Channel cross sections of Eq.~\eqref{eq:SigmaStack} are before loss correction (\S\ref{subsec:Losses}). }
\end{table*}

\subsection{Blind lines and triads}
\label{subsec:Blind}

We apply the matched filters to the stacked rest-frame spectra of the different catalogs and samples.
Using no rebinning, the filters pick up the cluster-frame peaks of Fig.~\ref{fig:Spectra} (cyan bars, of height $\propto Z_l$).
Table~\ref{tab:FeaturesS} lists the lines with $Z_l>1$, the three strongest triads, and the strongest nonad of \SAll.
Matched-filter results for each separate catalog and different subsamples are deferred to Appendix~\ref{app:SubSpectra}.

Triad channels are marked in figures using a magenta star for $\gamma\gamma$, a green circle for $\gamma Z$, and a blue square for $\gamma h$.
Figure \ref{fig:Spectra} shows these channels for the strongest blind triad (top row) and for the outer triads of the strongest blind nonad (middle and bottom rows; see \S\ref{subsec:Nonads}).
The a priori, mass-fixed channels of $T_{72}$ are also shown (dashed lines, in the channel colors; \S\ref{subsec:FixedMass}).

The blind triad scan of \SAll\ finds an $\overline{m}_\chi c^2=69.9_{-0.2}^{+0.2}\GeV$ triad with $Z_l\simeq3.3$, calibrated as $2.8\sigma$ in Monte Carlo, and as $2.7\sigma$ in random longitudes and $2.7\sigma$ in redshift permutations (Appendix~\ref{app:Nulls}).
The $\gamma\gamma$ and $\gamma h$ channels carry this triad ($Z_{\gamma\gamma}\simeq3.4$ and $Z_{\gamma h}\simeq2.0$), whereas its $\gamma Z$ channel is weak ($Z_{\gamma Z}\simeq0.3$).
We henceforth refer to this blind triad of \SAll\ as the nominal triad.

This strongest, nominal triad recurs in independent subsamples.
For instance, the strongest blind triads of \SE\ and \SW\ lie within $0.5\GeV$ of each other, at $69.9\GeV$ in \SE\ and $69.5\GeV$ in \SW; accordingly, the lines of \SE\ (\SW) include $69.4\GeV$ ($69.2\GeV$).
Randomly finding the strongest \SE\ and \SW\ blind triads within $0.5\GeV$ of each other, at any mass, occurs in only $3.3$--$4.5\%$ (depending on the treatment of scan-edge maxima) of random-longitude stacks.
In all four Galactic quadrants (east and west, north and south of the plane), the nominal triad is positive ($Z_l\simeq1.2$--$2.1$).

The strongest single line of \SAll, at $69.3\GeV$ ($Z_l\simeq3.5$; $2.1\sigma$), coincides with the $\gamma\gamma$ channel of the nominal triad.
Every other $Z_l>1$ line of \SAll\ (Table~\ref{tab:FeaturesS}) is also a triad channel: the $13.9\GeV$ line is the $\gamma h$ channel of the nominal triad, the $22.1\GeV$ line the $\gamma h$ channel of a $74.6\GeV$ (third-strongest) triad, and the $43.5\GeV$ line the $\gamma Z$ channel of $T_{72}$.
Line and coincident triad channels integrate the same photons, so these coincidences are not independent evidence for the triads.

More generally, the strongest line of each subsample except MCXC (poorest statistics; see Appendix~\ref{app:SubSpectra}) is a channel of a candidate triad. Depending on the sample, this line is the $\gamma\gamma$ ($69.2$--$69.3\GeV$) or $\gamma h$ ($13.8$--$14.0\GeV$) line of the nominal triad, the $43\GeV$ $\gamma Z$ line of $T_{72}$,
or the $22\GeV$ $\gamma h$ line of a $74.5\GeV$ triad, near the upper mass of the blind nonad below.

The nominal triad lies $3\%$ below the fixed mass of $T_{72}$, less than one native bin and within the IRF core, so this triad and $T_{72}$ share their $\gamma\gamma$ photons.
It is the second-strongest maximum of the \SAll\ triad scan that exactly coincides with $T_{72}$.
Found blindly at $72.0\GeV$ ($Z_l\simeq2.2$), it has
$Z_g\simeq1.1$ as a blind feature in Monte Carlo, and $1.0$ in random longitudes.
A $74.6\GeV$ triad follows with nearly the same $Z_l$, carried by its $\gamma\gamma$ and $\gamma h$ channels ($Z_l\simeq2.1$ and $1.6$; Table~\ref{tab:FeaturesS}).

\subsection{Blind nonads}
\label{subsec:Nonads}

Two cross-annihilating WIMP species would add an intermediate triad to their two self-annihilation triads, so we also scan blindly for such nonads.
The blind nonad scan of \SAll\ (see Table~\ref{tab:FeaturesS}) finds a $Z_l\simeq3.2$ nonad with $m_1c^2=69.8_{-2.3}^{+0.3}\GeV$ and $m_2c^2=74.5_{-0.7}^{+0.5}\GeV$, calibrated as $3.3\sigma$ by Monte Carlo, $3.3\sigma$ in permutations, and $3.2\sigma$ in random longitudes.
We refer to this blind nonad of \SAll\ as the nominal nonad.

Note that mock nonads lose more of their $Z_l$ to detrending than triads, and are recovered less often (see Appendix~\ref{app:Mock}), so a real nonad is even harder to find than a real triad of the same $Z_l$ before detrending.
Recall that we do not correct for this effect anywhere, in any local or global $Z$ score.

The \{low, intermediate, high\} triads of the nominal nonad approximately coincide with the \{nominal and strongest, fixed-mass and second-strongest, third-strongest\} triads.
The $72.2\GeV$ intermediate triad of the nonad lies slightly above the fixed and second-strongest $72\GeV$ triad and adds only a conditional significance $Z_c\simeq0.4$ beyond the $Z_l\simeq3.3$ of the outer triads, thus lowering the local nonad significance to $Z_l\simeq3.2$.
However, due to the $\gamma\gamma$ overlap, this does not indicate that the intermediate triad is necessarily weaker.

\begin{figure*}[tp]
    \centering
    \makebox[0.35\textwidth][l]{\hspace{0.066\textwidth}\makebox[0.268\textwidth]{\large\color{blue}Uncorrected}}%
    \makebox[0.32\textwidth][l]{\hspace{0.036\textwidth}\makebox[0.268\textwidth]{\large\color{blue}Corrected}}%
    \makebox[0.32\textwidth][l]{\hspace{0.036\textwidth}\makebox[0.268\textwidth]{\large\color{blue}Null dispersion}}\\[2pt]
    \includegraphics[width=0.35\textwidth]{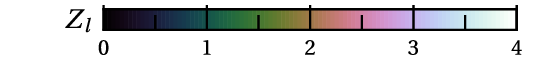}%
    \includegraphics[width=0.32\textwidth]{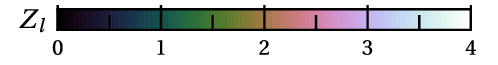}%
    \includegraphics[width=0.32\textwidth]{Figures/LAT_Stack_NonadMaps_Bar.eps}\\
    \MapPanelRow{0.35}{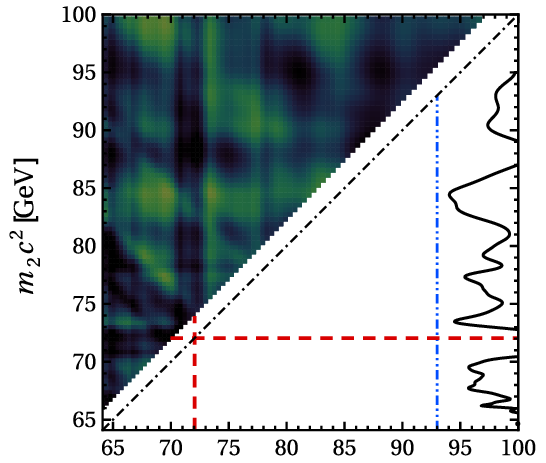}{0}{0}{$\delta\llangle\boldsymbol{I}\rrangle$\\(uncorrelated)}{\SCW}%
    \MapPanel{0.32}{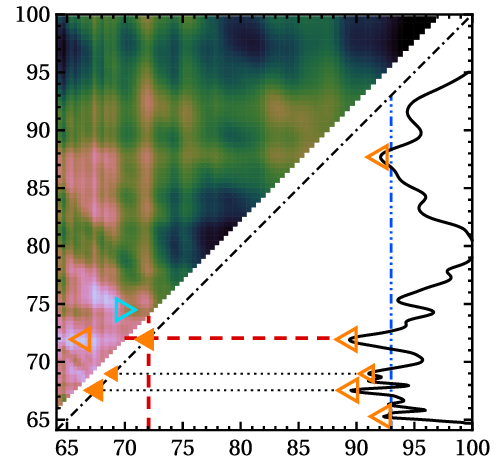}{-1.2}{-2.3}{$\delta I_\theta$\\(X-ray correlated)}%
    \MapPanel{0.32}{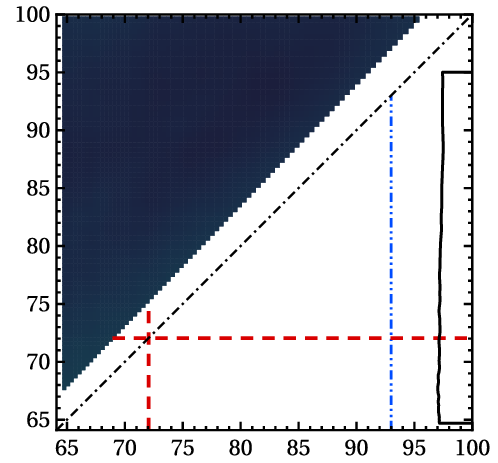}{0}{0}{$\sigma(\delta I_\theta)$}\\
    \MapPanelRow{0.35}{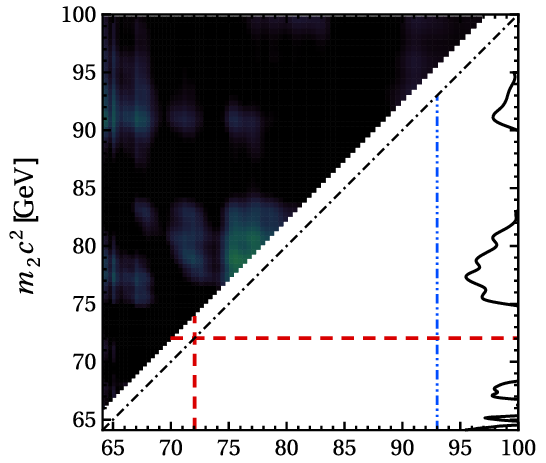}{-0.8}{-1.5}{$\delta I_0$\\(observer frame)}{\SAll}%
    \MapPanel{0.32}{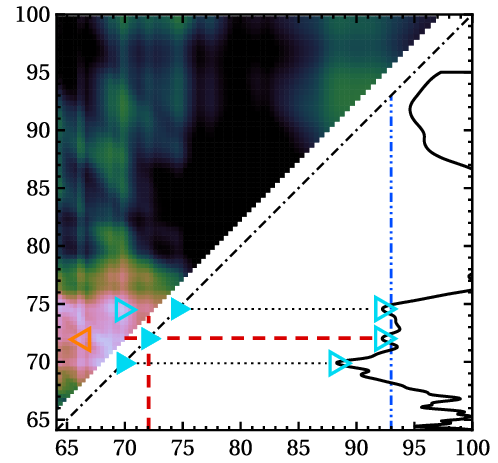}{0.2}{0.3}{$\delta I$\\(cluster frame)}%
    \MapPanel{0.32}{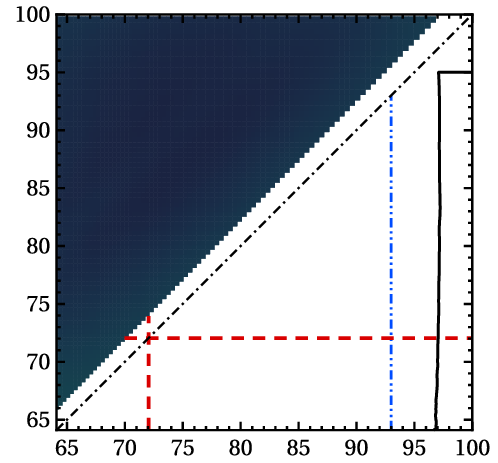}{0}{0}{$\sigma(\delta I)$}\\
    \MapPanelRow{0.35}{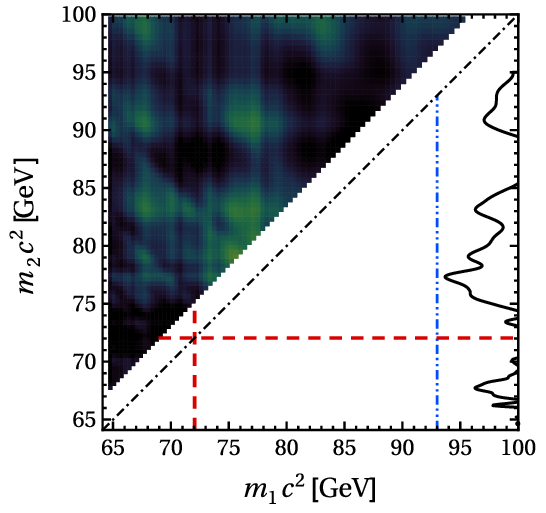}{0}{0}{\SJ$_0$ \\ ($\delta I_0$ + $\delta\llangle\boldsymbol{I}\rrangle$)}{\SJ}%
    \MapPanel{0.32}{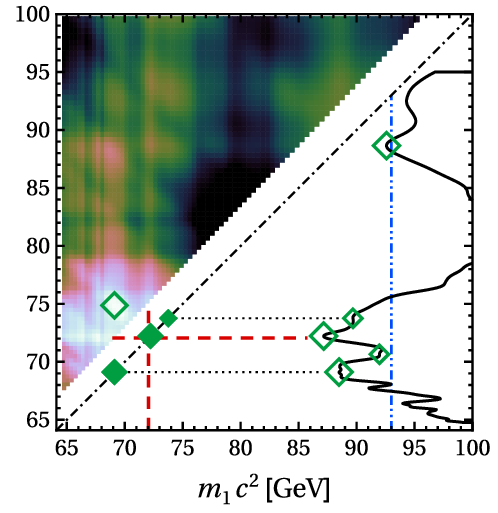}{0}{0}{\SJ \\ ($\delta I$ + $\delta I_\theta$)}%
    \MapPanel{0.32}{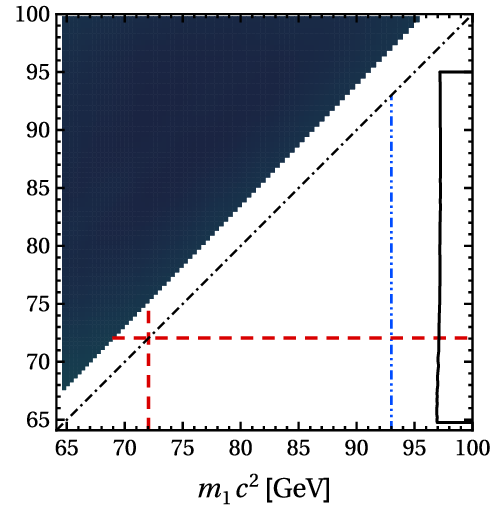}{0}{0}{$\sigma($\SJ$)$}\\
    \includegraphics[width=0.35\textwidth]{Figures/LAT_Stack_NonadMaps_BarL.eps}%
    \includegraphics[width=0.32\textwidth]{Figures/LAT_Stack_NonadMaps_Bar.eps}%
    \includegraphics[width=0.32\textwidth]{Figures/LAT_Stack_NonadMaps_Bar.eps}
    \caption{
    Composite matched-filter $\max(Z_l,0)$ scans with their strongest features (symbols) in X-ray-correlation (\SCW\ in observer frame; top row; orange left triangles), stacking (\SAll; middle row; cyan right triangles), and their joint (\SJ; bottom row; green diamonds) analyses, before (left column) vs. after (middle column) correcting for galaxy clusters, and compared to the standard deviation of the corrected $Z_l$ over null realizations (right column: a control of pipeline systematics).
    Nonad ($m_1<m_2$; common color scale in all panels) and triad (vs. $m_2$; black curve on right axis) scans highlight the strongest nonad (empty symbol on map) and three strongest triads (filled symbols on $m_1=m_2$ dot-dashed diagonal).
    All $Z_l>2$ triads marked with empty symbols (large for nominal $f=0.015$, small for $f=0.010$).
    The a priori fixed $72.05\GeV$ triad is also shown (dashed red lines).
    \label{fig:NonadMap}
    }
\end{figure*}

Figure~\ref{fig:NonadMap} visualizes the composite matched-filter scans, which combine lines into triads and triads into nonads.
Each panel shows the triad (black curve plotted against $m_2$ with $Z_l$ increasing leftward from the right axis) and nonad (shown in color for $m_1<m_2$) scans, compared for three different analyses: the \SCW\ X-ray correlation (top row), \SAll\ stacking (middle), and their joint \SJ\ combination (after masking \SCW; see \S\ref{subsec:JointNonads}).

Specifically, nonad $Z_l$ values are mapped over all mass pairs with $m_2\ge1.03\,m_1$, on a common grid of $0.5\%$ cells, irrespective of exclusion windows.
Nonad maxima are sought only at $m_2>1.04\,m_1$, outside the $f_t=0.02$ window.
Besides their nominal exclusion window ($f=0.015$), triads are also marked, for display only, using a narrower $f=0.01$ window to distinguish nearby features.

Evident structure with similar triads and nonads is found in all three analyses, but only after (middle column) and not before (left column) correcting for clusters, \ie shifting from the ambient to the X-ray-correlated spectrum, and from observer to cluster frame.
Strong triads of mass $m$ imprint vertical ($m_1=m$), horizontal ($m_2=m$), or anti-diagonal ($m_1+m_2=2m$) stripes, with nonads at pronounced stripe crossings.
The right column serves as a pipeline control, repeating the scans for many null realizations processed as the data.

In particular, the \SAll\ observer frame (middle row, left panel) scan shows none of the pronounced cluster-frame (middle panel) features; its strongest nonad reaches only $Z_l\simeq1.2$, while its null dispersion ($0.56$--$0.88$ across the map, not shown) matches that of the cluster frame (right panel).
The smooth, nearly uniform null-dispersion map of the latter, obtained from $2\times10^4$ Monte Carlo realizations,
shows neither the pronounced cluster-frame features, nor any
significant systematics in the
pipeline (e.g., from the binning, exposure, or detrending).

Table~\ref{tab:Nonads} lists the strongest blind nonad in each catalog and each sample.
As the table shows, the \SAll\ nonad is seen also in \SO, outside the \SCW\ region,
further showing that the stacked nonad (middle row of Fig.~\ref{fig:NonadMap}) does not hinge upon the \SCW\ sky region (top row).
While the \SE\ and \SW\ hemisphere stacking analyses indicate slightly different nonads, they both share the $70\GeV$ triad along with either the $75\GeV$ (\SE) or $72\GeV$ (\SW) triads of \SAll.

\begin{table}[h!]
    \centering
    \setlength{\tabcolsep}{3.5pt}
    \caption{Strongest blind nonad in each sample.\label{tab:Nonads}}
    \begin{tabular}{@{}l c c c c@{}}
        \toprule
        Sample & $m_1c^2$ & $\overline{m}_\chi c^2$ & $m_2c^2$ & $Z_l$ ($Z_g$) \\
         & (GeV) & (GeV) & (GeV) & \\
        \midrule
        {\bfseries\boldmath \SAll} & {\bfseries\boldmath $69.8_{-2.3}^{+0.3}$} & {\bfseries\boldmath $72.2_{-0.6}^{+0.3}$} & {\bfseries\boldmath $74.5_{-0.7}^{+0.5}$} & {\bfseries\boldmath 3.2 (3.3)} \\
        \SE & $65.1_{-0.7}^{+1.4}$ & $70.0_{-0.4}^{+0.6}$ & $74.9_{-1.4}^{+0.6}$ & 1.9 (1.3) \\
        \SW & $67.4_{-0.6}^{+2.5}$ & $69.5_{-0.4}^{+2.7}$ & $71.6_{-0.8}^{+3.3}$ & 2.9 (2.7) \\
        \SO & $69.8_{-1.6}^{+0.6}$ & $72.2_{-0.8}^{+0.4}$ & $74.5_{-0.6}^{+0.8}$ & 2.1 (1.6) \\
        MCXC & $73.4$ & $75.5$ & $77.5$ & 0.8 ($-0.7$) \\
        $\Delta$eROSITA & $66.1_{-0.9}^{+3.4}$ & $68.8_{-1.6}^{+2.4}$ & $71.6_{-1.3}^{+2.4}$ & 2.4 (2.1) \\
        $\Delta$DESI & $65.1_{-0.7}^{+0.4}$ & $69.8_{-0.2}^{+0.5}$ & $74.5_{-0.8}^{+0.8}$ & 2.7 (2.5) \\
        \bottomrule
    \end{tabular}
    \TblNote{
    Nominal sample \SAll\ in {\bf bold} text (henceforth).
    Shown are the strongest equal-weight nonads (\S\ref{subsec:MatchedFilters}), masses, including their derived mean $\overline{m}_\chi$, with profiled $\Delta\textrm{TS}=1$ bounds (except MCXC, whose $Z_l<1$), and $Z_g$ from Monte Carlo.
    }
\end{table}

\subsection{Lines and triads at the $\SCT$ fixed mass}
\label{subsec:FixedMass}

Fixing a $43.2\GeV$ line energy or a $72.1\GeV$ triad mass (see \S\ref{subsec:FixedTests}) turns some of the blind scans above into a priori tests of $L_{43}$, $T_{72}$, and the $C_{72}$ companions of $L_{43}$ in $T_{72}$, with no need for a mass scan and the associated multiple-trial penalty.
These tests are presented here; analogous tests of nonads that harbor $T_{72}$ are discussed next in \S\ref{subsec:FixedNonads}.
Each test is calibrated against the null distribution of its own fixed-energy or fixed-mass $Z_l$.

The blind \SAll\ scan above already finds this line: its fourth-strongest line, at $43.5_{-1.7}^{+1.9}\GeV$, is centered less than one native bin from the fixed energy.
Tested a priori, with its energy fixed, $L_{43}$ is calibrated as $1.1\sigma$ in \SAll.
Among the subsamples, $L_{43}$ is strong only in $\Delta$eROSITA ($Z_l\simeq3.2$), positive in \SW\ and \SO\ ($1.6$ and $0.3$), and negative in \SE, MCXC, and $\Delta$DESI ($-0.2$, $-1.1$, and $-0.6$);
see Appendix~\ref{app:Line43} for further detail.

In \SAll\ and in each of \SE, \SW, \SO, and the three catalogs, $T_{72}$ is positive,
exceeding unity in \SAll\ and in four of the six subsamples.
Similarly, $T_{72}$ is positive ($0.1$--$2.5$)
in all four Galactic quadrants of \SAll, consistent with a uniform signal (Appendix~\ref{app:SubSpectra}).
In \SAll, $T_{72}\simeq2.2$ is calibrated as $2.6\sigma$ by Monte Carlo ($p\simeq5\times10^{-3}$), $2.8\sigma$ by random longitudes, and $2.7\sigma$ by permutations.

The $\gamma\gamma$ channel dominates this triad ($Z_{\gamma\gamma}\simeq2.8$), whereas the $\gamma Z$ and $\gamma h$ channels are weak ($Z_{\gamma Z}\simeq0.9$ and $Z_{\gamma h}\simeq0.1$).
Combined, the $L_{43}$ companions in this triad give $C_{72}\simeq2.1$, calibrated as $2.2\sigma$.

Removing Virgo and Fornax (\S\ref{subsec:Samples}) from \SAll\ leaves its $T_{72}$ unchanged, so \SAll\ is practically independent of \SCT.
Further removing also the 121 clusters whose ROIs touch the regions of the \SCT\ clusters
still leaves $T_{72}\simeq2.2$ and its $2.6\sigma$ intact.
Fisher-combining with \SCT\ (\SCTb), $T_{72}$ in \SAll\ reaches $4.8\sigma$ ($3.5\sigma$), $C_{72}$ reaches $4.6\sigma$ ($3.3\sigma$), and the $43.2\GeV$ line measured in \SAll\ reaches $4.2\sigma$ ($2.7\sigma$).
Such Fisher combinations use the fixed-mass statistics only, and assume their inputs to be independent.

Sample \SAll\ is not independent of \SCW, as many of its western clusters lie in the \SCW\ sky region.
Sample \SO\ by definition shares no ROI photons with \SCW\ (\S\ref{subsec:Samples}), so its $T_{72}\simeq1.4$, calibrated as $1.7$--$2.0\sigma$ ($p\simeq0.042$ in Monte Carlo), can be combined with that of \SCW.
Fisher-combining this $p$-value with that of the $T_{72}$ of \SCW\ (with its templates at $z=0.03$; $p\simeq5\times10^{-4}$ at the mass fixed by $L_{43}$) yields $3.5\sigma$.

Adding the $L_{43}$ measurement to these two inputs
gives $5.3\sigma$ ($4.3\sigma$) using
\SCT\ (\SCTb).
This addition is possible as the above three inputs are practically independent: \SO\ excludes Virgo and Fornax, which lie in the \SCW\ sky region, as well as Ophiuchus, which lies below the latitude cut, whereas masking Virgo and Fornax changes the $T_{72}$ of \SCW\ (as observed) \cite{Keshet26Corr} only from $3.0$ to $2.9$.

As \SCW\ combines different LSS of unknown low ($z\lesssim0.1$) redshift, it does not have an obvious redshift correction.
Its reported $T_{72}$ ($3.0\sigma$) is computed with apertures centered at $z=0$, i.e., as observed, whereas the above Fisher combinations place its templates at $z=0.03$, the peak of the modeled distribution (see below).
The distribution of LSS emission along the line of sight entering the X-ray-correlated signal can be modeled\cite{Keshet26Corr}, so there are a few approximate ways to correct for the redshift.

Approximate treatments used in \SCW\ for redshift correction include broadening the spectral apertures uniformly over $0\le z\le\Delta z$, or shifting them by the offsets found in the \SCW\ mocks (indicating line energies $5$--$6\%$ higher than observed and, under simple line-ratio assumptions, $\sim0.9\%$ higher masses for triads and nonads),
or folding them over the redshift distribution modeled for the \SCW\ signal, which peaks at $z\simeq0.03$ (henceforth the modeled distribution) \cite{Keshet26Corr}.
Over these broadened, shifted, and folded treatments, $T_{72}$ gives $2.3$--$3.4\sigma$ in \SCW,
so combining \SO\ with \{\SCW, \SCW+\SCT, \SCW+\SCTb\} spreads \{$3.5\sigma$, $5.3\sigma$, $4.3\sigma$\} over \{$2.7$--$3.6\sigma$, $4.8$--$5.4\sigma$, $3.6$--$4.3\sigma$\}.
Co-adding instead the $Z_l$ of \SO\ and of \SCW\ (as for \SJ; \S\ref{sec:Joint}) gives \{$2.9$--$3.5\sigma$, $5.0$--$5.4\sigma$, $3.8$--$4.3\sigma$\}.

\subsection{Nonads at the fixed \SCT\ and \SCW\ masses}
\label{subsec:FixedNonads}

The nominal nonad (of \S\ref{subsec:Nonads}), which is the only $Z_l>1$ blind nonad of \SAll\ (Table~\ref{tab:FeaturesS}),
is essentially the fixed-mass nonad $N_{72}$ (see \S\ref{subsec:FixedTests}), with masses equal to within $<0.2\%$ (finer than the nonad grid): its intermediate triad nearly coincides with $T_{72}$, and its $\gamma Z$ line with $L_{43}$.
Fisher-combined with \SCT\ (\SCTb), the nominal nonad ($3.3\sigma$) reaches
$5.2\sigma$ ($4.0\sigma$).
Table~\ref{tab:FixedNonads} demonstrates $N_{72}$ nonads, of cross-annihilation triad fixed at $T_{72}$, for each sample.

\begin{table}[h]
    \centering
    \caption{The nonad $N_{72}$ in each sample.\label{tab:FixedNonads}}
    \begin{tabular}{l c c c}
        \toprule
        Sample & $m_1c^2$ & $m_2c^2$ & $Z_l$ ($Z_g$) \\
         & (GeV) & (GeV) & \\
        \midrule
         {\bfseries\boldmath \SAll} & {\bfseries\boldmath $69.7_{-0.5}^{+0.4}$} & {\bfseries\boldmath $74.4_{-0.4}^{+0.5}$} & {\bfseries\boldmath 3.1 (4.0)} \\
        \SE & $69.9_{-0.5}^{+0.4}$ & $74.2_{-0.4}^{+0.5}$ & 1.8 (2.0) \\
        \SW & $69.5_{-0.5}^{+0.5}$ & $74.6_{-0.5}^{+0.5}$ & 2.7 (3.3) \\
        \SO & $69.7_{-1.0}^{+0.4}$ & $74.4_{-0.4}^{+1.0}$ & 2.0 (2.4) \\
        MCXC & $68.8$ & $75.3$ & 0.3 ($-0.4$) \\
        $\Delta$eROSITA & $69.2_{-1.5}^{+1.4}$ & $74.9_{-1.4}^{+1.5}$ & 2.0 (2.3) \\
        $\Delta$DESI & $69.7_{-0.3}^{+0.3}$ & $74.4_{-0.3}^{+0.3}$ & 2.6 (3.1) \\
        \bottomrule
    \end{tabular}
    \TblNote{
    Same as Table~\ref{tab:Nonads}, but for $N_{72}$, a priori fixed with an intermediate $T_{72}$ triad.
    }
\end{table}

The blind nonad of \SO\ ($Z_g\simeq1.6$), which shares no ROI photons with \SCW, lies at the same masses (Table~\ref{tab:Nonads}). At these masses,
\SCW\ placed at $z=0.03$ gives $Z_l\simeq2.7$,
calibrated as $2.8\sigma$.
Fisher-combined, the two give $3.0\sigma$, and $5.0\sigma$ ($3.9\sigma$) when also combining \SCT\ (\SCTb).
Co-adding instead the $Z_l$ of \SO\ and of \SCW\ (as for \SJ; \S\ref{sec:Joint}), the strongest blind nonad reaches $3.0$--$3.3\sigma$ for different \SCW\ redshift corrections, and $5.0$--$5.2\sigma$ ($3.8$--$4.1\sigma$) when combined with \SCT\ (\SCTb).

Of the three readings of $T_{72}$ within an \SAll\ nonad (see \S\ref{subsec:FixedTests}),
the intermediate $N_{72}$ reading peaks at $m_1c^2=69.7_{-0.5}^{+0.4}\GeV$ and $m_2c^2=74.4_{-0.4}^{+0.5}\GeV$, with $Z_l\simeq3.1$, calibrated as $4.0\sigma$, as shown in the table.
The high reading ($m_2c^2=72.1\GeV$) gives $Z_l\simeq3.1$ at $m_1c^2\simeq68.1\GeV$, calibrated as $3.6\sigma$, whereas the low reading ($m_1c^2=72.1\GeV$) gives $Z_l\simeq1.9$ at $m_2c^2\simeq77.3\GeV$ ($2.2\sigma$).
The strongest of the three readings, $N_{72}^{\max}$, taken alike in the data and in each realization, gives $3.5\sigma$.

These \SAll\ results slightly favor the intermediate reading. In this $N_{72}$ reading, all samples except MCXC peak at nearly the same masses as \SAll\ ($m_1c^2\simeq 69.2$--$69.9\GeV$), at $2.0$--$3.3\sigma$ (see Table~\ref{tab:FixedNonads}); this includes \SO, which is outside the \SCW\ sky region, at $2.4\sigma$.
Fisher-combined
with \SCT\ (\SCTb),
$N_{72}$ reaches $5.7\sigma$ ($4.7\sigma$) in \SAll.
Alternatively, outside the \SCW\ sky region, the $N_{72}$ of \SO\ ($2.4\sigma$) lies at the masses of \SAll,
where \SCW, with its templates at $z=0.03$, gives $Z_l\simeq2.8$,
calibrated as $2.8\sigma$. Fisher-combined, the two give $3.5\sigma$, and $5.3\sigma$ ($4.3\sigma$) with \SCT\ (\SCTb).

At the masses $\{68.4, 74.1\}\GeV$ of the blind \SCW\ nonad scaled from the observer frame by $1.03$ to match the peak $z=0.03$ redshift, \SAll\ gives $Z_l\simeq2.8$, calibrated as $3.7\sigma$, Fisher-combined with \SCT\ (\SCTb) to $5.5\sigma$ ($4.3\sigma$).
The intermediate triad of this nonad has its $\gamma Z$ line at $42.1\GeV$, within one native bin of $L_{43}$.
Every subsample but MCXC gives $1.8$--$3.3\sigma$ at these masses,
including $2.2\sigma$ in \SO, outside the \SCW\ sky region.
The above scaled nonad is not identical to the $\{69.1, 75.6\}\GeV$ nonad ($2.6\sigma$; \S\ref{subsec:SumSig}) of \SCW\ with templates placed at $z=0.03$, with $3.5\sigma$ in \SAll, and with $2.1\sigma$ in \SO\ Fisher-combined with \SCW\ to $3.2\sigma$, and to $5.1\sigma$ ($4.0\sigma$) with \SCT\ (\SCTb).

Overall, the a priori nonads are more significant than $T_{72}$ alone, despite the stronger suppression of nonads by detrending (uncompensated; \S\ref{subsec:Nonads}). In \SAll, the nonads at masses fixed by \SCT\ or \SCW\ reach $3.7$--$4.0\sigma$ ($3.5\sigma$ for $N_{72}^{\max}$), against $2.6\sigma$ for $T_{72}$. With \SCT\ (\SCTb), $N_{72}$ reaches $5.7\sigma$ ($4.7\sigma$) and
the blind \SCW\ nonad scaled by 1.03 reaches
$5.5\sigma$ ($4.3\sigma$), against $4.8\sigma$ ($3.5\sigma$) for $T_{72}$.

The highest combined significance is that of $N_{72}$ in \SJ\ with \SCT,
giving $6.0\sigma$ (see \S\ref{subsec:JointFixed}).
Outside the \SCW\ sky region, \SO\ combined with \SCW\ gives
$5.0$--$5.3$$\sigma$ ($3.8$--$4.3\sigma$) with \SCT\ (\SCTb), for its blind nonad, its $N_{72}$, and the blind \SCW\ nonad, as shown above.

Of the three readings of $T_{72}$, the intermediate one, $N_{72}$, is marginally favored.
It gives the largest $Z_g$ in \SAll, in \SJ, and in four of the six subsamples; the high reading prevails only in $\Delta$eROSITA (Appendix~\ref{app:SubSpectra} and \S\ref{subsec:JointFixed}).
With no mass fixed, the blind nonads of \SAll, \SO, and \SJ\ also have their intermediate triads within $0.1\GeV$ of $T_{72}$,
whereas those of \SE, \SW, $\Delta$eROSITA, and $\Delta$DESI center on the nominal, lower energy triad.

The nominal nonad,
the blind \SCW\ nonad scaled by 1.03, and the \SCW\ nonad with templates placed at $z=0.03$
appear
to be one and the same, as the masses of the latter lie within the bounds of the former (Table~\ref{tab:Nonads}).
The joint analysis \SJ\ is therefore expected to find this nonad, consistent with both \SAll\ and the masses fixed by cross-annihilation ($N_{72}$),
and indeed we find it (in \S\ref{subsec:JointNonads} and \S\ref{subsec:JointFixed}).

The nominal nonad is robust, with its three triads found blindly as the three strongest triads of \SAll. The lower, $69.9\GeV$ triad is particularly strong: it is the strongest triad of \SAll, its $\gamma\gamma$ and $\gamma h$ lines are the two strongest lines of \SAll, and it is also the strongest triad of \SE, \SW, and $\Delta$DESI.
The intermediate triad is the a priori fixed $T_{72}$ to within $0.1\GeV$, and is also found blindly, as the second-strongest triad of \SAll, \SW, and $\Delta$eROSITA and the strongest triad of \SJ; the $\gamma Z$ line of $T_{72}$ is $L_{43}$, the strongest line of $\Delta$eROSITA, and its $\gamma\gamma$ line, placed at $z=0.03$, falls within one native bin of the strongest \SCW\ line\cite{Keshet26Corr}.
The upper, $74.6\GeV$ triad is nearly tied for the second triad of \SAll, is
the strongest triad of \SO, and the second-strongest of \SE\ and $\Delta$DESI.

\section{Signal properties and tests}
\label{sec:Tests}

This section quantifies basic properties of the signal obtained in \S\ref{sec:Results} and tests whether it is consistent with WIMP annihilation in clusters.
We test the response of the signal to the redshift correction and quantify its cluster-frame
redshift width in \S\ref{subsec:Alignment},
and follow this response across redshift tertiles, especially the most distant one, in \S\ref{subsec:Distant}.
Cross sections are inferred in \S\ref{subsec:Scaling}, and the consistency of the $\gamma\gamma$ cross section $\Sigma$ is
tested for $T_{72}$ across cluster subsets and against \SCW.
Additional controls
(moving the ROIs off the clusters and stacking the field alone, also under quiet conditions) are carried out in \S\ref{subsec:Controls}, and robustness tests (removing clusters or sky regions and varying the analysis) in \S\ref{subsec:RobustTests}.

Besides the cluster positions and redshifts used above, this section also incorporates the cataloged $M_{500}$ and $\theta_{500}$ values in the cross sections of Eq.~\eqref{eq:SigmaStack}, and uses these values as well as $L_{500}$ to rank, bin, select, or mask clusters.

\subsection{Cluster vs. observer frame}
\label{subsec:Alignment}

\begin{figure*}[t!]
    \centering
    \makebox[0.54\textwidth][l]{\hspace{0.077\textwidth}\makebox[0.451\textwidth]{\large\color{blue}\SAll}}%
    \makebox[0.46\textwidth][l]{\hspace{0.007\textwidth}\makebox[0.451\textwidth]{\large\color{blue}\SJ}}\\[2pt]
    \includegraphics[width=0.54\textwidth]{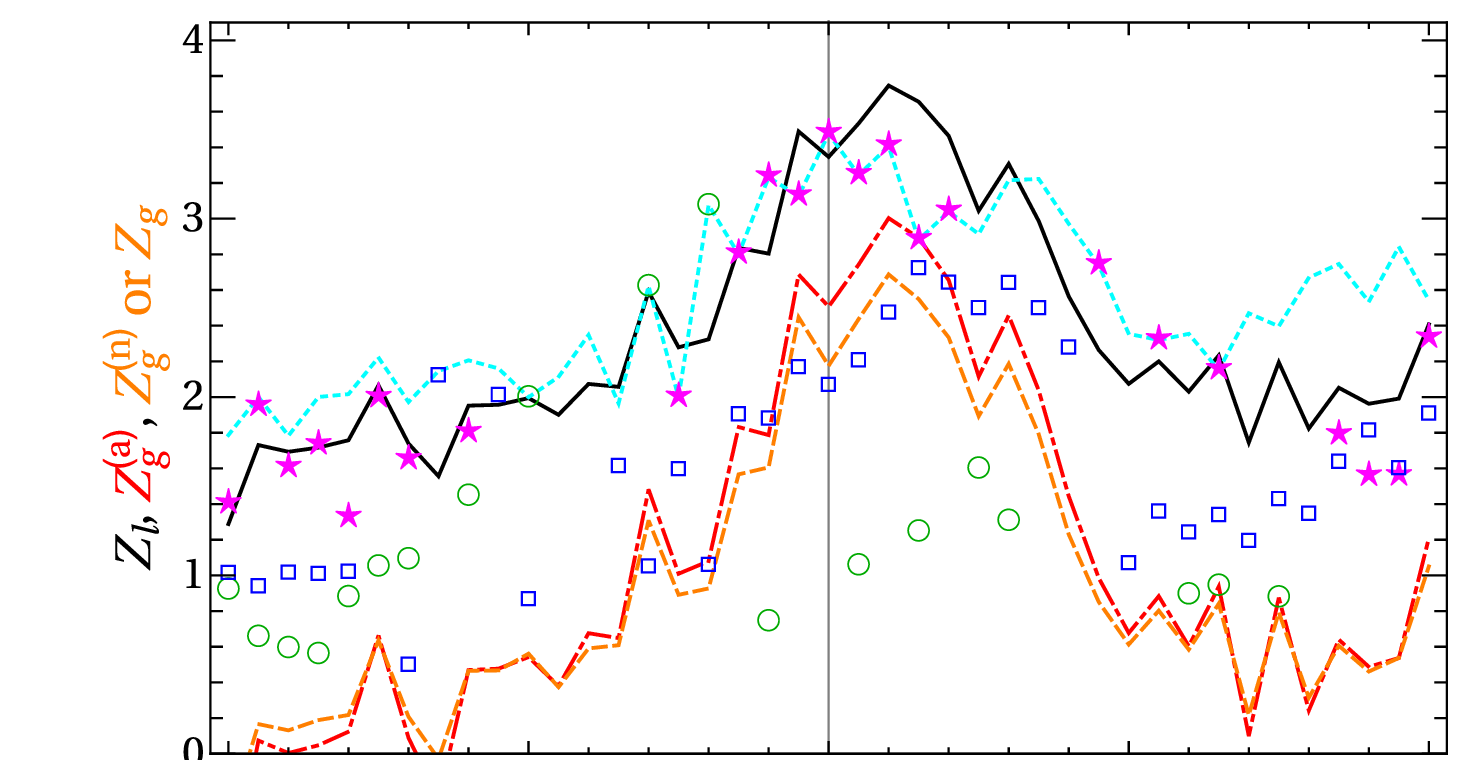}\hfill
    \includegraphics[width=0.46\textwidth]{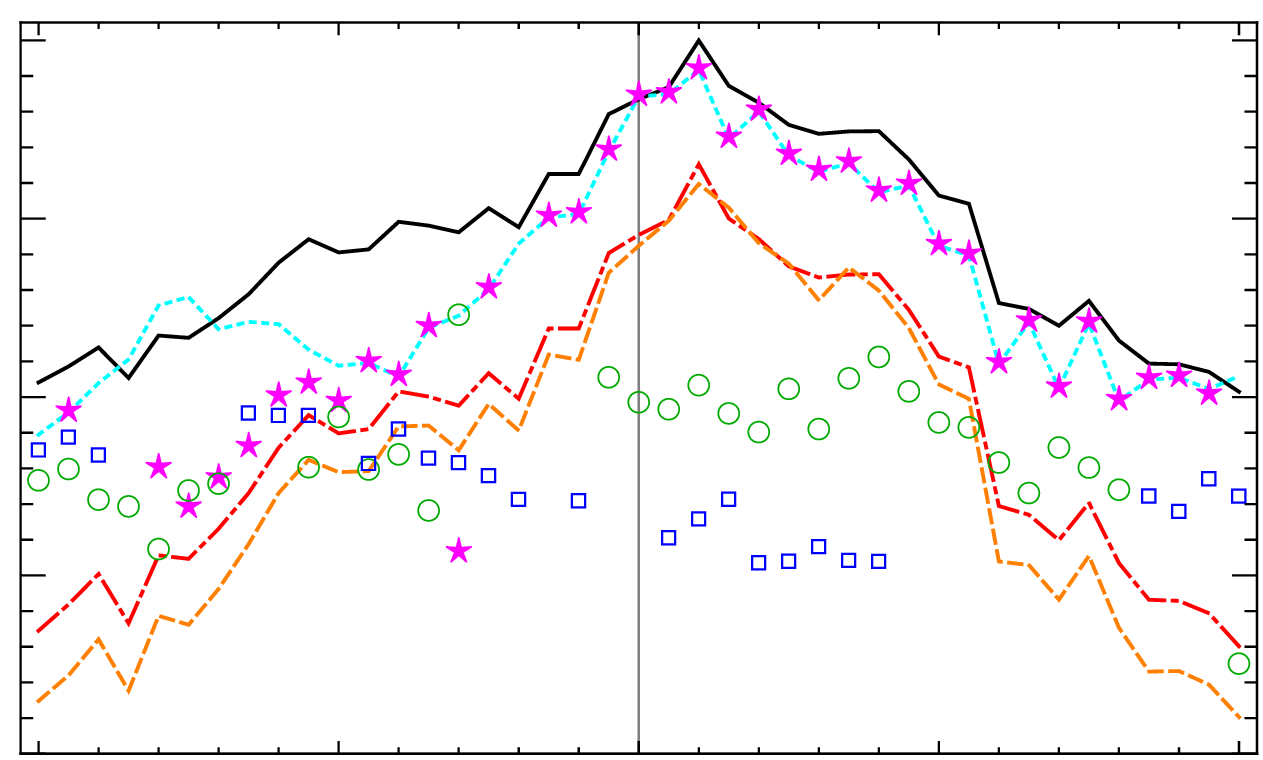}\\
    \includegraphics[width=0.54\textwidth]{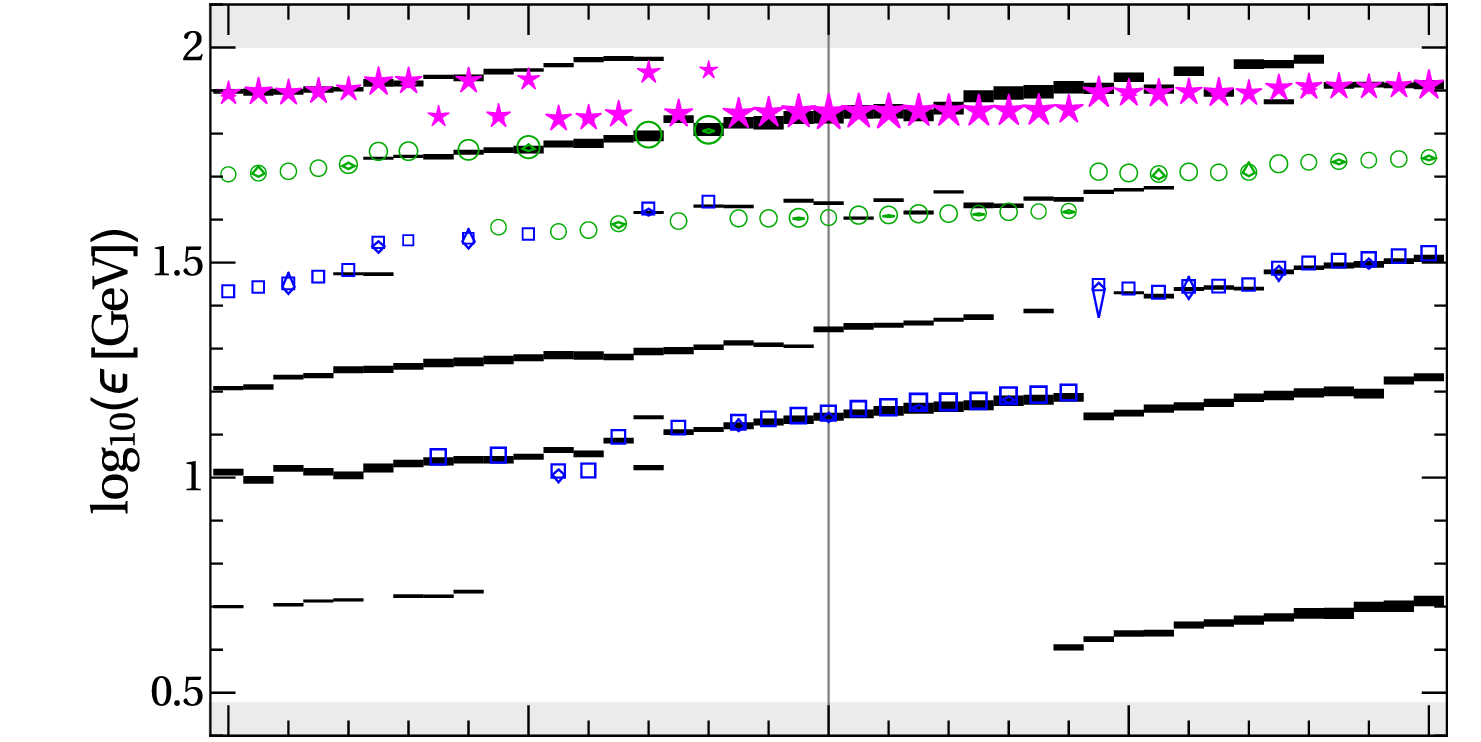}\hfill
    \includegraphics[width=0.46\textwidth]{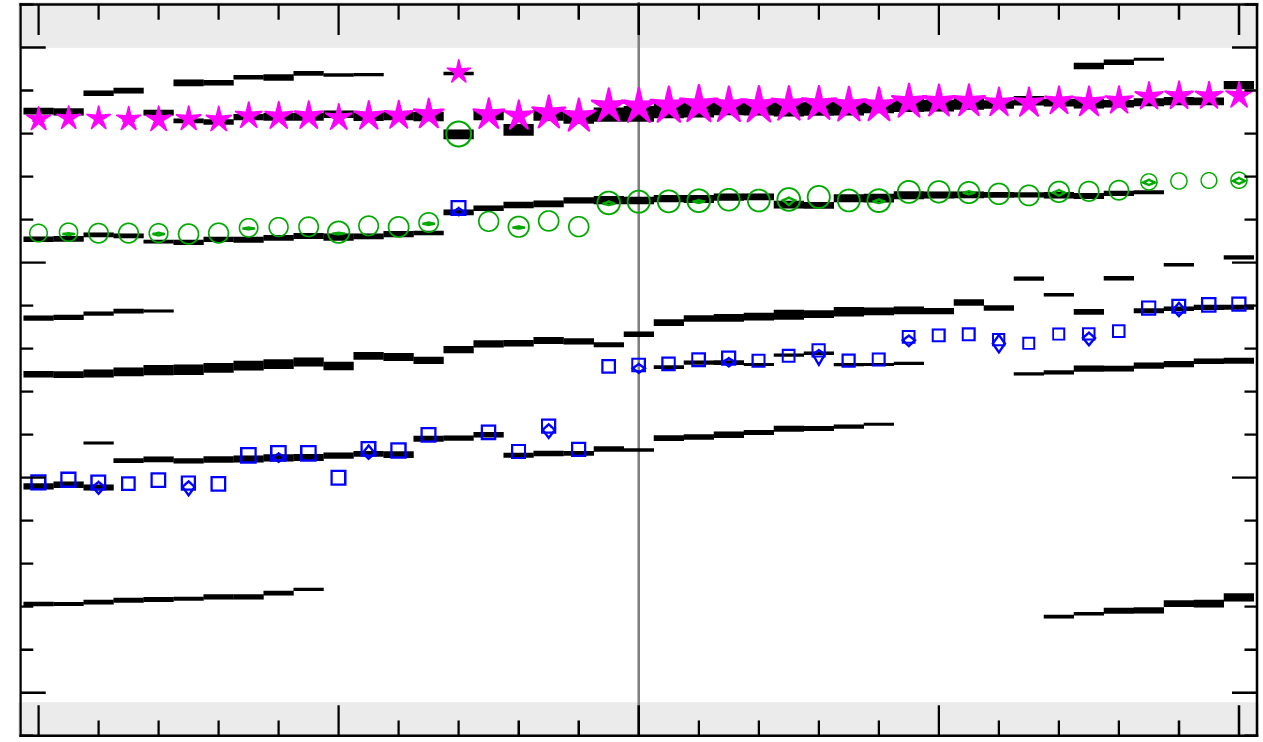}\\
    \includegraphics[width=0.54\textwidth]{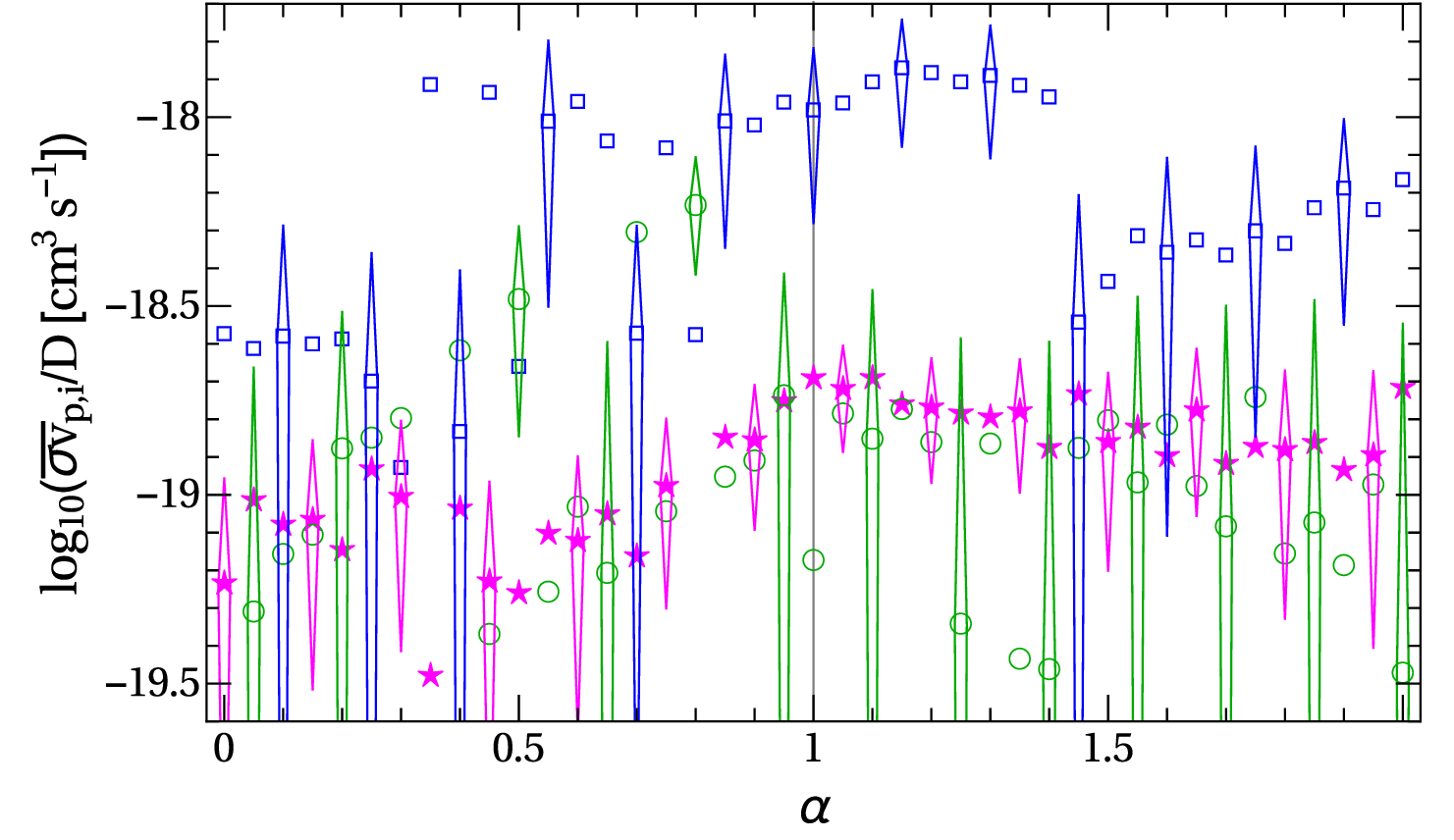}\hfill
    \includegraphics[width=0.46\textwidth]{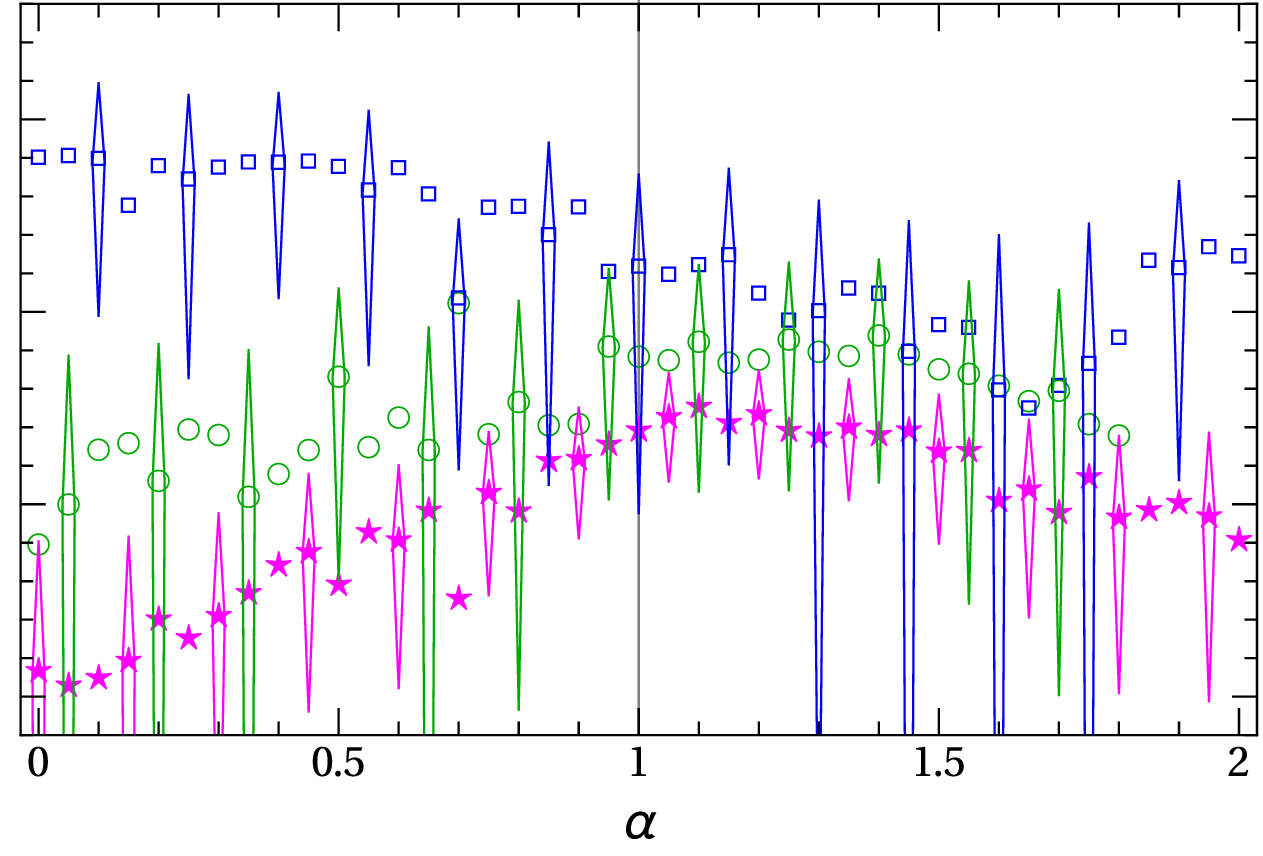}
    \caption{
        Response of \SAll\ (left) and \SJ\ (right) to redshift correction away from the $\alpha=1$ cluster frame (vertical line; the observer frame is $\alpha=0$).
        Top: $Z_l$ of strongest triad (solid black), line (dotted cyan), and triad channel-adjacent lines (symbols as in Fig.~\ref{fig:Spectra}). Triad global $Z$ scores shown per $\alpha$
        as $Z_g^{\smash[t]{(n)}}$ for \SAll\ and $Z_g$ for \SJ\ (dashed orange)
        and $Z_g^{\smash[t]{(a)}}$ (analytic; dot-dashed red).
        Middle: channel energies of the strongest triad (symbols of size increasing with channel $Z_l>0$, with $\Delta\textrm{TS}=1$ bounds as diamonds plotted every third $\alpha$ value even when smaller than symbol) and of single $Z_l>1$ lines (black lines of thickness linear in $Z_l$ within the unshaded scan range).
        Bottom: channel cross sections of the strongest triad as measured (before loss corrections; symbols), with their uncertainties (diamonds, as above; outliers cropped by the frame).
        The dependence on $\alpha$ is diluted by nearby clusters and is stronger (see Fig.~\ref{fig:AlphaFar}) for distant clusters.
        \label{fig:Alpha}
    }
\end{figure*}

Figure~\ref{fig:Alpha} presents the Z-scores, channel energies, and measured (not loss-corrected) cross sections
of the strongest lines and triad of \SAll\ (left column) as a function of the redshift scaling parameter $\alpha$; recall that $\alpha=1$ is the cluster frame and $\alpha=0$ is the observer frame.
Results shown for \SJ\ (right column) are discussed in \S\ref{subsec:JointSpectra}.

The numerical global scores in the top row (dashed orange) are calibrated as $Z_g^{\smash[t]{(n)}}$ by Gaussian realizations for \SAll\ and by null pairs with the shuffled-contrast \SMW\ skies as $Z_g$ for \SJ\ (see \S\ref{subsec:JointSpectra}), and do not multiple-trial correct for the scan over $\alpha$.
Symbols in the top row mark single lines near the strongest triad channels: for each channel, the line closest in energy among the ten strongest lines of the line scan at that $\alpha$ (away from the scan edges, with exclusion parameter $f=0.1$), provided that it lies within $5\%$ of the channel energy, plotted at its own $Z_l$.

The cross sections of this triad are seen to be a few $10^{-19}\cm^3\se^{-1}$ (before loss corrections), with $\gamma h$ stronger than the other channels, consistent with the \SCW\ analysis\cite{Keshet26Corr}.
The $\gamma\gamma$ and $\gamma Z$ cross sections appear comparable in \SAll, unlike in \SCW, but \SAll\ alone does not constrain $\gamma Z/\gamma\gamma$ well,
as its $\gamma Z$ cross sections are weak, with $\sim3$ times larger uncertainties than $\gamma\gamma$ (Table~\ref{tab:FeaturesS}).
The \SJ\ cross sections do show the same $\gamma h>\gamma Z> \gamma\gamma$ hierarchy as \SCW, but $\gamma Z> \gamma\gamma$ is carried by (the masked) \SCW.

The strongest \SAll\ triad reaches $Z_l\simeq3.3$ at $\alpha=1$ and peaks with $3.7$ at $\alpha\simeq1.1$, its mass drifting only from $69.5$ to $70.8\GeV$ for $0.85\le\alpha\le1.4$.
In \SAll, the fixed $T_{72}$ (not shown in the figure) rises from $(-0.2)$ in the observer frame to $2.2$ in the cluster frame, peaks at $2.4$ for $\alpha\simeq1.1$, and falls back to $(-0.5)$ at $\alpha=2$.
The preference of $T_{72}$ and of the strongest triad for $\alpha\simeq1.1$ over $\alpha=1$ is within the fluctuations between adjacent $\alpha$ values, as adjacent values shift different subsets of clusters by one native bin.
The $Z_l$ of the strongest blind nonad of \SAll\ at each $\alpha$ (see \S\ref{subsec:Nonads}) similarly peaks at $\alpha\simeq1$--$1.25$.

As in \SCW, small changes in $\alpha$ near the peak of the signal probe its spectral structure.
Near the cluster-frame redshift, features are robust to small changes in $\alpha$. However, jumps in channel energies and cross sections can be seen, in resemblance of the \SCW\ analysis, corresponding to switches of the strongest triad within the nonad outside $0.85\lesssim\alpha\lesssim1.4$ in \SAll\ and between its two leading triads at $\alpha\simeq0.95$ in \SJ.
Two separate weak ($Z_l\simeq1.0$--$1.5$) $\gamma Z$ lines $\sim8\%$ apart, too close for both to be extracted simultaneously, appear intermittently in \SAll\ as $\alpha$ varies:
the $43.5\GeV$ cross-annihilation $\gamma Z$ line (the a priori $L_{43}$ in our native binning)
appears for some values
including $\alpha=1$, whereas other values show the $\gamma Z$ line of the nominal triad, extrapolated to $40.1\GeV$ for $\alpha=1$.
Similarly, while the $\sim18\GeV$ cross-annihilation $\gamma h$ line is missing in \SJ\ at $\alpha=0$, it appears for every $1<\alpha<1.5$ value scanned.
More generally, the different lines in the second row
do not appear random; rather, they seem to belong to the nonad.

The observer frame ($\alpha=0$) provides a powerful control against most artifacts, whether instrumental (e.g., in the effective area or energy dispersion), foreground (Solar system or Galactic), or pipeline (binning, exposure, or detrending).
Such an artifact would typically be sharpest in the observer frame, and only smeared by the correction over different cluster redshifts (by up to $\sim11$ native bins for $z=0.5$).

With the same photons, errors, and pipeline as the cluster frame, the observer frame shows no feature at all.
Namely, it shows only noise levels of maximal rebinned pull $1.9$ (against $3.5$ in the cluster frame), nonad ($Z_l\simeq 1.2$ against $3.2$), triad ($1.3$ against $3.3$), and $T_{72}$ ($-0.2$ against $2.2$).
This artifact veto is even more powerful for \SJ\ before correction, which replaces the X-ray-correlated component of the masked \SCW\ by the $\sim2.4$ times more sensitive ambient sky\cite{Keshet26Corr}.

Appendix~\ref{app:SubSpectra} repeats the $\alpha$ scan for the hemispheres \SE\ and \SW.
Appendix~\ref{app:Dz} explores broadening (instead of shifting) the triad aperture in redshift.
The nominal triad is found to be sharpest for an unbroadened aperture, as expected for narrow lines emitted from low-dispersion WIMPs in the frame of each cluster.

\subsection{Distant clusters}
\label{subsec:Distant}

\begin{figure}[h!]
    \centering
    \includegraphics[width=\columnwidth,trim={1.8cm 0cm 0.1cm 0cm},clip]{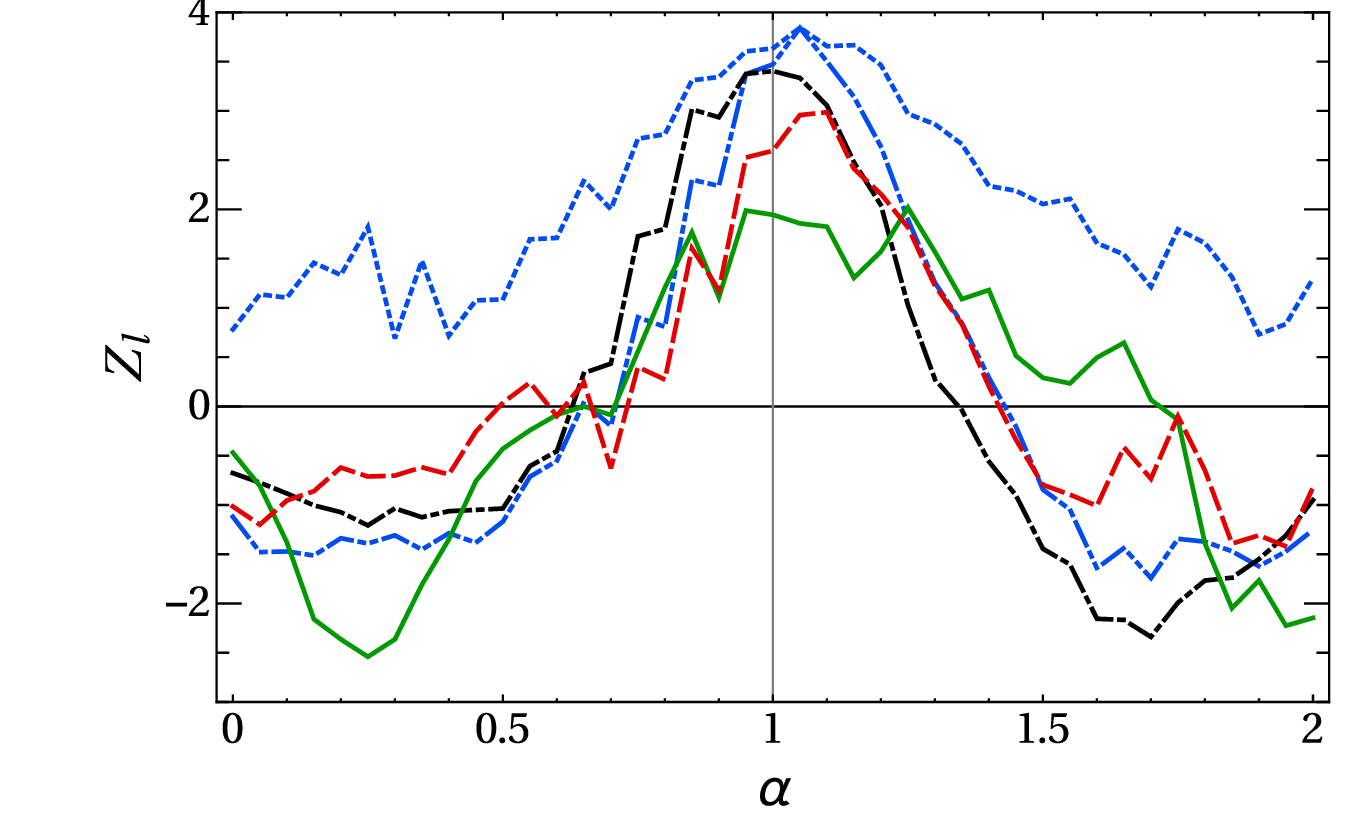}
    \caption{
        Local $Z_l$ as a function of $\alpha$ among the most distant ($0.38<z<0.50$) tertile of \SAll\ clusters, for fixed spectral features:
        the a priori $L_{43}$ (solid green), $T_{72}$ (dashed red), and the \SCW\ nonad (blind, masses scaled by 1.03; dot-dashed black).
        Also shown are the nominal nonad (blind $\{69.8,74.5\}\GeV$ of \SAll\ at $\alpha=1$; double-dot-dashed blue) and the strongest blind nonad at each $\alpha$ (dotted blue).
        \label{fig:AlphaFar}
    }
\end{figure}

While the spectral features in the top row of Fig.~\ref{fig:Alpha} clearly weaken away from $\alpha=1$, consistent with a cluster-frame signal, the effect is diluted by the inclusion of many low-redshift clusters.
To better probe the effect, we divide \SAll\ into three equal-sized groups (tertiles) of clusters by redshift.
The fractions of \{MCXC, $\Delta$eROSITA, $\Delta$DESI\} clusters are $\{16, 50, 34\}\%$ in the nearest ($z<0.24$) tertile, $\{3, 31, 65\}\%$ in the middle ($0.24<z<0.38$) one, and $\{1, 19, 79\}\%$ in the most distant ($0.38<z<0.50$) one.
Figure~\ref{fig:AlphaFar} plots $Z_l(\alpha)$ for a few spectral features among only the most distant, $\Delta$DESI-dominated tertile.

At $\alpha=1$, the blind nonad scan of this tertile finds a $\{68.8,74.5\}\GeV$ nonad with $Z_l\simeq3.6$, calibrated as $3.9\sigma$, stronger than the $Z_l\simeq3.2$ calibrated as $3.3\sigma$ for the nonad of the entire \SAll.
The former nonad drops sharply away from $\alpha=1$, whether we keep the nearly equal masses of the nominal nonad or scan for the strongest nonad at each $\alpha$.
Within the tertile, the nominal nonad is found in both northern and southern Galactic hemispheres ($Z_l\simeq2.8$ and $2.1$), above and below its median
$M_{500}$ ($2.9$ and $1.8$), and in both $\Delta$eROSITA and $\Delta$DESI ($2.0$ and $2.8$).

The \SCW\ nonad and $T_{72}$ each exceed $Z_l=2$ only within $0.85\le\alpha\le1.2$, peak at $\alpha\simeq1$--$1.1$, and fall below zero again by $\alpha\simeq1.45$, as does the nominal nonad (with $Z_l\simeq3.8$ at $\alpha\simeq1.05$).
The a priori $L_{43}$ ($Z_l\simeq1.1$ in \SAll) exceeds $1.5$ within $0.85\le\alpha\le1.3$, peaking at $2.0$.
The strongest blind nonad at each $\alpha$, a scan maximum and thus positive even without a signal,
exceeds $Z_l=2$ ($Z_l=3$) only for $0.65\le\alpha\le1.55$ ($0.85\le\alpha\le1.2$).

In this distant tertile, moving from the observer to the cluster frame raises the a priori \SCW\ nonad (either scaled by 1.03 or placed at $z=0.03$) from $(-0.7)$ to $3.4$, $T_{72}$ from $(-1.0)$ to $2.6$, and $L_{43}$ from $(-0.5)$ to $1.9$.
The nominal \SAll\ nonad and the strongest (not shown in figure) triad and line rise from $\{-1.1, -1.7, -0.6\}$ to $\{3.5,3.7,3.3\}$, but these features, blindly found at $\alpha=1$, are expected to peak there.
The other two tertiles gain less upon shifting from observer to cluster frames.
For instance, $T_{72}$, the nominal triad, and the $69.3\GeV$ line change from $\{0.4,0.4,1.6\}$ to $\{1.7,1.0,2.6\}$ in the nearest tertile, and from $\{0.2,-0.3,-1.0\}$ to $\{-0.5,1.0,0.0\}$ in the middle one.

The farthest tertile shows signals stronger than expected based on the cluster $\tilde{\mathcal{J}}$ factors (assuming common cross sections).
One anticipates, for example, a tertile distribution $\sim\{1.7,0.7,0.4\}$ of $T_{72}$, against the observed $\{1.7,-0.5,2.6\}$.
Here, the nearest tertile agrees, and the middle one falls short by statistically plausible $1.4$ null standard deviations, but the distant tertile exceeds expectations by $2.7$ null standard deviations ($p\simeq0.01$ for the three tertiles jointly).
Including the Poisson noise of the signal itself, simulations give $p\simeq0.03$, or $0.19$ after the multiple trials of nine subsample comparisons (Appendix~\ref{app:Samples}).

\subsection{Intrinsic cross section in cluster subsets}
\label{subsec:Scaling}

\begin{figure}[htb!]
    \centering
    \includegraphics[width=\columnwidth,trim={0.2cm 0cm 0.3cm 0cm},clip]{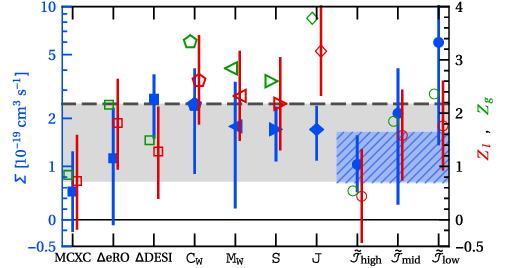}
    \caption{
        Intrinsic $\gamma\gamma$ cross section (measured $\Sigma$ before loss-correction; filled blue symbols, left axis) and $T_{72}$ scores (empty symbols, right axis) in the three catalogs (squares), the samples \SCW\ (pentagons), \SMW\ (left triangles), \SAll\ (right triangles), and \SJ\ (diamonds; \S\ref{subsec:JointFixed}), and tertiles of predicted signal (disks).
        Both local ($Z_l$ minus its null mean; red, with null standard-deviations error bars) and global (green without error bars) $T_{72}$ scores are shown.
        Correlation-based \SCW\ and \SMW\ are presented as $\Sigma_{\textrm{X}}$ with templates at $z=0.03$.
        Ranges of $\Sigma$ are given for tertiles (inverse-variance-weighted mean $\Sigma$ as hatched blue band) and of $\Sigma_{\textrm{X}}$ for different \SCW\ redshift models (Appendix~\ref{app:eROSITA}; gray band) ranging from nominal (dashed line) to uniform over $z<0.05$.
        \label{fig:Scaling}
    }
\end{figure}

If the reported spectral features, such as the a priori triad $T_{72}$, indeed arise from WIMP annihilation in galaxy clusters, then every subsample should agree on the same intrinsic cross sections.
Figure~\ref{fig:Scaling} presents the $\gamma\gamma$-channel cross section (as measured, before loss correction) $\Sigma$
and the $Z_l$ and $Z_g$ scores of $T_{72}$ in tertiles of the predicted
$\propto \smash{\tilde{\mathcal{J}}_\Myc}$ signal, in the three catalogs, \SCW, its masked \SMW, \SAll, and the latter two joint in \SJ.

As the figure shows, $\Sigma$ and $T_{72}$ are positive in all three tertiles, three catalogs, and four samples, and the measured $\Sigma$ is consistent across all cases.
In particular, the three catalogs (squares) give mutually consistent cross sections ($\chi^2\simeq2.6$ for 2 degrees of freedom).
The $\Sigma$ shown for \SCW\ and \SMW\ is based on the X-ray cross-correlation but expressed as the cross section that stacking would measure (\ie NFW mapping with pipeline losses), denoted $\Sigma_{\textrm{X}}$ and discussed in \S\ref{subsec:JointFixed}.

To construct the tertiles, the \SAll\ clusters of each catalog are ranked by their $\smash{\tilde{\mathcal{J}}_\Myc}$ factor and divided into three bins of equal numbers, joined across the catalogs ($\smash{\tilde{\mathcal{J}}}_{\textrm{high}}$, $\smash{\tilde{\mathcal{J}}}_{\textrm{mid}}$, and $\smash{\tilde{\mathcal{J}}}_{\textrm{low}}$ in Fig.~\ref{fig:Scaling}).
These factors are strongly peaked at nearby, extended clusters,
so for the subsets of this section we conservatively supplement the cross-catalog $<0\dgrdot02$ exclusion criterion by an additional, stricter, $<0.4\,\theta_{500}$ exclusion (for clusters within $0.01$ in $z$). This exclusion removes 160 $\Delta$eROSITA and $\Delta$DESI clusters, rendering the reduced sample less sensitive to any effect of overlapping or duplicate clusters on fine $\smash{\tilde{\mathcal{J}}_\Myc}$ binning, with a negligible effect on the stacked signal (\eg lowering $T_{72}$ by $0.07$,
similar to random removals of as many clusters).

Each 7050-cluster tertile holds a third of each catalog's clusters, with fractions $\{0.72, 0.20, 0.08\}$ of the predicted signal. The tertiles give $\Sigma/(10^{-19}\cm^3\se^{-1})$ of $1.0\pm0.5$, $2.2\pm1.9$, and $6.0\pm4.6$.
The inverse-variance-weighted mean, $\Sigma=(1.2\pm0.5)\times10^{-19}\cm^3\se^{-1}$ (hatched band in the figure), is more precise than $\Sigma=(1.7\pm0.6)\times10^{-19}\cm^3\se^{-1}$ found for \SAll\ itself, where all photons contribute equally, and favors a signal over its absence by $2.4$ standard deviations ($2.4\sigma$ in Monte Carlo).
The first, high-weight tertile gives the strongest measurement ($\Sigma$ divided by its error), consistent with expectations.
The three $\Sigma$ measurements are mutually consistent ($\chi^2\simeq1.5$ for 2 degrees of freedom), but suggestive of a trend.
Such a trend ($\Sigma$ rising toward low predicted signal) is expected in part from catalog mass uncertainties, as overestimated masses bias $\Sigma$ low in groups of strong predicted signal.

Modeling the line flux of each cluster as $\propto\smash{\tilde{\mathcal{J}}_\Myc^{\,\kappa}}$ (where $\kappa=1$ is the predicted scaling and $\kappa=0$ corresponds to an equal flux in each cluster), Poisson simulations give $\kappa=0.24\pm0.58$ (68\%) in the tertiles, disfavoring $\kappa=1$ at $1.3$ standard deviations
($p\simeq0.09$, multiple-trial corrected to $p\simeq 0.5$; Appendix~\ref{app:Samples}).
Alternative tertiles, quartiles, and quintiles give $\kappa\simeq-0.1$ to $0.4$
($p\simeq0.03$--$0.07$, or $\gtrsim0.3$ trial-corrected),
but they re-bin the same clusters; their $\Sigma$ are mutually consistent, as $\kappa=1$ predicts ($p\ge0.28$ in each set).
Note that a catalog $M_{500}$ scatter would raise $p$ (approximately doubling it for $30\%$ scatter); a signal more extended than modeled would also lower $\kappa$, as the fixed ROI captures less of the nearby, extended clusters.

The cross section of \SAll\ agrees with that of \SCW.
Before loss correction, stacking measures $\Sigma=(1.7\pm0.6)\times10^{-19}\cm^3\se^{-1}$ in \SAll\
and $(1.2\pm0.5)\times10^{-19}\cm^3\se^{-1}$ for the tertile mean, both within
the $\Sigma_{\textrm{X}}\simeq(0.7$--$2.5)\times10^{-19}\cm^3\se^{-1}$ that \SCW\ implies for stacking (gray band in Fig.~\ref{fig:Scaling}), for redshift distributions ranging from uniform over $z<0.05$ to the modeled one.
After loss correction (see \S\ref{subsec:Losses}), stacking gives $\Sigma_0=(2.7\pm1.0)\times10^{-19}\cm^3\se^{-1}$ in \SAll\
and $(1.9\pm0.8)\times10^{-19}\cm^3\se^{-1}$ for the tertile mean.
Both values lie within the $\Sigma_0\simeq(1.0$--$3.4)\times10^{-19}\cm^3\se^{-1}$ implied by \SCW\ (see Appendix~\ref{app:eROSITA}).

In summary,
the cross sections of the catalogs and of the tertiles are mutually consistent ($p\ge0.2$), and $\Sigma$ and $T_{72}$ are positive in each of the above subsets.
The tertile mean favors a signal at $2.4\sigma$, with a cross section consistent with \SCW, although the scaling with $\smash{\tilde{\mathcal{J}}_\Myc}$ is weaker than predicted.
The data place the intrinsic $\gamma\gamma$ cross section at $\Sigma_0\simeq(1$--$3)\times10^{-19}\cm^3\se^{-1}$ for the NFW profile, and $1.6$ times higher for the cored profile.

\subsection{Signal controls: localization and fields}
\label{subsec:Controls}

The signal is based on localizing the centers of clusters and removing their surrounding fields.
The controls below test these elements by
translating the ROI off the clusters and examining the stacked fields, also under quiet conditions.

\subsubsection{Off-center stacking}

\begin{figure}[h!]
    \centering
    \includegraphics[width=\columnwidth,trim={2.0cm 0cm 0.1cm 0cm},clip]{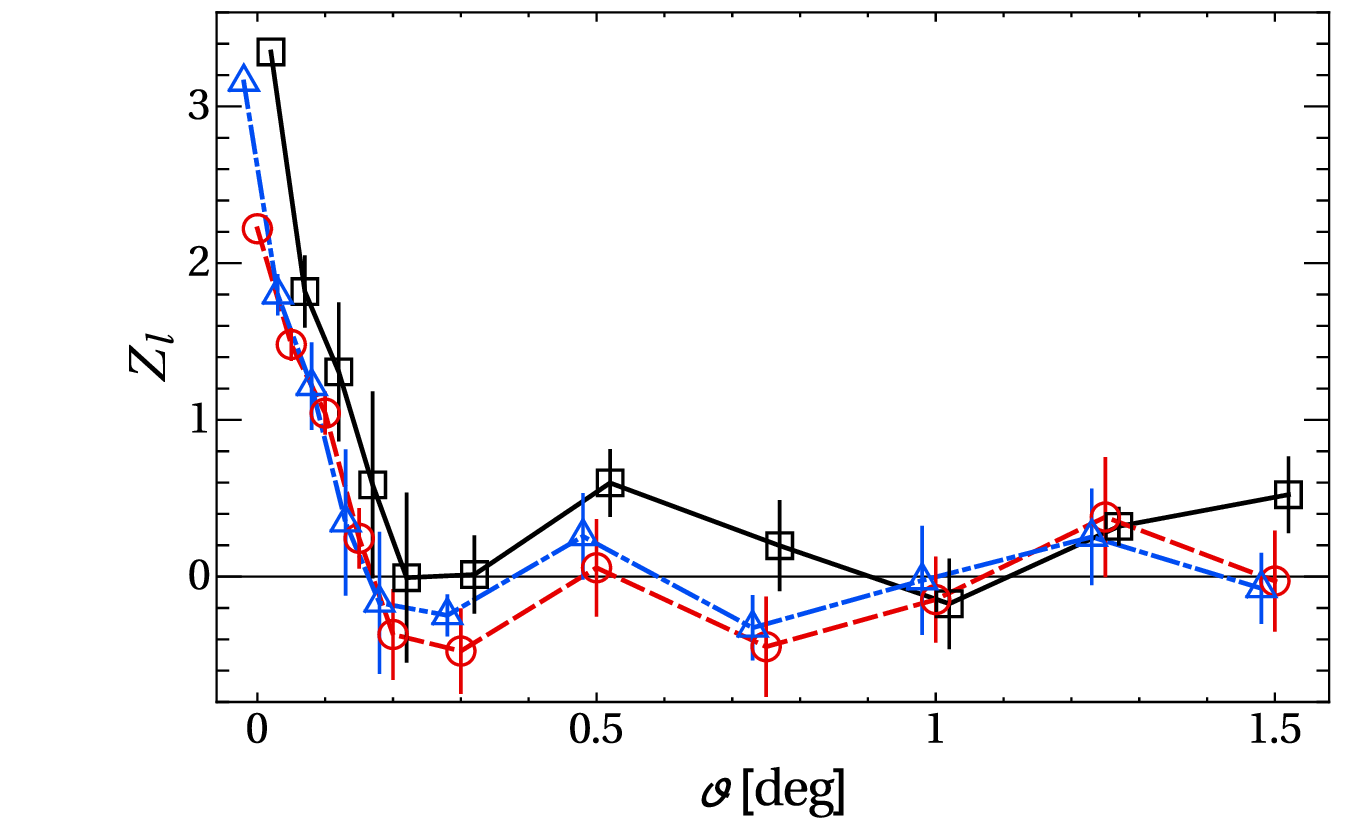}
    \caption{
        Mean $Z_l$, with its standard error, of $T_{72}$ (red circles with dashed line to guide the eye), the nominal triad (black squares, solid), and the nominal nonad (blue triangles, dot-dashed)
        over five random sets of directions of ROI translation by sky angle $\vartheta$.
        Nominal triad and nonad symbols are offset by $\pm0\dgrdot02$ in $\vartheta$ for visibility.
        \label{fig:OffCentre}
    }
\end{figure}

If the signal indeed arises from the very centers of clusters, then translating every ROI away from the center of its cluster by a fixed sky angle $\vartheta$, in random directions, should diminish the stacked signal once $\vartheta$ exceeds the ROI.
Figure~\ref{fig:OffCentre} shows the response of $T_{72}$, the nominal triad, and the nominal nonad to such translation.

Thus averaged over five sets of directions, the a priori triad $T_{72}$, nominal triad, and nominal nonad drop from
$\{2.2, 3.3, 3.2\}$ before offset to
$\{1.5, 1.8, 1.8\}$ at $\vartheta=0\dgrdot05$, and to
$\{1.0, 1.3, 1.2\}$ at $\vartheta=0\dgrdot1$.
Signals are no longer seen in the $0\dgrdot15\le\vartheta\le 1\dgrdot5$ range, where the above statistics span the ranges
$\{(-0.5)\mbox{--}0.4, (-0.2)\mbox{--}0.6, (-0.3)\mbox{--}0.3\}$.
We conclude that the signal is indeed confined to the cluster centers, approximately within the ROI.

\begin{figure*}[t]
    \centering
    \makebox[0.35\textwidth][l]{\hspace{0.069\textwidth}\makebox[0.265\textwidth]{\large\color{blue}$T_{72}$}}%
    \makebox[0.32\textwidth][l]{\hspace{0.039\textwidth}\makebox[0.265\textwidth]{\large\color{blue}$N_{72}$}}%
    \makebox[0.32\textwidth][l]{\hspace{0.039\textwidth}\makebox[0.265\textwidth]{\large\color{blue}Blind nonad}}\\[2pt]
    \includegraphics[width=0.35\textwidth]{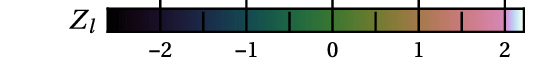}%
    \includegraphics[width=0.32\textwidth]{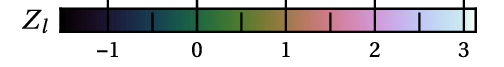}%
    \includegraphics[width=0.32\textwidth]{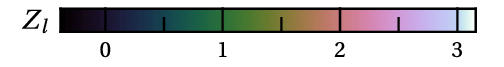}\\
    \includegraphics[width=0.35\textwidth]{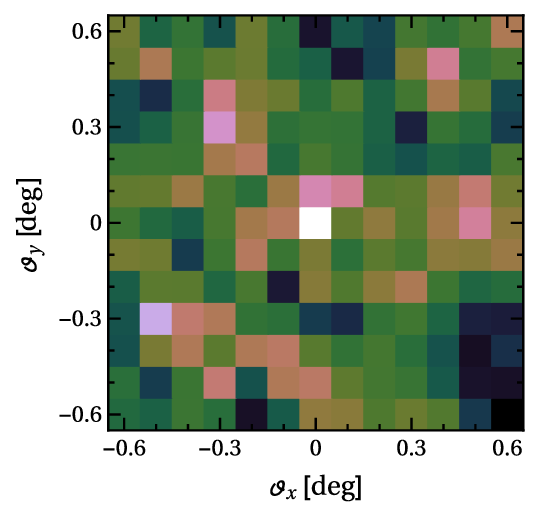}%
    \includegraphics[width=0.32\textwidth]{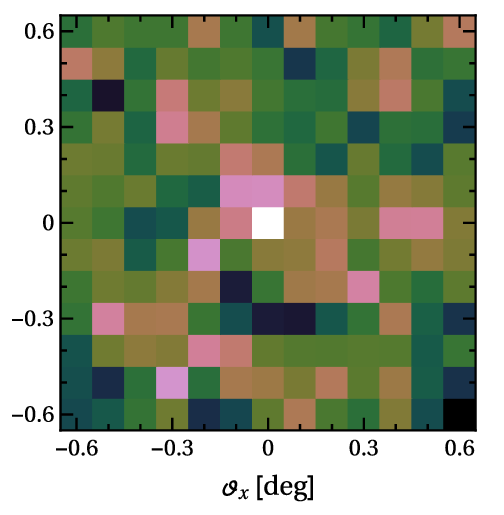}%
    \includegraphics[width=0.32\textwidth]{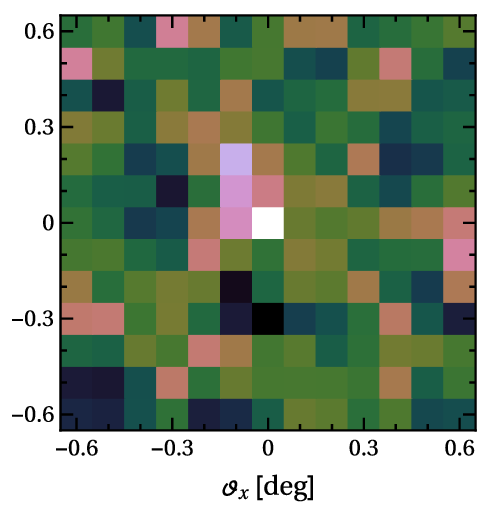}
    \caption{
        Stacked \SAll\ images of $T_{72}$ (a priori fixed mass), $N_{72}$ (one-mass scan), and a blind nonad (two-mass) scan.
        The nominal pipeline is applied on a $\{\vartheta_x, \vartheta_y\}$ grid in the tangent plane of each cluster, with $\vartheta_y$ increasing away from the Galactic plane and $\vartheta_x$ away from the Galactic-center meridian.
        Adjacent pixels, $0\dgrdot1$ apart, share ROI photons.
        Note that the matched filters in the left and right panels are entirely independent: the left one is a priori fixed by previous studies, while the right one scans two WIMP masses in each pixel.
        \label{fig:Images}
    }
\end{figure*}

The stacked images in Fig.~\ref{fig:Images} demonstrate such tests, for $T_{72}$, $N_{72}$, and a blind nonad scan.
Each pixel in these images fully repeats the nominal analysis, from local field subtraction to matched-filter application, but
at an angular offset (denoted $\{\vartheta_x, \vartheta_y\}$) from the center of each cluster.
The central pixel is our nominal result, and shows the maximal $Z_l$ in all three plots.

To test the pipeline's robustness against ambient gradients, the tangent plane of each cluster is oriented with $\vartheta_y$ increasing away from the Galactic plane, and $\vartheta_x$ away from the meridian of the Galactic center.
While Galactic foreground gradients then add up coherently in the stacked images, they are effectively eliminated by the local field subtraction, and no spurious effects are found in the images.
Orienting the tangent planes instead with Galactic north up, or rotating them randomly around each cluster, gives similar images, consistent with noise outside the central pixels.

\subsubsection{Stacked cluster fields}
\label{subsubsec:Stackedclusterfields}

Any spectral feature of the field, such as an instrumental or event-reconstruction artifact present throughout the sky, would enter the ROI and fields of clusters alike.
An additive feature, similarly enhancing the brightness of ROI and field, cancels by construction in the field subtraction of Eq.~\eqref{eq:DeltaN}. A multiplicative feature, acting on all photons alike (e.g., an error in the effective area or in the energy dispersion), would survive in $\delta I$ only in proportion to the cluster excess, but would be fully present in the field spectrum.

We therefore stack the field alone, i.e., use Eq.~\eqref{eq:deltaI} with $\Delta\Myn$ replaced by $\Myf$. This field spectrum is shifted to the cluster frame and detrended as $\delta I$, and its $Z_l$ is normalized by its null dispersion (including the covariance of photons shared by overlapping fields).
Averaged over $\sim100$ times the ROI area, the field measures the brightness of a line $\sim5$--$6$ times more sensitively than $\delta I$.
At the energies of the reported spectral features,
the field spectrum is $6$--$20$ times brighter
than the continuum excess of the clusters there, taken conservatively as the larger of the trend removed from $\Delta\tilde{I}$ by detrending and the $95\%$ upper bound of a power law fitted to $\Delta\tilde{I}$ at $3$--$100\GeV$.

Were the reported features of $\delta I$ multiplicative artifacts, the field would thus show them at $Z_l\simeq60$--$220$.
However, we find no field signals
corresponding to the lines, triads, or nonad of Table~\ref{tab:FeaturesS}, or to the a priori $T_{72}$, $C_{72}$, and $L_{43}$.
Hence, the field limits such artifacts to at most $4\%$ of the reported $Z_l$ of each of these features
($95\%$ confidence), and less than $0.9\%$ for the nominal triad.
The reported spectral features of $\delta I$ thus do not arise from artifacts acting on all photons.

\subsubsection{Stacking under quiet cluster fields}

We repeat the stacking analysis of both clusters and their fields, under stricter conditions on field contamination.
A quiet field, chosen both farther from the Galactic plane and cleaner of extended cluster emission, is obtained from the 19,632 clusters of \SAll\ whose fields lie entirely at $|b|>30\dgr$. From each field, we remove the disk of radius $\max(\theta_{\MyROI},\theta_{500})$ about every cluster of \SAll,
thus also removing the cluster emission beyond the ROI, which the nominal field retains.

The stacked clusters retain the features of the nominal analysis, while the fields still show nothing.
For instance, we find a $69.9\GeV$ blind triad with $Z_l\simeq 4.0$ and $T_{72}\simeq1.8$ (against $3.3$ and $2.2$ in \SAll).
Calibrated by Gaussian realizations of its noise (with the overlap covariance), scanned and selected as the data, the strongest blind line, triad, and nonad of this quiet field have $p\simeq0.22$, $0.44$, and $0.64$. The largest $Z_l$ of the quiet field at the eight features of Table~\ref{tab:FeaturesS} and the three fixed statistics ($T_{72}$, $C_{72}$, and $L_{43}$) has $p\simeq0.11$, calibrated likewise as the largest of them.

\subsection{Robustness to clusters and analysis choices}
\label{subsec:RobustTests}

The removal of Virgo and Fornax, the two \SCT\ clusters in \SAll\ (Ophiuchus lies below the latitude cut), was discussed in \S\ref{subsec:Samples} and \S\ref{subsec:FixedMass}.
No single cluster drives $T_{72}$. For fixed detrending, $T_{72}$ is a sum over clusters, granular because each cluster has few photons near the triad energies: in \SAll,
removing the five strongest contributors, a removal biased against the signal, still leaves $T_{72}\simeq1.9$.

Reasonable variations in analysis parameters change the results only gradually, and as expected from Poisson statistics.
For example, changes to the GPR correlation length ($0.20$--$0.35$ decades), linear-fit range (from $2$--$5\GeV$, up to $100$--$150\GeV$), the order of the field fit (quadratic to quartic), $\theta_f$ ($0\dgrdot8$--$1\dgrdot5$), native bin width ($1$--$2$ times $s_0$), and rest-frame alignment (split between adjacent channels), all leave $T_{72}$ in \SAll\ at $2.0$--$2.5$ ($2.3$--$3.0\sigma$) and the strongest blind triad at the nominal $69.9\GeV$ ($Z_l\simeq3.2$--$3.7$).
Variations of ROI radius ($0\dgrdot09$--$0\dgrdot15$), $\theta_{\min}$ ($0\dgrdot45$--$0\dgrdot75$), and catalog quality cuts ($0.4$--$0.6$ limits on \texttt{PCONT} and $30$--$60$ on $\lambda_{500}$) give similar results, scaled as expected with the number of photons retained (Appendix~\ref{app:Nulls}).
The signal is seen both before and after 2018.

Removing each of twelve sky regions in turn (four $90\dgr$ longitude ranges times three latitude bands) changes the $T_{72}\simeq2.2$ of \SAll\ to values between $1.6$ and $2.5$,
consistent with the $2.1\pm0.3$ expected for removing a twelfth of the statistical weight.
Restricting all catalogs to $|b|\ge20\dgr$ gives $T_{72}\simeq2.1$ ($2.4\sigma$) and a $69.9\GeV$ blind triad with $Z_l\simeq3.5$.
Removing the 425 clusters whose ROIs or fields reach within $8\dgr$ of the Large Magellanic Cloud or $3\dgr$ of the Small Magellanic Cloud
leaves $T_{72}\simeq2.4$ ($2.7\sigma$).
These results are all consistent with a signal distributed across clusters on the sky.

\section{Joint (\SJ) \SAll\ and \SMW\ (masked \SCW) analysis}
\label{sec:Joint}

Analyzing LAT data by stacking galaxy clusters or by cross-correlating with X-rays probes the same \gama-ray photons in complementary ways: \SAll\ focuses on the centers of clusters in our sample in their rest frames, while \SCW\ pulls the observer-frame overlap with integrated emission along the line of sight, primarily bremsstrahlung from extended LSS.
Both find triads and a nonad near $70\GeV$, at consistent masses (\S\ref{subsec:FixedNonads}) and cross sections (\S\ref{subsec:Scaling}).

Co-adding the $Z_l$ scores of the two analyses at each aligned energy or mass combines them into one dataset, which can be scanned and calibrated as a whole, and in which features add up only where they accurately match in both.
Such a co-addition requires the two datasets to share no ROI photons, so our joint analysis \SJ\ cannot pair \SAll\ with \SCW.
Instead, \SAll\ is paired with \SMW, defined as the \SCW\ cross-correlation recomputed after all pixels within $0\dgrdot1$ of any stacked cluster were masked.
The two datasets share photons in the stacked field annuli, but these are $>0\dgrdot1$ from the centers of the clusters and enter \SAll\ only by subtraction, averaged over $\sim100$ times the ROI area.

In order to derive \SMW\ and thus \SJ, here we incorporate the eROSITA map (see Appendix~\ref{app:eROSITA}) of the western Galactic sky, which was used until this point in the analysis only indirectly, through the outcomes of the \SCW\ analysis\cite{Keshet26Corr}.
The eROSITA-map pixels within $0\dgrdot1$ of the stacked clusters constitute $3.2\%$ of the analyzed pixels
and carry $6.9\%$ of the X-ray weight (\ie the exposure-weighted squared X-ray contrast).
Masking these pixels lowers the $T_{72}$ of \SCW\
from $3.0$ to $2.8$ (as observed; from $3.2$ to $2.9$ with its templates at $z=0.03$), as expected from this weight if the \SCW\ signal follows the X-rays.

Co-adding the cluster-frame \SAll\ with \SMW\ requires aligning the latter with an approximation of the rest frame, which is not well defined as the cross-correlated signal arises from a poorly constrained integrated range of redshifts.
The modeled distribution (\S\ref{subsec:FixedMass}), which weights halos by their X-ray surface brightness, peaks at $z\simeq0.03$, as nearby halos are rare in the analyzed sky and distant ones are unresolved by the LAT point-spread function\cite{Keshet26Corr}.
For simplicity, in the nominal \SJ, \SAll\ is thus combined with \SMW\ shifted to $z=0.03$; the results of other redshifted treatments are reported alongside.

Previous \SCW\ data support this nominal redshift offset.
With its templates placed at a fixed redshift $z$, the a priori $T_{72}$
is fairly uniform for $z=0.01$--$0.04$, peaking at $z\simeq0.02$.
Broadening the aperture uniformly over $0\le z\le\Delta z$ raises $T_{72}$
towards a $\Delta z\simeq0.05$ peak\cite{Keshet26Corr}, \ie a mean $z\simeq0.025$.
A similar behavior is found in \SMW, whose $T_{72}$ falls from $2.9$ when placed at $z=0.03$, to $2.2$ at $z=0.06$ and $1.0$ at $z=0.10$.
Note that a redshift offset of \SCW\ cannot be much larger than $0.05$, otherwise the line shifts would exceed the $\sim\pm4\%$ Gaussian core of the IRF,
and the lines would drop out of the a priori window.

In practice, \SJ\ sums, at each aligned energy or mass, the $Z_l$ of \SAll\ and of \SMW, and divides the sum by $\sqrt{2}$, as in Eq.~\eqref{eq:CoAddedZ}.
In the null pairs of \SJ,
each shuffled-contrast \SMW\ sky enters with its $Z_l$ normalized by its own errors.
Normalizing instead by the errors of the real data would raise the calibrated significance of the a priori \SJ\ tests by $\sim0.1\sigma$,
so the adopted normalization is conservative.

The \SJ\ spectra and their blind lines and triads are presented in \S\ref{subsec:JointSpectra}, whereas nonads are shown in \S\ref{subsec:JointNonads}.
Fixed mass results are discussed in \S\ref{subsec:JointFixed}, where they are also combined with \SCT.

\subsection{Spectra, lines, and triads of \SJ}
\label{subsec:JointSpectra}

Some of the above figures already show \SJ\ results: the raw vs. cluster-corrected spectra in Fig.~\ref{fig:Spectra} (right column), the triad and nonad scans in Fig.~\ref{fig:NonadMap} (bottom row), and the redshift-correction response in Fig.~\ref{fig:Alpha} (right column).

As \SJ\ combines the $Z$-scores of \SAll\ and \SMW, its spectra (in Fig.~\ref{fig:Spectra}) simply show the sum of
their normalized brightness $\delta I/\sigma$, divided by $\smash{\sqrt{2}}$.
The uncorrected spectra (red diamonds in Fig.~\ref{fig:Spectra}) and matched-filter scans (Fig.~\ref{fig:NonadMap}, left column)
lack a last analysis stage:
picking the cross-correlated component in \SCW, and boosting to the cluster frame in \SAll; accordingly, the uncorrected \SJ\ (denoted \SJ$_0$) combines the observer-frame \SAll\ with the ambient \SMW\ spectrum.
In the redshift-correction tests of \S\ref{subsec:Alignment} (and Fig.~\ref{fig:Alpha}), the \SMW\ templates are placed at $1+z=1.03^{\alpha}$, so like \SAll\ they coincide with the observer frame for $\alpha=0$, and with the nominal ($z=0.03$) rest frame for $\alpha=1$.

We begin with a blind matched-filter scan of the \SJ\ spectrum, essentially a no-overlap combination of the blind scans of \SAll\ (in \S\ref{sec:Results}) and of the earlier \cite{Keshet26Corr} \SCW\ (masked and redshift-corrected).
Table~\ref{tab:FeaturesJ} lists the resulting strongest blind (with no masses fixed a priori) spectral features of \SJ.
Blind scans for alternative, non-nominal treatments of the \SMW\ redshift distribution are deferred to Appendix~\ref{app:JointStats}.

\begin{table}[h]
    \centering
    \caption{Strongest blind spectral features of \SJ.\label{tab:FeaturesJ}}
    \setlength{\tabcolsep}{2.5pt}
    \begin{tabular}{@{}l c c c c@{}}
        \toprule
        Feature & $\epsilon$ & $Z_l$ & $Z_g$ & $\{\epsilon_{\gamma Z},\epsilon_{\gamma h}\}$ \\
         & (GeV) & (\SMW, \SAll) & ($Z_g^{(n)}$) & (GeV) \\
        \midrule
        \multirow{4}{*}{\shortstack[l]{Lines\\($f=0.1$)}} & $70.1_{-1.0}^{+1.4}$ & 3.7\,(1.8,\,3.4) & 2.4\,(2.4) & \\
         & $44.1_{-1.1}^{+0.8}$ & 2.0\,(1.7,\,1.0) & --- & \\
         & $21.5_{-0.9}^{+2.2}$ & 1.6\,(0.9,\,1.4) & --- & \\
         & $11.6_{-0.5}^{+2.4}$ & 1.1\,(0.7,\,0.9) & --- & \\
        \midrule
        \multirow{3}{*}{\shortstack[l]{Triads\\($f=0.015$)}} & $72.2_{-0.2}^{+0.2}$ & 3.7\,(3.0,\,2.2) & 2.9\,(2.9) & $\{43.4,\,18.0\}$ \\
         & $69.1_{-0.6}^{+0.8}$ & 3.3\,(2.1,\,2.6) & 2.3\,(2.4) & $\{39.0,\,12.4\}$ \\
         & $88.7_{-0.7}^{+0.7}$ & 2.1\,(1.8,\,1.2) & 0.4\,(0.8) & $\{65.2,\,44.4\}$ \\
        \midrule
        \multirow{2}{*}{\shortstack[l]{Nonad\\($f=0.05$)}} & $69.1_{-0.6}^{+1.0}$ & \multirow{2}{*}{3.9\,(2.6,\,3.0)} & \multirow{2}{*}{3.8\,(3.6)} & \\
         & $74.9_{-1.2}^{+0.8}$ & & & \\
        \bottomrule
    \end{tabular}
    \TblNote{Same as Table~\ref{tab:FeaturesS}, but for \SJ.
    In addition to the combined $Z_l$ score,
    those of \SMW\ and \SAll\ are also provided
    separately (in parentheses).
    In addition to the nominal $Z_g$ of null pairs with shuffled-contrast \SMW\ skies, $Z_g^{\smash[t]{(n)}}$ with Gaussian \SMW\ realizations is also provided (in parentheses; Appendix~\ref{app:JointStats}).
    }
\end{table}

The strongest blind triad of \SJ, at a global $2.9\sigma$, is found at $72.2\GeV$, coincident with the a priori $T_{72}$, which is the second-strongest triad in \SAll.
A close second in \SJ, at $2.3\sigma$, is a $69.1\GeV$ triad, consistent with the strongest, nominal triad of \SAll.
The third-strongest, higher-energy triad of \SAll\ is located on a broad $Z_l$ shoulder of \SJ, seen in Fig.~\ref{fig:NonadMap} (middle panel in bottom row). This weaker triad does not appear in Table \ref{tab:FeaturesJ} because it is masked by the nominal exclusion window;
reducing the latter to $f=0.01$ (small symbols in Fig.~\ref{fig:NonadMap})
uncovers a local $Z_l\simeq 2.9$ peak at $73.8\GeV$.

Some \SJ\ features, like its second, $69.1\GeV$ triad and its strongest, $\{69.1, 74.9\}\GeV$  nonad, have comparable contributions from \SAll\ and from \SMW.
Other features are less balanced, like the strongest \SJ\ line ($70.1\GeV$) arising mostly from \SAll, or the strongest triad ($72.2\GeV$) leaning mostly on \SMW.

The blind \SJ\ features depend weakly on the redshift correction applied to \SMW\ (see Appendix~\ref{app:JointStats}).
Over these treatments, the strongest \SJ\ triad reaches $2.6$--$3.1\sigma$, lying either at $72.0$--$72.3\GeV$ (where \SMW\ dominates) or at $\sim 69.7\GeV$ (where \SAll\ dominates);
a second triad then lies near the other of these two masses.
In every treatment, one of the two strongest triads of \SJ\ lies within one native bin of the nominal triad, while the other is consistent with $T_{72}$.

Like \SAll, the \SJ\ features in Fig.~\ref{fig:Alpha} (right column) are maximized near the approximate cluster frame.
The strongest triad, with $Z_l\simeq2.1$ ($0.3\sigma$) at $\alpha=0$, grows to $3.7$ ($2.8\sigma$) at $\alpha=1$, peaks with $4.0$ at $\alpha\simeq1.1$, and rapidly declines for larger $\alpha$.
This $\alpha$-dependence is largely carried by \SAll, and is especially attributed to its distant clusters shown in Fig.~\ref{fig:AlphaFar}.
With its low redshift constituents, \SMW\ carries \eg only about a fifth of the decline beyond $\alpha=1$.

The global $Z$-scores of \SJ\ in Fig.~\ref{fig:Alpha} are calibrated by at least $10^4$ null pairs at each $\alpha$, each pair consisting of a Monte Carlo realization and a shuffled-contrast \SMW\ sky. This calibration is sufficiently accurate for tracing the curves. For example, the $4\times10^5$ pairs of Table~\ref{tab:FeaturesJ} calibrate the $\alpha=1$ triad as $2.9\sigma$, only slightly more accurately than the figure's $2.8\sigma$.

The channel cross sections of \SJ\ (bottom right in Fig.~\ref{fig:Alpha}) approximately combine \SAll\ and \SMW, with weights inversely proportional to their uncertainties. Each \SMW\ channel is first expressed as the cross section that stacking would measure: its brightness is loss-corrected, converted from the X-ray-gas mapping\cite{Keshet26Corr} of \SCW\ to the NFW mapping and losses of stacking, tuned, for simplicity, for a $\sim72\GeV$ triad spread over the modeled redshift distribution
(see Appendix~\ref{app:JointStats}).
The result shows the $\gamma\gamma<\gamma Z<\gamma h$ hierarchy of \SCW\ but, as mentioned in \S\ref{subsec:Alignment}, with the lower, $\gamma\gamma<\gamma Z$ behavior carried only by \SMW.

\subsection{Nonads of \SJ}
\label{subsec:JointNonads}

The nonad scan of \SJ\ in Fig.~\ref{fig:NonadMap} (bottom row) essentially combines those of \SCW\ (top row, after masking to \SMW\ and a $z=0.03$ shift) and of \SAll\ (middle row), so we first discuss the latter two.

In \SAll, while the observer frame scan (left panel) is consistent with noise (right), with nonads reaching at most $Z_l\simeq1.2$, the cluster frame (middle) shows the pronounced $Z_l>2$ triads and $Z_l>3$ nonad of Table \ref{tab:FeaturesS}.
These features manifest in the $m_1$--$m_2$ plane as horizontal, vertical, and anti-diagonal stripes, strongest in the $66$--$75\GeV$ range.
The strongest nonad (nominal; empty right-triangle) is found at the crossings of such stripes.

The unmasked \SCW\ is shown in the top row as observed, \ie with no redshift correction.
While the much deeper, ambient spectrum (left panel) is consistent with noise, with nonads reaching at most $Z_l\simeq1.7$,
the X-ray-correlated spectrum (middle) shows the aforementioned (in \S\ref{sec:Intro}) $Z_l\gtrsim3$ triads and nonad\cite{Keshet26Corr}.
In the $m_1$--$m_2$ scan, these again appear as horizontal and vertical stripes, with an $(m_1+m_2)c^2/2\simeq 70\GeV$ ridge, meeting at the most likely \SCW\ nonad (empty left-triangle).
The \SCW\ features are broader, and appear $\sim2$--$3\%$ lower in mass than their \SAll\ counterparts.
Multiple thin stripes can be seen, suggesting separate LSS contributions along the line of sight.

The blind nonad scan of \SMW\ finds the same $\{66.4, 71.9\}\GeV$ (as observed) strongest nonad as the unmasked \SCW, at a global $1.6\sigma$ (in its $2\times10^4$ shuffled-contrast skies).
Scaled by 1.03,
its $\{68.4, 74.1\}\GeV$ masses lie within the bounds of the nominal nonad
(of \SAll; Table~\ref{tab:Nonads})
despite sharing
no ROI photons with \SAll.
At these scaled masses,
\SAll\ gives $3.7\sigma$,
and the two Fisher-combine to $3.8\sigma$.
With its templates placed at $z=0.03$, \SMW\ finds instead a $\{69.1, 75.6\}\GeV$ nonad ($1.9\sigma$), within $1.5\%$ of the nominal nonad, where \SAll\ gives $3.5\sigma$, and the two again Fisher-combine to $3.8\sigma$.
Therefore, the nominal nonad and the blind \SCW\ and \SMW\ nonads (scaled by $1.03$ or placed at $z=0.03$)
appear to be one and the same (see \S\ref{subsec:FixedNonads}).
At the masses of the nominal nonad, \SMW\ is carried mainly by the intermediate, $\sim72\GeV$ triad of this nonad.
At these masses, \SMW\ has $Z_l\simeq2.4$ at $z=0.03$, with a shallow maximum at $z\simeq0.06$.

More generally, the nonad maps of \SMW\ (placed at $z=0.03$) and of \SAll, which share no ROI photons, correlate with a Pearson coefficient of $0.59$ over the mass grid,
as only $2.5\%$ of pairs of independent null maps do.
This alone suggests that the two analyses are seeing the same signal, despite their different methodologies, pipelines and redshift treatments.
Placing \SMW\ at different redshifts maximizes this Pearson coefficient at $z\simeq0.023$, although the peak is shallow; by $z=0.1$ it drops to $0.32$.

The \SJ\ nonad scan in Fig.~\ref{fig:NonadMap} (middle panel of bottom row) shows the same expected combination of horizontal, vertical, and anti-diagonal stripes as in panels above it, but stronger. The scan apex (empty green diamond) overlaps those of \SAll\ (above it) and, after shifting to $z=0.03$, also of \SCW\ (above \SAll).
The bottom left panel shows the uncorrected \SJ$_0$, combining the observer-frame control of \SAll\ with the ambient (not X-ray-correlated; at $z=0.03$) \SMW\ spectrum.
Despite the much more sensitive ambient component, this panel shows nothing significant.

If the reported spectral features were artifacts (instrumental, foreground, or pipeline), they would in general be much clearer in this ambient spectrum, passing all photons in the \SMW\ sky region through the same instrument response, detrending, and matched filters.
However, the strongest \SJ$_0$ triad reaches only $Z_l\simeq1.8$, as noise typically does (bottom right panel) in such a scan.
A multiplicative artifact
would have a $T_{72}$ signal-to-noise ratio $\sim110$ times higher in the ambient spectrum than in its X-ray-correlated counterpart.
Had $T_{72}$ of \SMW\ been such an artifact, it would present with $Z_l\simeq300$ in the ambient spectrum, instead of its measured $Z_l\simeq-0.3$.
The ambient component of \SJ$_0$ thus limits such artifacts to $<0.5\%$ ($95\%$ confidence) of the $T_{72}$ in \SMW.
This conservative limit is
stronger than the $\le4\%$ that the stacked field imposes (in \S\ref{subsubsec:Stackedclusterfields}) on each feature of \SAll.

The strongest \SJ\ nonad (empty green diamond) reaches $Z_l\simeq3.8$--$4.0$ in every redshift treatment, calibrated as $3.8\sigma$, and as $3.9$--$4.0\sigma$ over the other treatments.
The masses of this nonad change only little with redshift treatment, coinciding within one native bin with the nominal nonad when \SMW\ is broadened, shifted, folded, or placed at $z=0.03$.
The cross-annihilation triad of this nonad thus coincides, within one native bin, with the a priori $72.1\GeV$ triad,
whose $\gamma Z$ line is the a priori $L_{43}$.

Consider the nominal (of \SAll) nonad in \SJ.
This nonad coincides with the blind \SJ\ nonad, within one native bin nominally or when placing \SMW\ at any $0.01\leq z\leq0.10$, and exactly for $0.04\leq z\leq 0.07$.
Its three triads, at $\{69.8, 72.2, 74.5\}\GeV$, reach $Z_l\simeq\{3.2, 3.7, 2.7\}$.
The low triad is carried mainly by \SAll, the intermediate mainly by \SMW, and the high triad weaker but carried by both.
Placing \SMW\ at a higher $z$ would strengthen its high-triad support, from $\{1.2,2.9,1.1\}$ at $z=0.01$ to $\{1.2,2.7,2.2\}$ at $z=0.05$; \SJ\ would reach a balanced $Z_l\simeq\{3.2, 3.4, 3.2\}$ if \SMW\ were placed at $z=0.05$.

Conditioned on its two companions, the three triads add $Z_c\simeq\{2.2, 2.1, 1.6\}$ nominally,
or $\{2.4, 1.5, 2.3\}$ for \SMW\ placed at $0.05$,
so each carries significance beyond the other two.
The expected $Z_c$ values depend on the WIMP abundances, cross sections, and line branching ratios.
If all nine nonad lines are equally bright, one would expect $Z_c\simeq\{2.1, 1.7, 2.0\}$, as adjacent triads share photons.
The six measured triads
(in \SAll\ and \SMW, each dataset scaled to its own strength)
are consistent with this pattern ($p\simeq0.16$ nominally; $0.35$ for $z=0.05$; see Appendix~\ref{app:JointStats}).

If the two species are equally abundant and with equal cross sections, then the intermediate, cross-annihilation triad should be twice as bright (\S\ref{subsec:FixedTests}), giving $Z_c\simeq\{1.4, 2.4, 1.4\}$, which the data mildly disfavor ($p\simeq0.045$; $0.09$).
If each triad split its $\gamma\gamma$:$\gamma Z$:$\gamma h$ lines in the best-fit $\sim1$:$2$:$8$ ratio derived in \S\ref{subsec:ModelTests}, one would expect $Z_c\simeq\{1.8, 1.7, 2.3\}$ for equally strong triads ($p\simeq0.11$; $0.27$) and $\{1.2, 2.4, 1.6\}$ for an intermediate triad twice as bright ($p\simeq0.035$; $0.07$).

The null dispersion scans in Fig.~\ref{fig:NonadMap} (right column) are all smooth, showing no significant structure or bias, as discussed for the case of \SAll\ in \S\ref{subsec:Nonads}.
Over $2000$ of the shuffled-contrast skies of \SCW, the standard deviation of the X-ray-correlated component (top right) lies in the range $0.51$--$0.79$.
Over null pairs combining \SAll\ with the shuffled-contrast \SMW\ skies, the \SJ\ map has a standard deviation of $0.57$--$0.82$ (bottom right).
These scans thus support the viability of the middle-column features.

In conclusion, the nonad scans of \SAll\ and \SMW, which share no ROI photons, are correlated over the mass grid, and find the same nonad (within one native bin throughout $0.015<z<0.07$).
This is also the strongest blind nonad ($3.8$--$4.0\sigma$) of their co-addition \SJ, in every redshift correction, to within one native bin, with its cross-annihilation triad matching the a priori $T_{72}$.
The uncorrected \SJ$_0$ shows nothing, which limits multiplicative artifacts to $<0.5\%$ of $T_{72}$ in \SMW.
With \SMW\ placed at $z=0.01$--$0.06$
(around the nominal $z=0.03$ peak of the model\cite{Keshet26Corr}), both datasets contribute to this nonad, and its three triads each add significance beyond their two companions.
The relative line strengths are consistent with several models, slightly favoring triads of equal intrinsic strength.

\subsection{Joint results at the fixed mass}
\label{subsec:JointFixed}

Next, we apply the a priori tests based on \SCT\ (see \S\ref{subsec:FixedTests}) to \SJ, analogous to the \SAll\ tests shown in \S\ref{subsec:FixedMass} and \S\ref{subsec:FixedNonads}.
The results are summarized in Table~\ref{tab:JointSummary}, along with the main blind features of \S\ref{subsec:JointSpectra} and \S\ref{subsec:JointNonads}, shown for comparison.
As \SJ\ uses \SMW, which derives from \SCW, the latter cannot be used as a prior (as done for \SAll); only $L_{43}$ can be used here to fix the triad mass of \SJ\ a priori.

\begin{table*}[t]
    \centering
    \caption{Strongest blind and fixed-mass \SJ\ features, alone and combined with \SCT\ or \SCTb.\label{tab:JointSummary}}
    \begin{tabular}{l c c c c c}
        \toprule
        Feature & $\epsilon$ or $mc^2$ (GeV) & $Z_l$ (\SMW, \SAll) & $Z_g$ ($Z_g^{\smash[t]{(n)}}$) & With \SCT & With \SCTb \\
        \midrule
        \multicolumn{6}{l}{Strongest blind features} \\
        Line & $70.1$ & 3.7 (1.8, 3.4) & 2.4 (2.4) & --- & --- \\
        Triad & $72.2$ & 3.7 (3.0, 2.2) & 2.9 (2.9) & 5.0 & 3.8 \\
        Nonad & $\{69.1, 74.9\}$ & 3.9 (2.6, 3.0) & 3.8 (3.6) & 5.6 (5.4) & 4.5 (4.3) \\
        \midrule
        \multicolumn{6}{l}{At fixed masses} \\
        $L_{43}$ & $43.2$ & 1.8 (1.4, 1.1) & 1.6 (1.8) & 4.3 (4.4) & 2.9 (3.0) \\
        $C_{72}$ & $17.7$ & 3.2 (2.5, 2.1) & 3.2 (3.3) & 5.2 & 4.0 (4.1) \\
        $T_{72}$ & $72.1$ & 3.6 (2.9, 2.2) & 3.8 (3.8) & 5.6 & 4.5 \\
        $N_{72}$ & $\{69.7, 74.4\}$ & 3.9 (2.4, 3.1) & 4.5 (4.2) & 6.0 (5.8) & 5.0 (4.8) \\
        $N_{72}^{\max}$\,$^\dagger$ & $\{69.7, 74.4\}$ & 3.9 (2.4, 3.1) & 4.1 (3.9) & 5.8 (5.6) & 4.7 (4.6) \\
        \bottomrule
    \end{tabular}
    \TblNote{Results shown for the nominal treatment of \SMW\ (placed at $z=0.03$; the others in Appendix~\ref{app:JointStats}). Columns as in Table~\ref{tab:FeaturesJ} ($Z_g$ with all $20000$ skies; Appendix~\ref{app:JointStats}), and $Z_g$ ($Z_g^{\smash[t]{(n)}}$ in parentheses, where different) Fisher-combined with the $L_{43}$ of \SCT\ ($Z_g\simeq4.3$) or \SCTb\ ($Z_g\simeq2.7$), for blind features only if they hold a $\gamma Z$ line at the energy of $L_{43}$.\\
    $^\dagger$The strongest of the three readings of $T_{72}$ (\S\ref{subsec:FixedTests}), taken alike in the data (where it is $N_{72}$) and in each null pair; its $Z_g$ is from the smooth version of the skies, as a quarter of the exceeding pairs of the skies themselves ($4.1\sigma$) come from one sky (Appendix~\ref{app:JointStats}).}
\end{table*}

For $T_{72}$, stacking \SAll\ gives $p\simeq0.005$ ($2.6\sigma$)
and \SMW\ $p\simeq0.002$ ($2.8\sigma$), so the two datasets contribute comparably.
Fisher-combined, the two give $3.6\sigma$ ($3.0$--$3.7\sigma$ over the different redshift treatments of \SMW).
In \SJ, which weights the two datasets locally,
$T_{72}$ becomes a single test, reaching $3.8\sigma$ ($3.2$--$3.8\sigma$ over the treatments).
The shuffled-contrast and Gaussian calibrations agree to within $0.15\sigma$ in every treatment (Appendix~\ref{app:JointStats}).
Fisher-combined, $T_{72}$ in \SJ\ reaches $5.6\sigma$ ($5.1$--$5.6\sigma$ over the treatments) with \SCT, or $4.5\sigma$ ($4.0$--$4.5\sigma$) with \SCTb.

We test the results when
removing from \SAll\ the 121 clusters (see \S\ref{subsec:FixedMass}), whose ROIs touch the regions of the
\SCT\ clusters, and masking these regions in \SMW.
Such masking changes the nominal $T_{72}$ of \SMW\ by only $0.03$, within the $\sim0.1$ fluctuation expected from masking $\sim1\%$ of its sky region, and leaves the above $T_{72}$ results in \SJ\ unchanged.

Within $T_{72}$, the companions $C_{72}$ of $L_{43}$ alone
reach $Z_l\simeq3.2$ in \SJ, calibrated as $3.2\sigma$,
exceeding
$L_{43}$, which reaches only $1.7$--$2.0$ over the treatments.
In \SJ, as in \SAll\ (see \S\ref{subsec:FixedMass}), $T_{72}$ thus does not rest on the $43\GeV$ line.
Combined with \SCT\ (\SCTb), $C_{72}$ gives $5.2\sigma$ ($4.0\sigma$) in \SJ.

The constrained nonad $N_{72}$, i.e., the intermediate reading of $T_{72}$ (see \S\ref{subsec:FixedTests}), peaks in \SJ\ with $Z_l\simeq3.9$ ($3.1$ in \SAll\ and $2.4$ in \SMW), calibrated as a highly significant $4.5\sigma$.
The indicated masses $m_1c^2=69.7_{-1.2}^{+0.4}\GeV$ and $m_2c^2=74.4_{-0.4}^{+1.2}\GeV$
are the same as those found for \SAll\ alone (Table~\ref{tab:FixedNonads}).
Fisher-combined with \SCT\ (\SCTb),
$N_{72}$ reaches $6.0\sigma$ ($5.0\sigma$) in \SJ.

While the intermediate reading is well justified by the \SCW\ results placed at any $0.025<z<0.07$, we also consider the agnostic $N_{72}^{\max}$.
In the data, this statistic is $N_{72}$ itself, but its null, the strongest of the three readings in every null pair, is higher,
so it gives $4.1\sigma$ in \SJ, and $5.8\sigma$ ($4.7\sigma$) with \SCT\ (\SCTb; Appendix~\ref{app:JointStats}).

A signal common to \SAll\ and \SMW\ should attribute consistent cross sections to the two datasets.
Before loss correction, \SAll\ measures $\Sigma=(1.7\pm0.6)\times10^{-19}\cm^3\se^{-1}$ (blue right triangle in Fig.~\ref{fig:Scaling}).
Expressed as the cross section that stacking would measure (\ie NFW mapping with pipeline losses; \S\ref{subsec:JointSpectra}, Appendix~\ref{app:JointStats}), \SMW\ indicates a consistent
$\Sigma_{\textrm{X}}=(1.8\pm1.5)\times10^{-19}\cm^3\se^{-1}$ for the modeled distribution, with its templates at $z=0.03$
(blue left triangle).
Combined with the weights of \SJ, the two give $\Sigma=(1.7\pm0.6)\times10^{-19}\cm^3\se^{-1}$ (blue diamond), within the range expected from \SCW\ (gray band).

Overall, with the triad mass fixed a priori by (Fisher-combined with) \SCT, \SJ\ finds $T_{72}$ at $3.8\sigma$ ($5.6\sigma$) and $N_{72}$ at $4.5\sigma$ ($6.0\sigma$).
These results are based on the \SAll\ and \SMW\ datasets that share no ROI photons, contribute comparably, and agree in mass and cross section.
The results do not depend strongly on the exact redshift placement of \SMW\ near its $z=0.03$ expectation; for example, $T_{72}$ of \SJ\ is $3.8$--$3.9\sigma$ for $z\le0.03$, $3.7\sigma$ at $z=0.04$ and $3.4\sigma$ at $z=0.05$ ($5.3$--$5.6\sigma$ Fisher-combined with \SCT\ for $0.01\le z\le 0.05$).
The shifted and folded treatments, which extend to redshifts far beyond the modeled peak, lower $T_{72}$ to $3.2\sigma$ ($5.1$--$5.2\sigma$ with \SCT). See Appendix~\ref{app:JointStats}.

\section{Two-species WIMP model}
\label{sec:Model}

The nominal nonad suggests two cross-annihilating WIMP species (\S\ref{subsec:MatchedFilters}).
This is the strongest nonad in \SAll, in the nominal \SMW, and in their joint \SJ, consisting of the three strongest blind triads of \SAll\ with the a priori $T_{72}$ as an intermediate triad.
The unique nonad pattern is not set by our pipeline or scan parameters.

Indeed, the nonad maps of Fig.~\ref{fig:NonadMap} involve no exclusion window, the three strongest blind triads of \SAll\ persist for $f<0.015$ triad windows, and $f_t$ excludes only narrower nonads whose outer triads nearly coincide ($m_2/m_1<1.04$).
The outer triads are not leakage of a single, intermediate species (as only $\sim35$--$50\%$ leaks out to $\sim2.3\GeV$, without producing local maxima), and there is no hint of sky or pipeline artifact in these masses (\eg random-longitude stacks have a $|Z_l|\lesssim0.1$ mean within $0.1\GeV$ of the three triad masses).

\begin{figure}[t]
    \centering
    \includegraphics[width=\columnwidth]{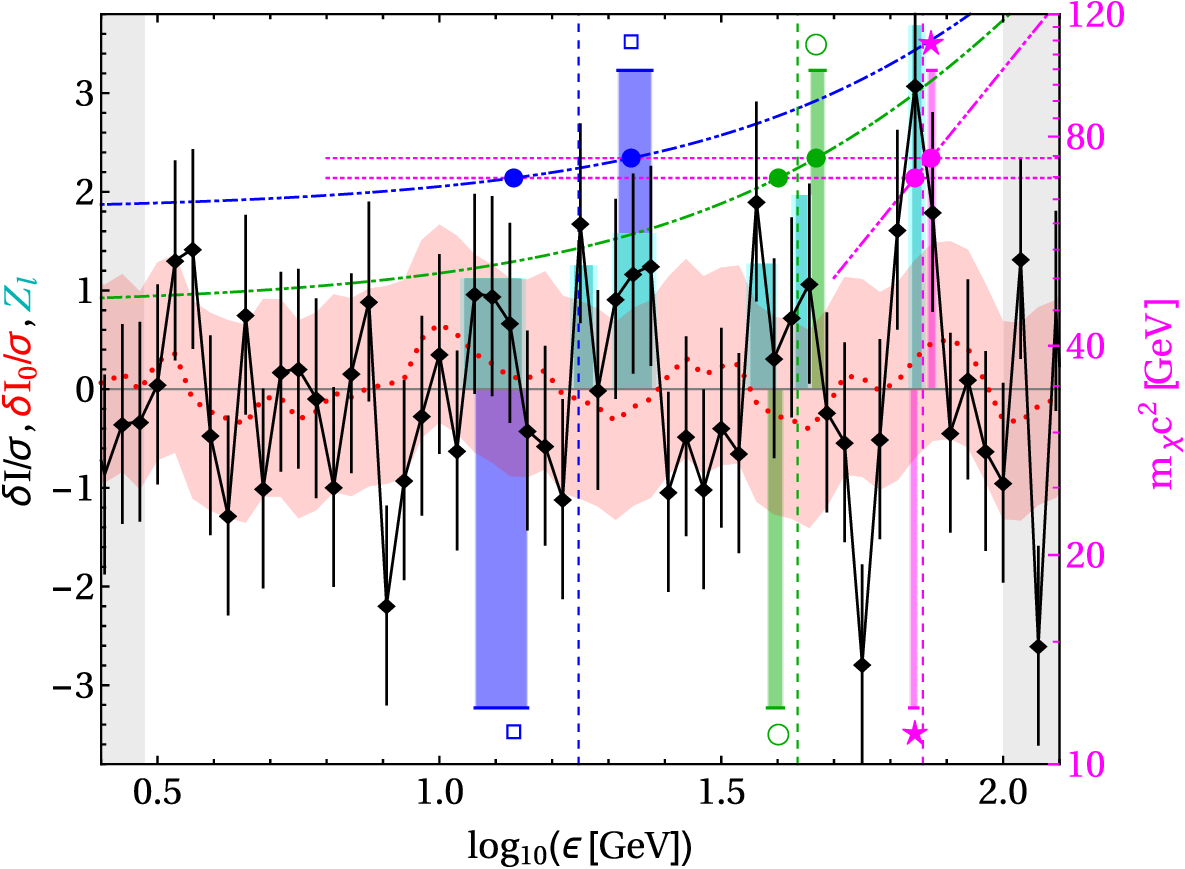}
    \caption{
        Best-fit two-species WIMP model illustrated over the medium-resolution \SJ\ spectrum of Fig.~\ref{fig:Spectra} (its right column, middle row; same notation, but here cyan bars reach their left-axis $Z_l$, and the red uncorrected spectrum is shown as a shaded $\pm1\sigma$ band over its running four-bin mean).
        The three Eq.~\eqref{eq:kin} channel energies are plotted against $\overline{m}_\chi c^2$ (dot-dashed curves against right axis), and intersect (disks) the best-fit masses (dotted horizontal magenta lines) at the six self-annihilation lines; the three cross-annihilation lines (between each pair of disks) nearly coincide with the a priori $72.1\GeV$ triad lines (dashed vertical lines).
        \label{fig:Model}
    }
\end{figure}

We therefore attempt to model all uncovered spectral features as the outcome of two WIMP species (denoted $\chi_1$ and $\chi_2$) that both self-annihilate and cross-annihilate.
Each annihilating pair (denoted $\chi_1\chi_1$, $\chi_1\chi_2$, and $\chi_2\chi_2$) generates its own
$\gamma\gamma$, $\gamma Z$, and $\gamma h$ channels, with photon energies dictated by Eq.~\eqref{eq:kin}, further constrained by the intermediate triad mass given by Eq.~\eqref{eq:nonad} as the mean of the two WIMPs.
Figure~\ref{fig:Model} illustrates how the two masses fix the nine lines, at their best-fit values (\S\ref{subsec:ModelMasses}), plotted over the \SJ\ spectrum.

Next, we estimate the best-fit masses of the model by combining all available data (\S\ref{subsec:ModelMasses}), test the model by varying the analysis and assumptions (\S\ref{subsec:ModelVariants}), and examine the implications of the model (\S\ref{subsec:ModelTests}).
Table~\ref{tab:NineLines} lists, for each of the model's nine lines, its rest-frame energy, $Z_l$ score in each dataset, cross sections in \SAll, in \SJ, and in \SAll\ and \SMW\ combined, and the channel branching ratios if assumed common to the three triads.

\subsection{Best-fit WIMP masses}
\label{subsec:ModelMasses}

We use three independent datasets to constrain the two WIMP masses: \SAll\ (cluster frame), \SMW\ (placed at $z=0.03$), and \SCT\ (or \SCTb).
In particular, we use the nonad scan maps of the former two, and the reported $L_{43}$ of the latter.
The three datasets share no ROI photons: \SAll\ drops here the 121 clusters whose ROIs touch the \SCT\ regions, and \SMW\ masks these regions in addition to the ROIs of all \SAll\ clusters (see \S\ref{subsec:JointFixed}).

A finer nonad scan grid, consisting of $\{m_1,m_2\}$ mass pairs in steps of $0.2\%$,
is corrected for the (modest) detrending bias of the masses by dividing each mass by the factor
inferred from injected mock nonads ($1.001$ for both \SAll\ and \SMW).
At each pair, the TS of each map is the square of its $Z_l$ where positive, and zero elsewhere.
For \SCT, we define $\mbox{TS}\equiv L_{43}^2-[(\epsilon-43.2\GeV)/\Delta\epsilon_{43}]^2$ where positive, with $L_{43}\simeq5.5$ from the published\cite{FanEtAl26} $\textrm{TS}\simeq30.1$ peak and its $\Delta\epsilon_{43}\simeq0.5\GeV$ energy uncertainty
($L_{43}\simeq3.2$ for \SCTb, at its fixed $43.2\GeV$, with $\Delta\epsilon_{43}$ scaled to $\sim0.86\GeV$),
$\epsilon$ being the nonad-line energy closest to $43.2\GeV$.

The model is inferred as the maximal sum of the three TS on the mass grid; all nine lines are within the $3.5$--$95\GeV$ range.
The bounds on each mass, the mean mass, and the mass difference lie where the summed TS, maximized over the other degrees of freedom, drops by unity.
The own TS maxima of each dataset reproduce the blind nonads of \SAll\ and \SMW, and the $43.2\GeV$ line of \SCT.
The $\mathcal{T}$ goodness of fit parameter\cite{MaltoniSchwetz03}, defined as the sum of the three own TS maxima minus the combined maximal TS, tests whether the three measurements agree on their masses.
For full model agreement, $\mathcal{T}$ would follow a $\chi^2$ distribution of $5-2=3$ degrees of freedom.

The best fit is $\{m_1,m_2\}c^2=\{69.7_{-1.0}^{+0.4},74.5_{-0.6}^{+1.1}\}\GeV$.
The mean mass $\overline{m}_\chi c^2=72.1_{-0.3}^{+0.2}\GeV$ is well constrained, only partly by $L_{43}$ (\SAll\ and \SMW\ alone give $72.1\pm0.4\GeV$), whereas the mass difference $(m_2-m_1)c^2=4.8_{-0.9}^{+2.1}\GeV$ is the least constrained.
The $L_{43}$ line is identified as the $\gamma Z$ channel of the intermediate triad (henceforth its host triad).
All nine lines of the model are positive in \SAll\ and in \SJ, and eight of the nine are positive in \SMW\ (Table~\ref{tab:NineLines}).

The datasets separately give $\{69.8_{-2.3}^{+0.3}, 74.5_{-0.7}^{+0.5}\}\GeV$ as the \SAll\ blind nonad (Table~\ref{tab:Nonads}),
$\{69.1_{-0.8}^{+0.4}, 75.6_{-0.5}^{+0.7}\}\GeV$ as the \SMW\ (at $z=0.03$) blind nonad, and $(72.1\pm0.4)\GeV$ as the triad with $L_{43}$ as its $\gamma Z$ channel.
Their disagreement with the model is $\mathcal{T}\simeq2.4$ ($p\simeq0.49$), almost entirely attributed to \SAll\ against \SMW.
Without \SCT, \SAll\ and \SMW\ reproduce the model masses, with $\mathcal{T}\simeq2.4$ ($p\simeq0.31$ for two degrees of freedom), and $L_{43}$ then lies only $0.1\GeV$ from the $43.3\GeV$ $\gamma Z$ line.

To test these bounds and the $\chi^2$ distribution of $\mathcal{T}$, we inject nonads within the model bounds into null realizations.
For simplicity, the nine lines are injected with equal $\mathcal{I}$, scaled to the measured $Z_l$ in each dataset.
These injections give $p\simeq0.50$ for $\mathcal{T}$, and recover the injected masses to within $0.2\GeV$ (median).
The $\Delta\textrm{TS}=1$ bounds contain the injected masses, their mean, and their difference in $66$--$71\%$ of the injections, thus supporting the viability of our analysis method.

\begin{table*}[t]
    \centering
    \caption{Best-fit model: properties of the nine annihilation lines of two cross-annihilating WIMPs.\label{tab:NineLines}}
    \setlength{\aboverulesep}{0pt}\setlength{\belowrulesep}{0pt}\setlength{\extrarowheight}{0.5ex}
    \begin{tabular}{l l c | c c c c | c c c c | c}
        \toprule
        Triad & Line & $\epsilon$ & \multicolumn{4}{c|}{$Z_l$} & \multicolumn{4}{c|}{$\overline{\sigma v}_{p,i}/D$ ($10^{-19}\cm^3\se^{-1}$)} & Branching ratio \\
        $\overline{m}_\chi c^2$ (GeV) & & (GeV) & \SAll & \SMW & \SJ & \SCT$^\S$ & \SAll & \SAll$^\ddagger$ & \SJ & (\SAll+\SMW)$^\ddagger$ & (\SAll+\SMW; common) \\
        \midrule
        $\chi_1\chi_1$ & $\gamma\gamma$ & $69.7_{-1.0}^{+0.4}$ & $3.4$ & $1.7$ & $3.7$ & $\lesssim0.2$ & $2.0\pm0.6$ & $3.2\pm0.9$ & $1.9\pm0.6$ & $3.1\pm0.9$ & $1$ \\
        $69.7_{-1.0}^{+0.4}$ & $\gamma Z$ & $39.9_{-1.5}^{+0.6}$ & $0.3$ & $1.2$ & $1.1$ & $1.7^\dagger$ & $0.6\pm2.1$ & $1.0\pm3.4$ & $1.2\pm1.8$ & $1.7\pm2.9$ & $1.8_{-0.9}^{+1.0}$ \\
         & $\gamma h$ & $13.5_{-1.9}^{+0.7}$ & $1.9$ & $-0.6$ & $0.9$ & $\lesssim0.2$ & $9.5\pm5.1$ & $12.4\pm6.7$ & $2.2\pm3.6$ & $2.9\pm4.7$ & $8.3_{-2.8}^{+3.7}$ \\
        \midrule
        $\chi_1\chi_2$ & $\gamma\gamma$ & $72.1_{-0.3}^{+0.2}$ & $2.8$ & $2.1$ & $3.4$ & $\lesssim0.2$ & $1.7\pm0.6$ & $2.7\pm1.0$ & $1.7\pm0.6$ & $2.7\pm0.9$ & $1$ \\
        $72.1_{-0.3}^{+0.2}$ & $\gamma Z$ & $43.3_{-0.4}^{+0.3}$ & $0.9$ & $1.5$ & $1.7$ & $5.4$ & $1.9\pm2.0$ & $3.1\pm3.3$ & $2.2\pm1.8$ & $3.5\pm2.9$ & [$1.8_{-0.9}^{+1.0}$] \\
         & $\gamma h$ & $17.8_{-0.5}^{+0.4}$ & $0.1$ & $1.6$ & $1.2$ & $0.5$ & $0.5\pm4.1$ & $0.7\pm5.3$ & $3.6\pm3.0$ & $4.5\pm4.0$ & [$8.3_{-2.8}^{+3.7}$] \\
        \midrule
        $\chi_2\chi_2$ & $\gamma\gamma$ & $74.5_{-0.6}^{+1.1}$ & $2.1$ & $1.5$ & $2.6$ & $\lesssim0.2$ & $1.4\pm0.6$ & $2.2\pm1.0$ & $1.3\pm0.6$ & $2.2\pm1.0$ & $1$ \\
        $74.5_{-0.6}^{+1.1}$ & $\gamma Z$ & $46.6_{-0.8}^{+1.5}$ & $0.1$ & $0.5$ & $0.5$ & $2.4^\dagger$ & $0.2\pm2.0$ & $0.4\pm3.3$ & $0.3\pm1.8$ & $0.5\pm2.8$ & [$1.8_{-0.9}^{+1.0}$] \\
         & $\gamma h$ & $21.9_{-1.0}^{+1.8}$ & $1.6$ & $0.6$ & $1.6$ & $0.9$ & $5.6\pm3.6$ & $7.3\pm4.7$ & $3.8\pm2.7$ & $5.1\pm3.5$ & [$8.3_{-2.8}^{+3.7}$] \\
        \bottomrule
    \end{tabular}
    \TblNote{
    The nine annihilation lines of the two-species model at the best-fit combined masses of \S\ref{subsec:ModelMasses}: rest-frame energies $\epsilon$, local $Z_l$ scores for best-fit masses, cross sections (see \S\ref{subsec:Losses}), and channel ratios if assumed common to the three triads (derived once and repeated in brackets).\\
    $^\S$Based on the published\cite{FanEtAl26} rest-frame TS spectrum (their Fig.~1), TS$^{1/2}$ interpolated linearly in $\ln\epsilon$; $\textrm{TS}\lesssim0.05$ (where the one-sided TS is given at zero) taken as $Z_l\lesssim0.2$.\\
    $^\dagger$On the flank of the $L_{43}$ peak, $8\%$ from $43.2\GeV$, so not independent of $L_{43}$.\\
    $^\ddagger$Loss-corrected; for \SAll+\SMW, the inverse-variance mean of the two loss-corrected datasets, as in the common-ratio fit (\S\ref{subsec:ModelTests}).}
\end{table*}

\subsection{Model variants and checks}
\label{subsec:ModelVariants}

Various modifications to the model are considered.
Placing \SMW\ at different $z=0.01$--$0.06$, replacing \SCT\ by \SCTb, omitting the detrending mass correction, or restoring both the 121 clusters in \SAll\ and the \SCT\ regions in \SMW, all shift the masses by $\lesssim0.7\GeV$, within their bounds, and the mean mass by $\lesssim0.1\GeV$.

There are alternative, secondary model maxima, as expected for the noisy scans in Fig.~\ref{fig:NonadMap}.
In particular, summed TS values nearly as high as its maximum are found near a wider, $\{69.2, 74.9\}\GeV$ nonad solution, within the bounds of the model. This solution is preferred by some of the alternative checks, such as when placing \SMW\ at $z\le0.02$.

While the $L_{43}$ line is nominally identified as the $\gamma Z$ channel of the intermediate triad, an alternative upper triad host would lower the summed TS by only $0.4$, shifting the nonad to lower, $\{67.8, 72.2\}\GeV$ masses. This shift approximately steps the nonad down by one triad.

The analysis can be repeated with $Z_l$ first standardized in each dataset, by the mean and standard deviation of each pair in the null realizations (Poisson Monte Carlo for \SAll\ and shuffled-contrast skies for \SMW; see \S\ref{subsec:Significance}).
The maximum then shifts to $\{69.0, 74.9\}\GeV$, still within the nominal bounds, with a higher $\mathcal{T}\simeq4.7$ ($p\simeq0.20$).
Here, $L_{43}$ is robustly associated with the intermediate triad, with the best alternative $\sim4$ times less likely ($\Delta\textrm{TS}\simeq2.9$).

The LAT energy scale is calibrated\cite{FermiCalibration12} within $[-5,+2]\%$, systematically affecting \SAll, \SMW, and $L_{43}$ alike.
Simply rescaling the WIMP masses does not correct for an overall energy shift, due to the non-linear mass-dependence of different Eq.~\eqref{eq:kin} channels.
Instead, an energy shift distorts the line complex, worsening the agreement among datasets.
The summed TS peaks for a LAT energy offset of $(-1.5)_{-2.2}^{+1.7}\%$; conversely,
assuming the model, the data limit the LAT offset to $[-5.9,+3.5]\%$ at $\sim95\%$ confidence ($\Delta\textrm{TS}=4$).
If the LAT energies are $5\%$ too low, the best fit moves up to $\{70.4, 75.9\}\GeV$, with $L_{43}$ still the $\gamma Z$ line of the intermediate triad, and $\mathcal{T}$ rises to $\sim6.9$ ($p\simeq0.07$).
If the LAT scale is $2\%$ too high, the best fit moves down to $\{67.7, 71.7\}\GeV$ ($\mathcal{T}\simeq3.4$, $p\simeq0.33$), with $L_{43}$ as the $\gamma Z$ line of the upper triad, close to the aforementioned nonad stepped down by one triad.

The three triads of the model are equally spaced, at $69.7\GeV$ plus integer multiples of $(m_2-m_1)c^2/2\simeq2.4\GeV$, so it is natural to ask if there are additional triads along this sequence.
Nearby $65.0$--$85\GeV$ triads in this sequence have $Z_l\lesssim1.2$ in \SAll\ (against $Z_l\gtrsim2.2$ for the three strongest triads), and none of them are positive in the triad and both its $\gamma Z$ and $\gamma h$ putative channels.

\subsection{Model tests and implications}
\label{subsec:ModelTests}

The model, fitted to the nonad scans of \SAll\ and \SMW\ (which share no ROI photons) and to $L_{43}$, inherits the agreement of the nominal nonad with the blind triads of the stacking subsamples (\S\ref{subsec:FixedNonads}), and its three triads are all positive in each of \SMW, \SJ, \SO, \SW\ ($Z_l\simeq1.1$--$3.6$), and even \SE\ ($Z_l\gtrsim 0.6$).

Through these triads and nonads, the model accounts for nearly all blind features, single lines included.
We count a feature as accounted for if its energy lies within $1.5\%$ (less than half a native bin) of the energy range of a model line.
By this criterion, the model accounts for all the lines and triads of \SAll\ in Table~\ref{tab:FeaturesS}, whose single lines are channels of the triads identified in \S\ref{subsec:Blind}, and for all the features of \SJ\ in Table~\ref{tab:FeaturesJ} except its $88.7\GeV$ triad (which is the $\gamma h$ reading of $L_{43}$, disfavored by the model).

In every subsample but MCXC, the strongest single line is one of the nine (Appendix~\ref{app:SubSpectra}).
As the windows around the nine lines cover $\sim23\%$ of the $3$--$100\GeV$ scan range in $\ln\epsilon$, however, single lines impose a far weaker test on the model with respect to triads and nonads.

At its best-fit masses, the model leaves free the cross sections of its three triads and the channel branching ratios within each triad.
The present data cannot fix these ratios for each triad separately, as the $\gamma\gamma$ lines of the three triads, $3\%$ apart, blend within the LAT resolution.
The better separated $\gamma Z$ and $\gamma h$ lines could fix their ratio within each triad, but the $\gamma Z$ lines are too faint.
We therefore fit ratios common to the three triads, and test their consistency across triads by the $\chi^2$ of this fit.

A joint fit of the nine measured cross sections of each of \SAll\ and \SMW, through the responses of their filters to loss-corrected single lines,
gives common ratios $\gamma\gamma:\gamma Z:\gamma h$ of $1:1.8_{-0.9}^{+1.0}:8.3_{-2.8}^{+3.7}$ (Table~\ref{tab:NineLines}; Appendix~\ref{app:BR}), consistent with both datasets ($\chi^2\simeq6.6$ for $10$ degrees of freedom).
Stacking \SAll\ alone gives nearly the same ratios, $1:1.7_{-0.9}^{+1.1}:8.7_{-2.9}^{+3.7}$ ($\chi^2\simeq4.3$ for $4$ degrees of freedom).
A joint fit of \SO\ and the unmasked \SCW\ gives consistent, $1:2.2_{-1.4}^{+2.0}:9.9_{-4.4}^{+7.5}$ ratios ($\chi^2\simeq6.8$ for $10$ degrees of freedom; Appendix~\ref{app:BR}).
In all three fits, the three $\gamma h$ lines carry most ($\sim75\%$) of the line cross section, as in the $\gamma h>\gamma Z>\gamma\gamma$ hierarchy reported\cite{Keshet26Corr} for the $T_{72}$ of \SCW\ (with loss-corrected ratios $1:2.6_{-2.6}^{+12}:4.2_{-3.3}^{+19}$).
As expected, branching ratio uncertainties are considerable in present data;
while
$\gamma h>\{\gamma Z,\gamma\gamma\}$ is indicated at $p\simeq0.006$ (and $p\simeq0.004$ for \SAll\ alone), each fit favors the $\gamma Z>\gamma\gamma$ hierarchy only at $p\simeq0.2$.

In \SCT, the model manifests (see Table~\ref{tab:NineLines}) its intermediate triad through its $\gamma Z$ line, $L_{43}$, which dominates the nine lines ($Z_l\simeq5.4$ according to the published\cite{FanEtAl26} rest-frame TS spectrum), with the $\gamma Z$ lines of the outer triads on its flank.
The $\gamma h$ line of $\chi_2\chi_2$ is also positive ($Z_l\simeq0.9$ at the best fit, and $2.6$ at $23.3\GeV$, within its energy range).

The three $\gamma\gamma$ lines of the model, however, are absent from \SCT\ ($\mbox{TS}\lesssim0.05$).
Photon counts in Virgo and Fornax (Appendix~\ref{app:Line43}) put the missing $72.1\GeV$ $\gamma\gamma$ line $\sim2$ standard deviations short
(for the $T_{72}$ channel ratios of \SCW\ as observed and for the common ratios above alike).
However, with only $\sim10$ line photons in $L_{43}$, this shortfall does not exclude the model.

\section{Summary and discussion}
\label{sec:Summary}

Stacking LAT photons from the centers of 23,343 redshift $z<0.5$ clusters, drawn from the largest available MCXC, eROSITA, and DESI catalogs (\S\ref{subsec:Samples}), shows a featureless spectrum in the observer frame; the Galactic plane is excluded and all known discrete sources are masked (\S\ref{subsec:FermiLAT}).
However, spectral features emerge (Figs.~\ref{fig:Spectra}--\ref{fig:Alpha}) once the stacked spectrum is corrected for the redshift of each cluster (\S\ref{subsec:Stacking}); the effect is particularly striking for the more distant clusters (Fig.~\ref{fig:AlphaFar}).
Such an emergence of spectral features only once corrected for cluster emission was also seen\cite{Keshet26Corr} when cross-correlating LAT data with the eROSITA X-ray map of the western Galactic hemisphere (\SCW; Fig.~\ref{fig:NonadMap}).

Single-line and composite matched filters applied to the high-resolution spectra (Fig.~\ref{fig:NonadMap}) show features similar to those found in \SCW, also after the overlapping ROIs are masked.
Blind scans of the stacked (\SAll) spectrum find three triads (the strongest, nominal $70\GeV$ triad at $2.8\sigma$; \S\ref{subsec:Blind}),
with nearly equally spaced WIMP masses $\overline{m}_\chi$.
Accordingly, a blind scan finds a $\{70, 74\}\GeV$ nonad (the nominal nonad; $3.3\sigma$; \S\ref{subsec:Nonads}) containing these three triads; its intermediate triad is (within $0.1\GeV$) the a priori $72.1\GeV$ triad inferred from previous studies\cite{FanEtAl26, Keshet26Corr}.
With the more careful, channel-masked detrending (\S\ref{subsec:Detrending}; henceforth), these triad and nonad are calibrated as $2.8\sigma$ and $3.4\sigma$ (the $72$ and $74.6\GeV$ triads as $1.3\sigma$ and $1.2\sigma$).

Jointly (\SJ; Figs.~\ref{fig:Spectra}--\ref{fig:Alpha}) analyzing \SAll\ and \SMW\ (the \SAll-ROI-masked \SCW)
finds a similar, $72.2\pm0.2\GeV$ triad ($2.8\sigma$) and a stronger, $\{69, 75\}\GeV$ nonad ($4.0\sigma$).
These global Z-scores, which take into account all multiple-trial corrections, agree among calibrations based on Monte Carlo, random-longitude, and redshift-permutation tests (for \SAll), the Monte Carlo coupled with shuffled-contrast skies, Gaussian realizations, and different redshift treatments (for \SMW).
However, the earlier \SCT\ provides a prior, raising the \SJ\ a priori triad $T_{72}$ and nonad $N_{72}$ to $4.0\sigma$ and $4.6\sigma$. Fisher-combined with \SCT\ (with \SCTb), $T_{72}$ and $N_{72}$ strengthen further to $5.7\sigma$ ($4.6\sigma$) and $6.1\sigma$ ($5.1\sigma$).

All spectral features are accounted for by two WIMP species, of masses $\{m_1,m_2\}c^2=\{69.7_{-1.0}^{+0.4},74.5_{-0.6}^{+1.1}\}\GeV$, that both self- and cross-annihilate into nine lines; see Table \ref{tab:NineLines} and Fig.~\ref{fig:Model}.
In this model, inferred from \SAll, \SMW, and \SCT\ combined (\S\ref{subsec:ModelMasses}), the cross-annihilation triad $\overline{m}_\chi c^2=72.1_{-0.3}^{+0.2}\GeV$ is strongly constrained, whereas the mass difference $(m_2-m_1)c^2=4.8_{-0.9}^{+2.1}\GeV$ is less certain.

There is substantial evidence supporting the interpretation of the above spectral features as true WIMP lines from galaxy clusters, as outlined next in \S\ref{subsec:Evidence}. We then discuss their combined significance (\S\ref{subsec:SumSig}), main uncertainties (\S\ref{subsec:Uncertainties}), and inferred cross sections (\S\ref{subsec:SumCross}).

\subsection{Evidence for WIMP lines from clusters}
\label{subsec:Evidence}

Several properties of the above spectral features support their identification as WIMP annihilation lines from clusters, disfavoring alternative interpretations such as instrumental, foreground, or pipeline artifacts.
In particular, the features:

\begin{enumerate}[label=\arabic*., nosep, wide=0pt, itemsep=3pt, topsep=3pt]

\item  Emerge from apparent noise only when boosted to the cluster frame --- artifacts don't do that. The features peak at $\alpha\simeq1$ (Fig.~\ref{fig:Alpha}), most strongly in the more distant clusters (Fig.~\ref{fig:AlphaFar}), and vanish for permuted cluster redshifts (Appendix~\ref{app:Nulls}).
    The observer frame shows no features (\S\ref{subsec:Alignment}), thus imposing an upper limit on any sky-artifact
    ($\lesssim2\%$ of $T_{72}$ in \SJ\ when folding in the ambient \SJ$_0$ sky;
    \S\ref{subsec:JointNonads}).

\item  Peak at the centers of the stacked images, for different and independent matched filters (Fig.~\ref{fig:Images}), vanishing when the ROIs are moved off the clusters beyond the expected $\sim0\dgrdot15$ offset (Fig.~\ref{fig:OffCentre}).

\item Are absent from the stacked field despite its higher sensitivity (\S\ref{subsec:Controls}), so the contribution of sky artifacts to the
reported features is limited to $<4\%$. Stacked against quiet fields, the clusters retain the reported features, while the fields again show nothing.

\item Present across the sky, throughout the redshift range, carried by numerous clusters. Removing a subset of strongly contributing clusters, some sky regions, or some mass tertile, lowers the signal as expected from the reduced number of photons (\S\ref{subsec:RobustTests} and Appendix~\ref{app:Samples}).

\item Present the highly-restrictive kinematic sub-structure (lines combined into triads) and structure (triads combined into a nonad) expected from annihilating WIMPs. The wide energy span of the restricted apertures acts as a natural veto against instrumental and other artifacts localized in energy.

\item  Are consistent with narrow annihilation lines broadened only by the IRF. In particular, the nominal triad is shown to be sharpest for an unbroadened aperture (in Appendix~\ref{app:Dz}). In contrast, the \SCW\ features, which arise from the integrated LSS along the line of sight, do strengthen\cite{Keshet26Corr} when the aperture is broadened, peaking at $\Delta z\simeq0.05$, as expected from theoretical modeling.

\item  Are consistent in mass and in cross section with previous studies of independent photons: \SAll\ with \SMW\ and
\SCT\ (or \SO\ with \SCW\ and \SCTb, after removing Virgo and Fornax; \S\ref{subsec:Scaling} and Appendices~\ref{app:eROSITA}, \ref{app:Line43}). The entire nonad maps of \SAll\ and \SMW\ strongly correlate (\S\ref{subsec:JointNonads}, Fig.~\ref{fig:NonadMap}).

\item  Recur in disjoint catalogs and subsamples. For example, the three triads of the model are positive in each of \SO, \SW, and \SE, $T_{72}$ is positive also in all three catalogs and all four Galactic quadrants, and the strongest \SE\ and \SW\ blind triads agree
    (\S\ref{subsec:Blind}, \S\ref{subsec:FixedMass}, and \S\ref{subsec:ModelTests}).

\item Give mutually consistent cross sections in the three catalogs, in different sub-samples, and in the tertiles (and other binnings) of predicted $\propto\smash{\tilde{\mathcal{J}}}$, with $\Sigma$ and $T_{72}$ positive in each (Fig.~\ref{fig:Scaling}).

\item Are robust to pipeline changes: variations of detrending, field subtraction, errors, binning and alignment, ROI, source mask, and matched filters leave the results intact or scale them as expected with the number of photons retained (\S\ref{subsec:RobustTests} and Appendix~\ref{app:Nulls}).

\item  Are accounted for by a simple two-mass model, approximately given by the blind nonad or the three blind triads of \SAll, accounting
for all the lines and triads of \SAll\ and
\SJ\ (\S\ref{subsec:ModelVariants} and \S\ref{subsec:ModelTests}).

\end{enumerate}

Three tests are notable for mild tension (before, but not after multiple-trial correction) with a uniform signal that follows its predicted per-cluster scaling:

\begin{enumerate}[label=\arabic*., nosep, wide=0pt, itemsep=3pt, topsep=3pt]
\item
The scaling of the $T_{72}$ cross section with the individual cluster $\smash{\tilde{\mathcal{J}}_\Myc}$ is weaker than expected.
For a line flux $\propto\smash{\tilde{\mathcal{J}}_\Myc^{\,\kappa}}$, Poisson simulations give $\kappa=0.24\pm0.58$ in the nominal tertiles of predicted signal, disfavoring the predicted $\kappa=1$ at $p\simeq0.09$ (multiple-trial corrected: $p\simeq 0.5$); other binnings give $\kappa\simeq-0.1$ to $0.4$ ($p\gtrsim0.3$ trial-corrected).
The tension would be partly relieved by a scatter in cataloged $M_{500}$ (a $30\%$ scatter $\sim$doubles the uncorrected $p$), or by a signal more extended than modeled.
A simple calibration of MCXC and $\Delta$DESI masses to the eROSITA scale (Appendix~\ref{app:MassCal}) only worsens the tension.

\item
The signal is unevenly distributed in redshift. The nominal redshift tertiles give $T_{72}\simeq\{1.7,-0.5,2.6\}$, against the predicted $\sim\{1.7,0.7,0.4\}$, inconsistent at $p\simeq0.03$ (multiple-trial corrected: $0.19$); other binnings show a similar or stronger pattern ($p\simeq0.12$ trial-corrected). The tension would be partly relieved if the signal is more extended than modeled (favoring distant clusters against the fixed ROI).

\item
The $T_{72}$ of \SE\ and \SW, of nearly equal expected sensitivity, differ ($0.6$ and $2.4$) by $1.5$ standard deviations, mainly in the $\gamma Z$ channel.
However, their $\gamma\gamma$ channels agree, the four Galactic quadrants are consistent with a uniform signal ($\chi^2\simeq3.5$ for three degrees of freedom),
and all three triads of the model are still positive in \SE\ (\S\ref{subsec:ModelTests} and Appendix~\ref{app:SubSpectra}).
A future eROSITA catalog of the eastern hemisphere may test this difference.
\end{enumerate}

\subsection{Significance}
\label{subsec:SumSig}

Nominally, stacking is calibrated by Poisson Monte Carlo simulations, \SCW\ and \SMW\ by shuffled-contrast skies, and \SJ\ (or \SO+\SCW) by null pairs of a Monte Carlo realization of \SAll\ (\SO) and a shuffled-contrast \SMW\ (\SCW) sky.
The Gaussian check $\smash{Z_g^{(n)}}$ uses Gaussian noise instead of shuffled-contrast skies, and instead of Poisson noise in unpaired stacking.

Table~\ref{tab:MainFeatures} summarizes the nominal global significance $Z_g$ of the main spectral features after channel-masked detrending: the strongest blind line, triad, and nonad, and the features fixed a priori by $L_{43}$.
Each of these features is given for \SAll, \SCW, \SO\ combined with \SCW, and \SJ, both alone and Fisher-combined with the $L_{43}$ of either \SCT\ or \SCTb.

\begin{table*}[t]
    \centering
    \caption{Energies and $Z_g$ of the main features, alone and Fisher-combined with \SCT\ or \SCTb.\label{tab:MainFeatures}}
    \newcommand{\TwoM}[3]{{\renewcommand{\arraystretch}{0.85}\begin{tabular}[c]{@{}c@{}}#1\\#2\end{tabular}}\ #3}
    \setlength{\tabcolsep}{3pt}
    \begin{tabular}{@{}l l l l l@{}}
        \toprule
         Feature & \SAll & \SCW & \SO+\SCW & \SJ \\
         & \multicolumn{4}{c}{$\epsilon/\mbox{GeV}$ or $m_\chi c^2/\mbox{GeV}$ \quad ($Z_g$ alone, with \SCT, or with \SCTb)}\\
        \midrule
        Line & $69.3_{-1.2}^{+1.1}$ ($2.3\sigma$) & $71.5_{-2.0}^{+1.5}$ ($0.7\sigma$) & $69.7_{-1.3}^{+2.5}$ ($1.5\sigma$) & $70.1_{-1.0}^{+1.4}$ ($2.5\sigma$) \\ \addlinespace[4pt]
        Triad & $69.9_{-0.2}^{+0.2}$ ($2.8\sigma$) & $72.3_{-0.3}^{+0.2}$ ($2.3\sigma$, $4.6\sigma$, $3.3\sigma$) & $72.1_{-0.2}^{+0.3}$ ($2.4\sigma$, $4.7\sigma$, $3.4\sigma$) & $72.2_{-0.2}^{+0.2}$ ($2.8\sigma$, $5.0\sigma$, $3.7\sigma$) \\ \addlinespace[4pt]
        Nonad & \TwoM{$69.8_{-2.3}^{+0.3}$}{$74.5_{-0.7}^{+0.5}$}{($3.4\sigma$, $5.3\sigma$, $4.2\sigma$)} & \TwoM{$69.1_{-0.8}^{+0.3}$}{$75.6_{-0.4}^{+0.8}$}{($2.6\sigma$, $4.8\sigma$, $3.6\sigma$)} & \TwoM{$69.1_{-0.8}^{+0.9}$}{$74.9_{-0.8}^{+1.2}$}{($3.3\sigma$, $5.2\sigma$, $4.1\sigma$)} & \TwoM{$69.1_{-0.6}^{+1.0}$}{$74.9_{-1.2}^{+0.8}$}{($4.0\sigma$, $5.7\sigma$, $4.6\sigma$)} \\ \addlinespace[4pt]
        $L_{43}$ & $43.2\pm0.5$ ($1.2\sigma$, $4.2\sigma$, $2.7\sigma$) & $43.2\pm0.5$ ($1.3\sigma$, $4.2\sigma$, $2.8\sigma$) & $43.2\pm0.5$ ($1.0\sigma$, $3.9\sigma$, $2.5\sigma$) & $43.2\pm0.5$ ($1.7\sigma$, $4.4\sigma$, $2.9\sigma$) \\ \addlinespace[4pt]
        $C_{72}$ & $17.7\pm0.6$ ($2.4\sigma$, $4.7\sigma$, $3.4\sigma$) & $17.7\pm0.6$ ($2.7\sigma$, $4.9\sigma$, $3.6\sigma$) & $17.7\pm0.6$ ($3.1\sigma$, $5.1\sigma$, $3.9\sigma$) & $17.7\pm0.6$ ($3.4\sigma$, $5.3\sigma$, $4.1\sigma$) \\ \addlinespace[4pt]
        $T_{72}$ & $72.1\pm0.4$ ($2.7\sigma$, $4.9\sigma$, $3.6\sigma$) & $72.1\pm0.4$ ($3.3\sigma$, $5.2\sigma$, $4.1\sigma$) & $72.1\pm0.4$ ($3.6\sigma$, $5.4\sigma$, $4.3\sigma$) & $72.1\pm0.4$ ($4.0\sigma$, $5.7\sigma$, $4.6\sigma$) \\ \addlinespace[4pt]
        $N_{72}$ & \TwoM{$69.7_{-0.5}^{+0.4}$}{$74.4_{-0.4}^{+0.5}$}{($4.0\sigma$, $5.7\sigma$, $4.7\sigma$)} & \TwoM{$68.5_{-0.2}^{+0.3}$}{$75.6_{-0.3}^{+0.2}$}{($3.2\sigma$, $5.2\sigma$, $4.0\sigma$)} & \TwoM{$69.7_{-1.0}^{+0.4}$}{$74.4_{-0.4}^{+1.0}$}{($3.7\sigma$, $5.5\sigma$, $4.4\sigma$)} & \TwoM{$69.7_{-1.2}^{+0.4}$}{$74.4_{-0.4}^{+1.2}$}{($4.6\sigma$, $6.1\sigma$, $5.1\sigma$)} \\
        \bottomrule
    \end{tabular}
    \TblNote{
    Line, triad, and nonad: the strongest blind features.
    Fisher combinations
    shown when relevant.
    Uncertainties are $\Delta\textrm{TS}=1$ bounds for scanned values,
    and $\pm\sim0.5\GeV$ for values fixed by $L_{43}$.
    Values are in the cluster frame (\SMW\ and \SCW\ placed at $z=0.03$).
    The $Z_g$ are calibrated by Monte Carlo (\SAll, \SO), shuffled-contrast skies (\SCW), and null pairs with such skies (\SJ), with stacking channel-masked detrended (Appendix~\ref{app:Nulls}). Energies and bounds are those of the nominal scans.
}
\end{table*}

The strongest blind features in the joint \SJ\ are a $72\GeV$ triad ($2.8\sigma$) and a $\{69,75\}\GeV$ nonad ($4.0\sigma$), both harboring the $43\GeV$ $\gamma Z$ line.
Using $L_{43}$ only as a prior based on earlier work, \SJ\ finds the triad $T_{72}$ at $4.0\sigma$ and the nonad $N_{72}$ at $4.6\sigma$.
Finding a significant nonad is particularly notable, as detrending suppresses the signal of a nonad more than that of a triad.
In mocks of \SAll\ (Appendix~\ref{app:Mock}), nonads
of the model lose a median $\sim50\%$ of their injected $Z_l$, compared with $\sim30\%$ for their intermediate triads injected into the same noise, and are recovered less often by the blind scans.
Mocks of the cross-correlation analysis\cite{Keshet26Corr} show the same behavior.
Nevertheless, the nonad $N_{72}$
exceeds the triad $T_{72}$ in \SAll\ and in \SJ, and matches it in \SO+\SCW,
with and without \SCT\ or \SCTb\ (Table~\ref{tab:MainFeatures}).

Our global false-alarm $Z_g$ calibrations understate the signal strength: as detrending suppresses the signal (coherently) more than it narrows the null $Z_l$ distribution (in quadrature), this signal loss lowers not only $Z_l$ but also $Z_g$, even though the null realizations pass through the same pipeline.
The lost signal is restored only in the loss-corrected cross sections (\S\ref{subsec:Losses}).
Had we compensated for the loss also in $Z_g$, using the median loss of the mocks,
the nominal \SAll\ blind triad would rise from $2.8\sigma$ to $\sim4.4\sigma$,
and its blind nonad would rise from $\sim3.3\sigma$ to $>7\sigma$ (Appendix~\ref{app:Mock}).
These values do not use $L_{43}$; a similarly high compensated significance is found in \SJ.

Part of this error in significance is avoidable, as detrending does not have to fit the continuum also through the tested lines.
Channel-masked detrending (\S\ref{subsec:Detrending}) thus corrects the error in part, without extrapolation.
For example, when thus detrended, the $T_{72}$ brightness in \SAll\ rises by $32\%$, while its null dispersion grows by $25\%$ (Appendix~\ref{app:Nulls}).
In general, channel-masked detrending (used in this \S\ref{subsec:SumSig}) changes $Z_g$ to within $[-0.1,+0.3]$, typically raising it by $\sim0.1$.

Without channel masking or extrapolated loss compensation, the global significance still exceeds $5\sigma$ when the earlier $L_{43}$ studies are folded in.
Fisher-combined with the $L_{43}$ of \SCT, the nonad $N_{72}$ reaches $5.7\sigma$ in \SAll\ and $6.0\sigma$ in \SJ,
while the triad $T_{72}$ reaches $4.8\sigma$ in \SAll\ and $5.6\sigma$ in \SJ.
Combined with the lower-significance \SCTb, these two features still reach $5.0\sigma$ and $4.5\sigma$, respectively, in \SJ.

\subsection{Uncertainties}
\label{subsec:Uncertainties}

The quoted uncertainties of line energies and WIMP masses are the $\Delta\textrm{TS}=1$ likelihood bounds of the matched filters, which are not standard Gaussian uncertainties.
Mock signals injected with $3\le Z_l\le5$ are recovered with a negligible bias, and with a dispersion of $1.5$--$2.6\%$ in line energy and $0.4$--$0.9\%$ in triad mass (Appendix~\ref{app:Mock}).

Quoted uncertainties of the intrinsic cross sections $\overline{\sigma v}_p$, like the errors that normalize $Z_l$, propagate the brightness errors, and are therefore closer to Gaussian.
However, the pipeline lowers both the cross sections and $Z_l$: mock lines lose $(13\pm25)\%$ of their injected values (noise-free loss $\pm$ scatter with noise; Appendix~\ref{app:Mock}) and triads lose $(37\pm20)\%$.
The correction factors $\mathcal{F}_i$ (of \S\ref{subsec:Losses}) correct the cross sections for these losses, as Table \ref{tab:SumCross} shows.

The dominant systematic uncertainty in the line energies and WIMP masses is the global $[-5,+2]\%$ uncertainty in the absolute LAT energy scale.
This uncertainty is common to the previous \SCW, \SCT, and \SCTb, and to the present \SAll\ and its subsamples, all based on LAT data, so such an offset  does not affect their comparison or co-addition.
However, as the channel energies~\eqref{eq:kin} depend nonlinearly on WIMP mass, an offset distorts the line complex and worsens the agreement between datasets, rather than merely rescaling the masses.
Nevertheless, the line structure and agreement among datasets persist for $(-1.5)_{-2.2}^{+1.7}\%$ offsets.
Conversely, assuming the model, the data thus limit the offset to $[-5.9,+3.5]\%$ at $\sim95\%$ confidence (\S\ref{subsec:ModelVariants}).

\begin{table*}[t]
    \centering
    \caption{Cross sections of $T_{72}$.\label{tab:SumCross}}
    \begin{tabular}{l c c c}
        \toprule
        Dataset & $\overline{\sigma v}_{p,\gamma\gamma}$ & $\overline{\sigma v}_{p,\gamma Z}$  & $\overline{\sigma v}_{p,\gamma h}$  \\
         & \multicolumn{3}{c}{$(10^{-19}D\cm^3\se^{-1}$)} \\
        \midrule
        \SO & $1.1\pm0.7$ ($1.9\pm1.2$) & $0.5\pm2.5$ ($0.8\pm4.1$) & $3.8\pm4.9$ ($4.8\pm6.3$) \\
        \SCW$^\dagger$ & $2.5\pm1.6$ ($4.0\pm2.5$) & $3.0\pm3.3$ ($4.9\pm5.5$) & $5.2\pm4.4$ ($6.8\pm5.7$) \\
        \SO+\SCW & $1.6\pm0.7$ ($2.6\pm1.2$) & $1.6\pm2.0$ ($2.6\pm3.4$) & $4.7\pm3.4$ ($6.0\pm4.4$) \\[-0.8ex]
        \multicolumn{4}{@{}c@{}}{\leavevmode\xleaders\hbox{\rule[0.5ex]{3pt}{0.4pt}\kern3pt}\hfill\kern0pt} \\[-0.8ex]
        \SAll & $1.7\pm0.6$ ($2.7\pm1.0$) & $1.8\pm2.0$ ($3.0\pm3.3$) & $0.6\pm4.1$ ($0.8\pm5.3$) \\
        \SMW$^\dagger$ & $1.8\pm1.5$ ($2.9\pm2.5$) & $2.8\pm3.4$ ($4.7\pm5.5$) & $6.4\pm4.6$ ($8.3\pm5.9$) \\
        \SJ$=$\SAll$+$\SMW\ & $1.7\pm0.6$ ($2.8\pm1.0$) & $2.2\pm1.8$ ($3.7\pm2.9$) & $3.3\pm3.0$ ($4.3\pm4.0$) \\[-0.8ex]
        \multicolumn{4}{@{}c@{}}{\leavevmode\xleaders\hbox{\rule[0.5ex]{3pt}{0.4pt}\kern3pt}\hfill\kern0pt} \\[-0.8ex]
        Model & $1.7\pm0.6$ ($2.7\pm0.9$) & $2.1\pm1.7$ ($3.5\pm2.9$) & $3.4\pm3.0$ ($4.5\pm4.0$) \\
        \bottomrule
    \end{tabular}
    \TblNote{
    Cross sections of the three channels of $T_{72}$ for the NFW profile, as measured and (in parentheses) loss-corrected.\\
    $^\dagger$Expressed as the cross sections that stacking would measure (\ie NFW mapping with pipeline losses; Appendix~\ref{app:eROSITA}) for templates placed at $z=0.03$ (in parentheses, loss-corrected by $\mathcal{F}_i$, like \SAll).
    }
\end{table*}

Other systematics are smaller.
Errors in the LAT event reconstruction and energy dispersion that shift the line energies like an overall offset are part of the above uncertainty.
Mock tests indicate that binning and the $\pm s_0/2$ alignment, IRF mismatch, detrending, and the filters bias triad masses by $\lesssim0.3\%$.
Random errors in the catalog redshifts broaden the lines rather than shift them (Appendix~\ref{app:Dz}).
A common bias of the eROSITA photometric redshifts, even as large as the quoted $\delta z/(1+z)\lesssim0.005$ accuracy\cite{KlugeEtAl24}, would shift the masses by only $\lesssim0.2\%$, as $\Delta$eROSITA contributes about a third of \SAll\ ($\Delta$DESI redshifts are spectroscopic).
Our analysis choices are also deliberately conservative, \eg the large $\theta_{\min}=\smash[t]{0\dgrdot5}$ source mask and the non-overlapping exclusion windows.

\subsection{Cross sections}
\label{subsec:SumCross}

Table~\ref{tab:SumCross} summarizes the cross sections of the three channels of $T_{72}$, measured and combined in three different (but not mutually independent; separated by dashed lines) ways:
measured in \SO, inferred from \SCW, and combined in \SO+\SCW;
measured in \SMW\ and \SAll\ and combined in \SJ;
and from the model of \S\ref{sec:Model}.
The \SCW\ and \SMW\ templates are placed at $z=0.03$ here and (unless otherwise stated) throughout the analysis.

The $\gamma\gamma$ channel, which dominates the significance of $T_{72}$, shows consistent cross sections in all datasets, whereas the $\gamma Z$ and $\gamma h$ cross sections are less constrained.
The combined datasets and the model agree, giving a loss-corrected $\overline{\sigma v}_{p,\gamma\gamma}/(10^{-19}D\cm^3\se^{-1})= 2.7\pm1.0$, positive at $2$--$3$ times its uncertainty, whereas the $\gamma Z$ and $\gamma h$ channels, $3\pm3$ and $5\pm4$, are individually consistent with zero in $T_{72}$.
These two lines are better constrained by the three triads of the model together: in their common-ratio fit (below), a vanishing $\gamma h$ ($\gamma Z$) ratio
is disfavored at $p\simeq3\times10^{-4}$ ($p\simeq0.03$).

The loss-corrected tertile mean, which weights clusters by their predicted $\propto \smash{\tilde{\mathcal{J}}_\Myc}$ signal,
favors $\Sigma_0=(1.9\pm0.8)\times10^{-19}\cm^3\se^{-1}$, and other binnings of predicted signal give $(1.2$--$1.7)\times10^{-19}\cm^3\se^{-1}$ (Appendix~\ref{app:Samples}). These values, lower than in \SAll\ with all photons contributing equally, reflect
the weaker-than-predicted scaling of the data with $\smash{\tilde{\mathcal{J}}_\Myc}$ (see \S\ref{subsec:Scaling}).

These inferred values of $\overline{\sigma v}_{p,i}/D$ depend on the reference DM model to which $D$ is normalized:
the cored profile would lower $D$, and thus raise these values, by a factor of $1.6$. The \SCW\ analysis normalizes $D$ instead to DM that traces the X-ray-emitting gas, with $\beta^2=10^{-5}$, which for the same signal lowers $D$, and raises these values, by a factor $K\simeq4.3$. The \SCW\ and \SMW\ rows of Table~\ref{tab:SumCross} are converted to the NFW mapping (Appendix~\ref{app:eROSITA}).

The hadronic $Z$ and $h$ boson decays of the $\gamma Z$ and $\gamma h$ channels should produce up to $\sim15\%$ (near $10\GeV$) of the smooth excess brightness that detrending removes, and $<1\%$ at the $\gamma Z$ and $\gamma\gamma$ line energies, independent of $D$.
Such a continuum is consistent with the data, and cannot be measured in the present analysis; its detrending residuals change $Z_l$ by $\lesssim0.04$.

A joint fit of \SAll\ and \SMW\ at the model best-fit masses gives common loss-corrected cross-section ratios $\gamma\gamma:\gamma Z:\gamma h$ of $1:1.8_{-0.9}^{+1.0}:8.3_{-2.8}^{+3.7}$ for its three triads ($\chi^2\simeq6.6$ for 10 degrees of freedom; \S\ref{subsec:ModelTests}).
The analogous fit of \SO\ and the unmasked \SCW\ gives $1:2.2_{-1.4}^{+2.0}:9.9_{-4.4}^{+7.5}$ ($\chi^2\simeq6.8$ for 10 degrees of freedom).
The results are consistent with the $T_{72}$ ratios of \SCW, weakly constrained as $1:2.6_{-2.6}^{+12}:4.2_{-3.3}^{+19}$ as observed\cite{Keshet26Corr}, or $1:1.7_{-1.9}^{+4.4}:2.7_{-2.3}^{+6.0}$ with its templates at $z=0.03$ (Appendix~\ref{app:eROSITA}).
In all cases, branching ratios are poorly constrained.

The loss-corrected cross sections of stacking, like those of the \SCW\ analysis\cite{Keshet26Corr}, are comparable to the previous continuum upper limits.
Canonical WIMPs such as neutralinos annihilate into continuum $\sim10^3$ times more often than into lines\cite{BuchmullerGarny12}, in which case the continuum limits would be in tension with present results by $\gtrsim2$ orders of magnitude (independent of $D$).
The line cross sections we infer from stacking and X-ray correlations\cite{Keshet26Corr} therefore favor models with an enhanced line branching, such as inert doublet\cite{GustafssonEtAl07} or Rayleigh\cite{WeinerYavin12} DM, or a suppressed continuum, such as leptophilic\cite{FoxPoppitz09} or secluded\cite{PospelovEtAl08} DM.

\acknowledgements

This research received funding from ISF grant No. 2126/22.

This work is
based on data from eROSITA, the soft X-ray instrument aboard SRG, a joint Russian-German science mission supported by the Russian Space Agency (Roskosmos), in the interests of the Russian Academy of Sciences represented by its Space Research Institute (IKI), and the Deutsches Zentrum f\"ur Luft- und Raumfahrt (DLR). The SRG spacecraft was built by Lavochkin Association (NPOL) and its subcontractors, and is operated by NPOL with support from the Max Planck Institute for Extraterrestrial Physics (MPE). The development and construction of the eROSITA X-ray instrument was led by MPE, with contributions from the Dr.~Karl Remeis Observatory Bamberg \& ECAP (FAU Erlangen-Nuernberg), the University of Hamburg Observatory, the Leibniz Institute for Astrophysics Potsdam (AIP), and the Institute for Astronomy and Astrophysics of the University of T\"ubingen, with the support of DLR and the Max Planck Society. The Argelander Institute for Astronomy of the University of Bonn and the Ludwig Maximilians Universit\"at Munich also participated in the science preparation for eROSITA.

\bibliography{Virial}

\appendix

\section{Notation}
\label{app:Notation}

Tables~\ref{tab:Notation}--\ref{tab:Symbols} define the samples, datasets, treatments, and recurring symbols used in the analysis.
Where the symbol of a spectral feature (such as $L_{43}$, $T_{72}$, or $C_{72}$) is quoted as a value, it refers to the $Z_l$-score of that feature.

\begin{table*}[tp!]
    \centering
    \caption{Samples, datasets, treatments, cluster properties, and spectra.\label{tab:Notation}}
    \begin{tabularx}{\textwidth}{@{}l X l@{}}
        \toprule
        Symbol & Meaning & Defined in \\
        \midrule
        \multicolumn{3}{l}{\bf Stacking samples} \\
        \SAll & Nominal sample: all 23,343 stacked clusters & \S\ref{subsec:Samples} \\
        \SE, \SW & Clusters in the eastern, western Galactic hemispheres & \S\ref{subsec:Samples} \\
        \SO & Clusters outside the \SCW\ region & \S\ref{subsec:Samples} \\
        MCXC & Clusters of the MCXC catalog & \S\ref{subsec:Samples} \\
        $\Delta$eROSITA or $\Delta$eRO & eROSITA clusters beyond $0\dgrdot02$ of every MCXC cluster & \S\ref{subsec:Samples} \\
        $\Delta$DESI & DESI clusters beyond $0\dgrdot02$ of every MCXC or eROSITA cluster & \S\ref{subsec:Samples} \\
        \texttt{PCONT}, $\lambda_{500}$ & Contamination probability in eROSITA; richness in DESI & \S\ref{subsec:Samples} \\
        $b$, $|b|_{\min}$ & Galactic latitude; latitude cut & \S\ref{subsec:Samples} \\
        $\smash{\tilde{\mathcal{J}}}_{\textrm{high}}$, $\smash{\tilde{\mathcal{J}}}_{\textrm{mid}}$, $\smash{\tilde{\mathcal{J}}}_{\textrm{low}}$ & Tertiles of predicted signal, each a third of every catalog & \S\ref{subsec:Scaling} \\
        $\smash{\tilde{\mathcal{J}}}_1$--$\smash{\tilde{\mathcal{J}}}_5$ & Five bins of predicted signal, each a fifth of every catalog & Appendix~\ref{app:Samples} \\
        $\mathcal{M}_1$--$\mathcal{M}_5$ & As $\smash{\tilde{\mathcal{J}}}_1$--$\smash{\tilde{\mathcal{J}}}_5$, but in each catalog and redshift slice & Appendix~\ref{app:Samples} \\
        $\mathcal{N}_1$--$\mathcal{N}_5$ & Five bins of equal predicted signal, each a fifth of the total & Appendix~\ref{app:Samples} \\
        \midrule
        \multicolumn{3}{l}{\bf Other datasets and their combination} \\
        \SCW & LAT cross-correlation with the western eROSITA map; also its sky region & \S\ref{sec:Intro} \\
        \SMW & \SCW\ but masked within $0\dgrdot1$ of the stacked clusters & \S\ref{subsec:Samples} \\
        \SCT & Stacking of Virgo, Fornax, and Ophiuchus, specifically $L_{43}$ reports & \S\ref{sec:Intro} \\
        \SCTb & Alternative analysis of the \SCT\ clusters & \S\ref{sec:Intro} \\
        \SJ & \SAll\ co-added with \SMW\ (the latter nominally with templates placed at $z=0.03$) & \S\ref{sec:Joint} \\
        \SJ$_0$ & \SJ\ before correction: the observer-frame \SAll\ with the ambient \SMW & \S\ref{subsec:JointSpectra} \\
        Nominal & Defaults: NFW mapped \SAll, \SCW\ with templates at $z=0.03$, catalog masses & \S\ref{sec:Methodology} \\
        \midrule
        \multicolumn{3}{l}{\bf Treatments of the \SCW\ signal redshift} \\
        $z=0.03$ & All templates at the modeled\cite{Keshet26Corr} peak redshift; $z=0.01$--$0.05$ as checks & \S\ref{sec:Joint} \\
        $\Delta z$ & Apertures broadened uniformly over redshifts between $0$ and $\Delta z$ & \S\ref{subsec:FixedMass} \\
        Shifted & Templates shifted by the offsets found in the \SCW\ mocks & \S\ref{subsec:FixedMass} \\
        Folded & Templates integrated over the modeled distribution & \S\ref{subsec:FixedMass} \\
        Modeled & Templates distributed with the redshift model of the \SCW\ signal\cite{Keshet26Corr}, peaking at $z\simeq0.03$ & \S\ref{subsec:FixedMass} \\
        Uniform & Templates uniformly distributed over $z<0.05$ & Appendix~\ref{app:eROSITA} \\
        \midrule
        \multicolumn{3}{l}{\bf Clusters} \\
        $M_{500}$, $R_{500}$, $\theta_{500}$ & Mass, radius, and opening angle of mean density $500$ times the critical & \S\ref{subsec:Samples} \\
        $\theta_{\MyROI}$, $\theta_f$, $\theta_{\min}$ & ROI radius, outer radius of the field annulus, source-mask radius & \S\ref{subsec:FermiLAT}, \S\ref{subsec:Stacking} \\
        $\Myc$, $\MyC$ & A cluster; a sample of clusters & \S\ref{subsec:Stacking} \\
        $N_c$ & Number of clusters in a sample & \S\ref{subsec:Stacking} \\
        $z_\Myc$, $\xi_\Myc$ & Redshift of cluster $\Myc$; its rest- to observer-frame energy ratio $\sim1+z_\Myc$ & \S\ref{subsec:Stacking} \\
        $\mathcal{E}_\Myc$ & Exposure of cluster $\Myc$ summed over its unmasked ROI pixels & Eq.~\eqref{eq:Jtilde} \\
        $\mathcal{J}_\Myc$ & Kinematic $\mathcal{J}$-factor of cluster $\Myc$ & Eq.~\eqref{eq:Jc} \\
        $\mathcal{J}$, $\tilde{\mathcal{J}}$, $\tilde{\mathcal{J}}_\Myc$ & Sum of $\mathcal{J}_\Myc$, its effective mean in stacking, and the share of cluster $\Myc$ in $\tilde{\mathcal{J}}$ & Eq.~\eqref{eq:Jtilde} \\
        $\kappa$ & Exponent of a putative line flux $\propto\smash{\tilde{\mathcal{J}}_\Myc^{\,\kappa}}$, a tested behavior (predicted: $\kappa=1$) & \S\ref{subsec:Scaling} \\
        $\psi_\Myc$ & Net ROI signal brightness of cluster $\Myc$ over its flux per ROI area & \S\ref{subsec:CrossSections} \\
        \midrule
        \multicolumn{3}{l}{\bf Spectra and detrending} \\
        $k$, $k'$ & Photon energy channel in cluster frame, in observer frame & \S\ref{subsec:Stacking} \\
        $\epsilon$, $\epsilon_\Myfr$, $\delta\epsilon_\Myfr$ & Photon energy, mean energy of channel $\Myfr$, width of channel $\Myfr$ & \S\ref{subsec:Stacking} \\
        $s$, $s_0$ & Fractional width of a rebinned bin, of a native bin & \S\ref{subsec:FermiLAT} \\
        $\Myh$, $\delta\Omega$, $\MyH$ & HEALPix pixel, pixel solid angle, set of unmasked pixels in cluster ROI & \S\ref{subsec:Stacking} \\
        $\Myn$, $\Myf$, $\Delta\Myn$ & Detected, field-expected, and excess photon counts & Eq.~\eqref{eq:DeltaN} \\
        $\alpha$ & Scaling of the redshift correction ($0$: observer frame; $1$: cluster frame) & \S\ref{subsec:Stacking} \\
        $\Delta I_{\Myfr'}(\MyC)$ & Mean excess brightness of sample $\MyC$, in observer-frame channels & Eq.~\eqref{eq:ExcessIj} \\
        $\smash{\Delta\tilde{I}_\Myfr(\MyC)}$ & Aligned excess brightness: binned by emitted energy, in cluster-frame channels & Eq.~\eqref{eq:deltaI} \\
        $\sigma_\Myfr$ & Null error of channel $\Myfr$, before detrending & Eq.~\eqref{eq:Var} \\
        $\delta$, $\mathcal{D}$ & Detrending operator, and its matrix for a sample; $\delta I=\mathcal{D}\,\Delta\tilde{I}$ & \S\ref{subsec:Detrending} \\
        $\delta I_0$, $\delta I$ & Detrended $\Delta I_{\Myfr'}(\MyC)$ in observer frame, and $\Delta\tilde{I}_\Myfr(\MyC)$ in cluster frame & \S\ref{subsec:Stacking} \\
        $\delta\llangle\boldsymbol{I}\rrangle$, $\delta I_\theta$ & Detrended ambient, X-ray-correlated spectra of the \SCW\ region & Appendix~\ref{app:eROSITA} \\
        \bottomrule
    \end{tabularx}
    \end{table*}

\begin{table*}[tp!]
    \centering
    \caption{Features, fixed-mass tests, filters, statistics, and cross sections.\label{tab:Symbols}}
    \begin{tabularx}{\textwidth}{@{}l X l@{}}
        \toprule
        Symbol & Meaning & Defined in \\
        \midrule
        \multicolumn{3}{l}{\bf Features and masses} \\
        Triad & The $\gamma\gamma$, $\gamma Z$, and $\gamma h$ lines of a WIMP pair of mean mass $\overline{m}_\chi$ & Eq.~\eqref{eq:kin} \\
        Nonad & Triads of two species, $m_1<m_2$, and of their cross-annihilation & Eq.~\eqref{eq:nonad} \\
        Nominal triad, nonad & The strongest blind triad ($69.9\GeV$) and nonad ($\{69.8, 74.5\}\GeV$) of \SAll & \S\ref{subsec:Blind}, \S\ref{subsec:Nonads} \\
        $m_\chi$ & Mass of a single WIMP species & Eq.~\eqref{eq:Lchi} \\
        $\overline{m}_\chi$ & Mean mass of the two (with identical or different) annihilating WIMPs & Eq.~\eqref{eq:kin} \\
        $\epsilon_0$ & Line energy & \S\ref{subsec:MatchedFilters} \\
        $\chi_1$, $\chi_2$ & The two species of the model (best fit: $m_1c^2\simeq69.7$ and $m_2c^2\simeq74.5\GeV$) & \S\ref{sec:Model} \\
        \midrule
        \multicolumn{3}{l}{\bf The $43\GeV$ line and fixed-mass tests} \\
        $L_{43}$ & The $43.2\GeV$ line reported in \SCT\ ($Z_g\simeq4.3$; $2.7$ in \SCTb) & \S\ref{sec:Intro} \\
        Reading of $L_{43}$ & $L_{43}$ as a $\gamma Z$ (nominal) or a $\gamma h$ line (of an $87.8\GeV$ triad) & \S\ref{subsec:FixedTests} \\
        $T_{72}$ & The $72.05\GeV$ triad fixed a priori by the $\gamma Z$ reading of $L_{43}$ & \S\ref{subsec:FixedTests} \\
        $C_{72}$ & The $\gamma\gamma$ and $\gamma h$ companions of $L_{43}$ in $T_{72}$ & \S\ref{subsec:FixedTests} \\
        Readings of $T_{72}$ & $T_{72}$ as the lower (low), upper (high), or intermediate triad of a nonad & \S\ref{subsec:FixedTests} \\
        $N_{72}$ & Fixed-mass nonad: the strongest nonad with intermediate triad $T_{72}$ & \S\ref{subsec:FixedTests} \\
        $N_{72}^{\max}$ & The strongest nonad over the three readings of $T_{72}$ & \S\ref{subsec:FixedTests} \\
        \midrule
        \multicolumn{3}{l}{\bf Matched filters} \\
        $\mathbb{M}_i$, $\mathcal{I}_i$ & Matched filter of line $i$, and the integrated brightness it measures; $\mathcal{I}_i=\mathbb{M}_i[\delta I]$ & Eq.~\eqref{eq:MatchedFilter} \\
        $f$ & Fractional half-width of the exclusion window of a feature & \S\ref{subsec:MatchedFilters} \\
        $f_t$ & Window keeping apart the outer triads of a nonad & \S\ref{subsec:MatchedFilters} \\
        \midrule
        \multicolumn{3}{l}{\bf Local and global significance} \\
        $Z_l$ & Local significance of a matched filter (for triads and $C_{72}$, Eq.~\eqref{eq:CoAddedZ}) & \S\ref{subsec:MatchedFilters} \\
        $Z_{\gamma\gamma}$, $Z_{\gamma Z}$, $Z_{\gamma h}$ & The $Z_l$ values of channels within a triad & \S\ref{subsec:MatchedFilters} \\
        $Z_c$ & Significance of a nonad component beyond its overlap with its companions & Eq.~\eqref{eq:Zc} \\
        $Z_g$ & Global significance, calibrated by null realizations or as reported & \S\ref{subsec:Significance} \\
        $\smash{Z_g^{(a)}}$ & Analytic (\smash{\v{S}id\'{a}k}) estimate of $Z_g$ & Appendix~\ref{app:Nulls} \\
        $\smash{Z_g^{(n)}}$ & $Z_g$ with normal (Gaussian) nulls instead of shuffled skies or (stacking alone) Poisson & Appendix~\ref{app:Nulls} \\
        $\smash{Z_g^{(r)}}$, $\smash{Z_g^{(z)}}$ & $Z_g$ calibrated by random-longitude stacks, or by redshift permutations & \S\ref{subsec:Significance} \\
        $p$ & A $p$-value & \S\ref{subsec:Significance} \\
        \midrule
        \multicolumn{3}{l}{\bf Fits and null realizations} \\
        TS, $\Delta\textrm{TS}=1$ & Test statistic ($Z_l^2$; summed over datasets for the model), unit-drop bounds & \S\ref{subsec:MatchedFilters} \\
        $\mathcal{T}$ & Goodness of fit: sum of own TS maxima minus the maximal summed TS & \S\ref{subsec:ModelMasses} \\
        Shuffled-contrast sky & \SCW\ or \SMW\ null realization: the X-ray contrast shuffled among the pixels\cite{Keshet26Corr} & \S\ref{subsec:Significance} \\
        Null pair & Null combining Poisson stacking and shuffled-contrast (or Gaussian) \SMW\ or \SCW & \S\ref{subsec:Significance} \\
        \midrule
        \multicolumn{3}{l}{\bf Cross sections} \\
        $\overline{\sigma v}_p$ & Intrinsic $p$-wave cross section & Eq.~\eqref{eq:EffCS} \\
        $\beta^2$ & Squared line-of-sight velocity dispersion, dimensionless & Eq.~\eqref{eq:EffCS} \\
        $B$ & Substructure boost factor & Eq.~\eqref{eq:EffCS} \\
        $\Omega_m$, $\Omega_b$, $\Omega_c$ & Density parameters of matter, baryons, and cold DM & \S\ref{subsec:Samples} \\
        $\Omega_\chi$ & Density parameter of WIMP species $\chi$ & Eq.~\eqref{eq:Lchi} \\
        $\mathcal{S}$ & Shape parameter ($\mathcal{S}_{\text{NFW}}$ or $\mathcal{S}_{\textrm{core}}$) & Eq.~\eqref{eq:Lchi} \\
        $\overline{\sigma v}_{p,i}/D$ & $p$-wave cross section of channel $i$, scaled by the dark-sector factor $D$ & \S\ref{subsec:CrossSections} \\
        $\Sigma_0$ & Intrinsic $\gamma\gamma$ cross section $\overline{\sigma v}_{p,\gamma\gamma}/D$ of a triad & Eq.~\eqref{eq:Sigma0} \\
        $\Sigma$ & Stacking measurement of $\Sigma_0$ before loss correction ($\Sigma\simeq\mathcal{F}_{\gamma\gamma}\Sigma_0$) & Eq.~\eqref{eq:Sigma} \\
        $\mathcal{F}_i$ & Ratio of pipeline-measured to true cross section of line $i$, for an equal-brightness triad & \S\ref{subsec:Losses} \\
        $K$ & Ratio of the NFW-normalized $D$ to the gas-normalized $D$ of \SCW, for a given signal & Appendix~\ref{app:eROSITA} \\
        $\Sigma_{\textrm{X}}$ & Stacking-like (NFW mapping with stacking losses) $\Sigma$ estimated from \SCW\ or \SMW\ & Appendix~\ref{app:eROSITA} \\
        $\smash{\mathcal{F}^{\textrm{X}}_i}$ & Ratio of stacking-like to true X-ray-correlated cross sections of line $i$ & Appendix~\ref{app:eROSITA} \\
        $\mathcal{F}_b$ & Fraction of the X-ray variance in \SCW\ due to bremsstrahlung & Appendix~\ref{app:eROSITA} \\
        \bottomrule
    \end{tabularx}
    \end{table*}

\section{Composition of the samples}
\label{app:Samples}

Table~\ref{tab:Samples} lists the numbers of clusters of \SAll, of its subsamples (\S\ref{subsec:Samples}), and of the bins of predicted signal, in total and in each catalog.
The hemispheres \SE\ and \SW\ lie at Galactic longitudes $l$ (taken mod $360\dgr$) below and above $180\dgr$.
Besides the tertiles of \S\ref{subsec:Scaling}, which rank the clusters of each catalog by their $\smash{\tilde{\mathcal{J}}_\Myc}$ factor (predicted signal), three sets of five bins examine the dependence of $\Sigma$ on predicted signal more finely.

The $\smash{\tilde{\mathcal{J}}}_1$--$\smash{\tilde{\mathcal{J}}}_5$ (highest first) are constructed like the $\smash{\tilde{\mathcal{J}}}$ tertiles, each holding a fifth of every catalog.
The $\mathcal{N}_1$--$\mathcal{N}_5$ (smallest bin first) rank all clusters together, each bin carrying a fifth of their total predicted signal.
The $\mathcal{M}_1$--$\mathcal{M}_5$ are constructed like $\smash{\tilde{\mathcal{J}}}_j$,
but separately
in each stratum of one catalog and one redshift slice a factor $1.25$ wide in $z$,
so they rank clusters mainly by mass.
The bins of predicted signal exclude the 160 nearby clusters of \S\ref{subsec:Scaling} and the 2032 other clusters with fully masked ROIs.

\begin{figure}[h!]
    \centering
    \includegraphics[width=\columnwidth]{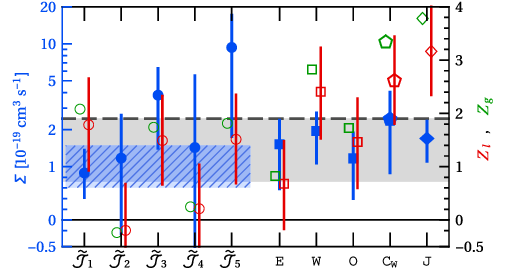}\\[2pt]
    \includegraphics[width=\columnwidth]{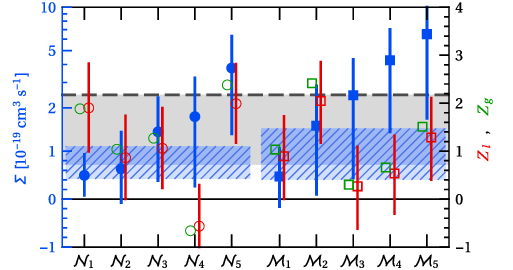}
    \caption{
        Same as Fig.~\ref{fig:Scaling}, different samples and in finer bins of predicted signal (labels), with the inverse-variance-weighted mean $\Sigma$ of each set of bins (hatched band).
        \label{fig:Bins5}
    }
\end{figure}

Figure~\ref{fig:Bins5} shows $\Sigma$ and $T_{72}$ in these bins.
Within each set ($\smash{\tilde{\mathcal{J}}}_j$, $\mathcal{N}_j$, and $\mathcal{M}_j$), the cross sections are found to be mutually consistent ($\chi^2\simeq2.7$, $2.8$, and $4.1$ for 4 degrees of freedom), with inverse-variance-weighted means $\Sigma/(10^{-19}\cm^3\se^{-1})$ of $1.0\pm0.4$, $0.75\pm0.33$, and $0.92\pm0.52$.
In Monte Carlo, these means favor a signal at $2.3\sigma$, $2.0\sigma$, and $1.8\sigma$, below the ratio of $\Sigma$ to its error, where a few highly weighted clusters dominate a bin and widen its Poisson null.
As in the tertiles, $\Sigma$ rises toward low predicted signal.
Poisson simulations of a line flux $\propto\smash{\tilde{\mathcal{J}}_\Myc^{\,\kappa}}$ (\S\ref{subsec:Scaling}) give $\kappa\simeq-0.1$ to $0.4$ for the binnings of the $\smash{\tilde{\mathcal{J}}}_j$ and $\mathcal{N}_j$ kinds (tertiles to quintiles), and $\kappa\simeq-0.3$ to $0$ for those of the $\mathcal{M}_j$ kind.
The binnings disfavor $\kappa=1$ at $p\simeq0.03$--$0.09$, or at $p\simeq0.05$--$0.2$ if the catalog masses scatter about the true masses by $0.3$--$0.4$ in $\ln M_{500}$.

The same simulations correct for the multiple trials of nine comparisons among cluster subsamples: the $\kappa$ fit of the tertiles, the $\Sigma$ of the catalogs, of each set of bins, and of the mass tertiles, and the $T_{72}$ of the redshift tertiles, hemispheres, and quadrants.
The correction raises the $p$ of the least consistent comparison, the $T_{72}$ of the redshift tertiles, from $0.03$ to $0.19$, and that of the $\kappa$ fit of the tertiles from $0.09$ to $0.52$.
For an extended list adding the other binnings and redshift quintiles, the correction gives $p\simeq0.12$ for the redshift quintiles and $p\gtrsim0.3$ for the $\kappa$ fits of all other binnings.

\begin{table}[h!]
    \centering
    \setlength{\tabcolsep}{2.5pt}
    \caption{Cluster samples.\label{tab:Samples}}
    \begin{tabular}{l l r r r r}
        \toprule
        Sample & Selection & $N_c$ & MCXC & $\Delta$eRO & $\Delta$DESI \\
        \midrule
        {\bfseries\boldmath \SAll} & {\bfseries\boldmath all clusters} & {\bfseries\boldmath 23343} & {\bfseries\boldmath 1628} & {\bfseries\boldmath 7833} & {\bfseries\boldmath 13882} \\
        \SE & $0<l<180\dgr$ & 10274 & 895 & 2 & 9377 \\
        \SW & $180\dgr\le l<360\dgr$ & 13069 & 733 & 7831 & 4505 \\
        \SO & outside \SCW & 14992 & 1226 & 3188 & 10578 \\
        $\smash{\tilde{\mathcal{J}}}$ tertiles & thirds$^\dagger$ & $\sim$7050 & $\sim$468 & 2359 & 4223 \\
        $\smash{\tilde{\mathcal{J}}}_{1\textrm{--}5}$ & fifths$^\dagger$ & $\sim$4230 & 281 & $\sim$1415 & $\sim$2534 \\
        $\mathcal{M}_{1\textrm{--}5}$ & fifths$^\ddagger$ & $\sim$4230 & 281 & $\sim$1415 & $\sim$2534 \\
        $\mathcal{N}_1$ & highest fifth$^\S$ & 357 & --- & --- & --- \\
        $\mathcal{N}_2$ & second fifth & 1187 & --- & --- & --- \\
        $\mathcal{N}_3$ & third fifth & 2321 & --- & --- & --- \\
        $\mathcal{N}_4$ & fourth fifth & 4711 & --- & --- & --- \\
        $\mathcal{N}_5$ & lowest fifth & 12575 & --- & --- & --- \\
        \bottomrule
    \end{tabular}
    \TblNote{Numbers of clusters in the samples and in the bins of predicted signal (Appendix~\ref{app:Samples}), in total ($N_c$) and in each catalog. The bins exclude the 160 nearby clusters of \S\ref{subsec:Scaling} and the clusters with fully masked ROIs. \\
    $^\dagger$Of the clusters of each catalog, ranked by predicted signal. \\
    $^\ddagger$Of the clusters of each catalog--redshift stratum, likewise ranked. \\
    $^\S$Of the predicted signal of all clusters.
    }
\end{table}

\begin{figure}[tbp]
    \centering
    \includegraphics[width=\columnwidth]{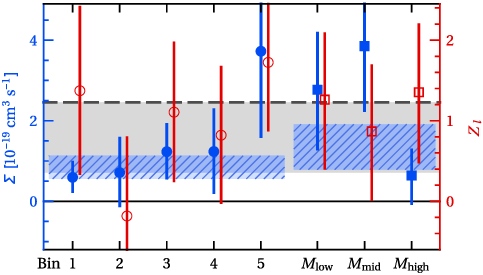}
    \caption{
        Same as Fig.~\ref{fig:Scaling}, for all clusters of \SAll, in five bins of predicted $\propto L_{500}$ signal per unmasked ROI pixel (disks; bin 1 highest) and in tertiles of catalog mass (squares), with the inverse-variance-weighted mean $\Sigma$ of each group (hatched bands).
        \label{fig:ScalingAlt}
    }
\end{figure}

Tertiles of catalog $M_{500}$ (Fig.~\ref{fig:ScalingAlt}) show no rise of $\Sigma$ with mass, which a cross section depending on velocity more steeply than $\beta^2$ would cause.
Removing each tertile in turn lowers $T_{72}$ only to $1.8$--$2.1$, as expected for removing a third of the statistical weight.

The predicted signal with $\mathcal{J}_\Myc$ replaced by the X-ray flux $L_{500}/(4\pi d_L^2)$ relates the line brightness to the X-ray luminosity.
Five bins of this signal per unmasked ROI pixel, each carrying a fifth of the total (Fig.~\ref{fig:ScalingAlt}), give mutually consistent cross sections ($\chi^2\simeq2.8$ for 4 degrees of freedom).
For clusters with no catalog $L_{500}$, we adopt $\log_{10}(L_{500}/\textrm{erg}\,\se^{-1})=43.04+1.36\log_{10}[M_{500}/(10^{14}M_\odot)]+4.69\log_{10}E(z)$, with $E(z)\equiv H(z)/H_0$, fitted to $\Delta$eROSITA.

\section{Spectral detrending}
\label{app:Detrending}

This appendix details the two steps of detrending (\S\ref{subsec:Detrending}) and the contributions omitted by the null variance of Eq.~\eqref{eq:Var}.
Detrending uses all rest-frame energy channels $\Myfr$ with a null error $\sigma_\Myfr>0$ and a positive stacked field brightness.

In the first step, the continuum is estimated by Gaussian-process regression (GPR), applied to the ratio $U_\Myfr\equiv\Delta\smash{\tilde{I}}_\Myfr/P_\Myfr$ of the aligned excess brightness of channel $\Myfr$ (Eq.~\eqref{eq:deltaI}) to a fit $P_\Myfr$ of the stacked field brightness, which is defined like $\Delta\smash{\tilde{I}}_\Myfr$ but with the field-expected counts $\Myf$ in place of the excess counts $\Delta\Myn$.
The logarithm of $P_\Myfr$ is the cubic polynomial in $x_\Myfr\equiv\ln\epsilon_\Myfr$ fitted by least squares to the logarithm of the stacked field brightness, with weights $\sum_{\Myc}\Myf_{\Myfr'}(\Myc,\MyH)$ (summed over the ROI pixels $\MyH$ of each cluster), proportional to the inverse Poisson variance of this logarithm.
As the fields exclude the ROIs of all stacked clusters, $P_\Myfr$ is independent of the ROI photons.

For a squared-exponential channel $\Myfr$, $j$ prior covariance $\mathcal{K}_{\Myfr j}=A\exp[-(x_\Myfr-x_j)^2/(2x_0^2)]$ of amplitude $A$ and correlation length $x_0$, the smoothed ratio is
\begin{equation}\label{eq:GPR}
\hat{U}=\bar{U}+\mathcal{K}\left[\mathcal{K}+\textrm{diag}\left(\sigma_\Myfr^2/P_\Myfr^2\right)\right]^{-1}\left(U-\bar{U}\right) \coma
\end{equation}
where $U$ and $\hat{U}$ are channel vectors, $\bar{U}$ is the mean of $U_\Myfr$ weighted by $P_\Myfr^2/\sigma_\Myfr^2$, and $\textrm{diag}(\sigma_\Myfr^2/P_\Myfr^2)$ is the noise covariance of $U$.
Equation~\eqref{eq:GPR} is the mean of a Gaussian process of mean $\bar{U}$ and covariance $\mathcal{K}$, conditioned on the measured $U$.
The estimated continuum $P_\Myfr\hat{U}_\Myfr$ is then subtracted from $\Delta\smash{\tilde{I}}_\Myfr$.

The correlation length, fixed a priori to match previous work\cite{Keshet26Corr} at $0.27$ dex in energy, i.e., $x_0=0.27\ln10\simeq0.62$ in $x=\ln\epsilon$, is $\sim9$ times the $\sim7\%$ half-width, in $\ln\epsilon$, of the LAT $68\%$ energy containment of a line.
The amplitude $A$ maximizes the marginal likelihood of $U$, on a broad grid spanning negligible to dominant priors.
Varying $x_0$ over $0.20$--$0.35$ decades, or omitting the GPR altogether (the second step only), changes $T_{72}$ and the $Z_l$ of the strongest triad of \SAll\ by $\le0.1$.
The linear fit of the second step uses the residuals $\Delta\smash{\tilde{I}}_\Myfr-P_\Myfr\hat{U}_\Myfr$ at $3<\epsilon_\Myfr/\mbox{GeV}<100$.

With $A$ held at its value for the data, both steps are linear in $\Delta\smash{\tilde{I}}$ and together define the fixed matrix $\mathcal{D}$ of \S\ref{subsec:Detrending}, as their weights, set by $\sigma_\Myfr$ and $P_\Myfr$, do not depend on the ROI photons.
Applied to the Poisson Monte Carlo realizations, which share the errors and field fit of the data, $\mathcal{D}$ neglects only the small variation of $A$ among them, whereas the random-longitude stacks and redshift permutations rebuild $\mathcal{D}$ for each realization.

The null variance of Eq.~\eqref{eq:Var} omits the variance of any cluster excess and the covariance between the field estimates of clusters with overlapping annuli, which would raise the variance of the field estimates to $\lesssim2\%$ of the cluster-frame null variance.
The null variance also omits all covariances between channels, which arise from photons shared by ROIs whose spectra are shifted by different numbers of channels, and from detrending.
Of these omissions, the Poisson Monte Carlo includes all but the noise of the field estimates, the redshift permutations include all, and the random-longitude stacks, whose ROIs avoid the clusters, include all but the variance of the cluster excess.
The calibrated $Z_g$ thus include nearly all the contributions that the errors behind $Z_l$ omit.

\section{Mock matched-filter tests}
\label{app:Mock}

\begin{figure*}[t!]
    \centering
    \includegraphics[width=0.36\textwidth]{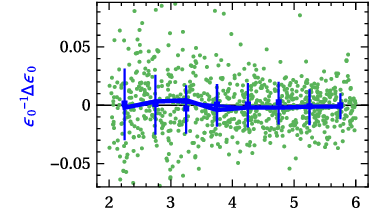}\hfill
    \includegraphics[width=0.31\textwidth]{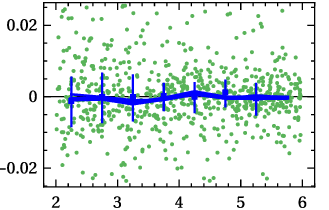}\hfill
    \includegraphics[width=0.31\textwidth]{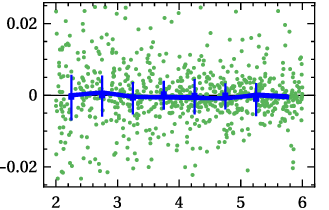}\\[2pt]
    \includegraphics[width=0.36\textwidth]{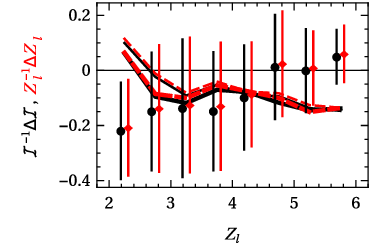}\hfill
    \includegraphics[width=0.31\textwidth]{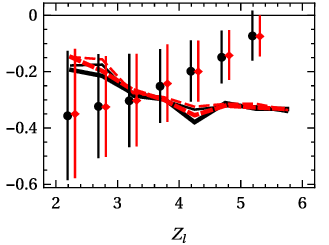}\hfill
    \includegraphics[width=0.31\textwidth]{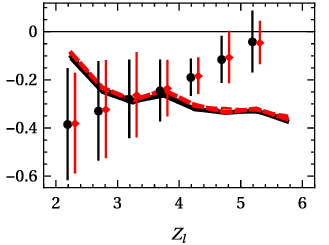}
    \caption{
        Recovery of mock lines (left column) and triads of equal channel $Z_l$ (middle) or brightness $\mathcal{I}$ (right), injected into \SAll\ with an equal line flux per cluster, binned in injected $Z_l$ (curves) or in recovered $Z_l$ (symbols and error bars).
Top: fractional error $\epsilon_0^{-1}\Delta\epsilon_0$ of the energy (triads: $\overline{m}_\chi$), binned (blue squares) and for each recovered mock (green points, at their injected $Z_l$).
        Bottom: fractional errors $\mathcal{I}^{-1}\Delta\mathcal{I}$ (triads: summed over channels; black circles, solid curves) and $Z_l^{-1}\Delta Z_l$ (red diamonds, dashed curves).
        Shown are both standard (means: thin curves) and robust (medians and $1.483\times$MAD: thick curves and error bars) statistics; circles and diamonds are offset by $\mp0.06$ for clarity.
\label{fig:Mock}
    }
\end{figure*}

Mock lines and triads, injected into the Poisson Monte Carlo backgrounds of \SAll\ and scanned like the data, test the matched filters (\S\ref{subsec:MatchedFilters}).
Mock cluster lines are introduced at the observed energy $\epsilon_0/(1+z_\Myc)$, dispersed by the LAT response, and the mock then passes, like a real signal, through the same pipeline as the data.
The line flux is taken as equal for all clusters or, as a check giving nearly identical statistics, $\propto\mathcal{J}_\Myc$.

Lines (at $5$--$90\GeV$) and triads ($66$--$95\GeV$) of equal channel $Z_l$ or brightness are injected at a prescribed pre-detrending $Z_l$.
An injected feature counts as found if it lies within $10\%$ (lines) or $3\%$ (triads) of one of the scan's three strongest blind features.
At injected $Z_l=3$--$5$, at least $84\%$ of the mock features are found, with a negligible ($\lesssim0.3\%$) bias and robust ($1.483\times$MAD) dispersions of $1.5$--$2.6\%$ in line energy and $0.4$--$0.9\%$ in triad mass. See Fig.~\ref{fig:Mock}.

Even without noise, brightness and $Z_l$ fall below their injected values by $(13\pm4)\%$ for lines and by $(37\pm3)\%$ for triads due to detrending.
At injected $Z_l=3$--$5$, noise adds a scatter of $\sim25\%$ (lines) and $\sim20\%$ (triads) about these losses.
Without the coupling of triad channels by detrending, the noise-free triad $Z_l$ would fall only $\sim15\%$ below its injected value.
For the triad injected into an empty sky to measure $\mathcal{F}_i$ (\S\ref{subsec:Losses}), detrending alone retains $\{0.60, 0.60, 0.76\}$ of the channel brightness measured before it, and the cross sections inferred before detrending exceed the true ones by $\{2.9, 1.8, 0.1\}\%$, giving $\mathcal{F}_i\simeq\{0.62, 0.61, 0.77\}$.

We compare model mock nonads injected at $2<Z_l<6$ with mocks of their intermediate triad injected at the same $Z_l$ into the same noise.
The nonads lose a median $51\%$ of their $Z_l$, against $32\%$ for the triads, at the injected masses, where the detrended signals still peak.
Blind scans recover nonads in only $30$--$62\%$ of the mocks, rising with injected $Z_l$, against $93$--$100\%$ for triads.
Random instead of model masses give similar results, as do the mocks of the cross-correlation analysis\cite{Keshet26Corr}.

Compensated for these median losses, the blind triad and nonad of \SAll\ would reach $Z_l\simeq4.9$ and $6.4$, too high to calibrate with the $2\times10^5$ null scans.
Their $Z_g$ is instead bounded by summing, over the scanned masses, the null probabilities of exceeding these values at each mass, extrapolated from the Monte Carlo mean and dispersion of the linear $Z_l$ with Poisson tails, which gives $Z_g>4.4$ and $>7.2$, and likewise $\sim4$ and $>7$ for \SJ.

\section{Null distributions and variants}
\label{app:Nulls}

\begin{figure*}[t!]
    \centering
    \includegraphics[width=\columnwidth]{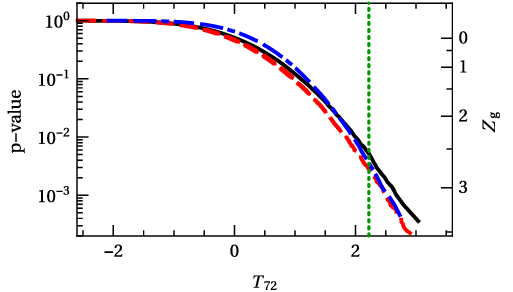}\hfill
    \includegraphics[width=\columnwidth]{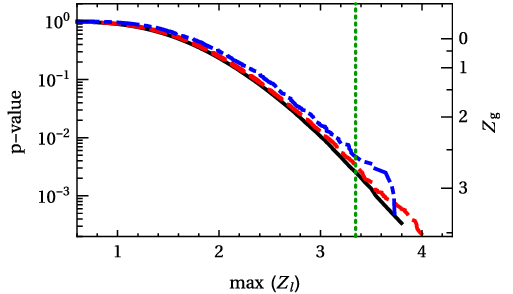}
    \caption{
        Null distributions for \SAll: the fraction of null stacks ($p$-value, left axis, and the corresponding $Z_g$, right axis) reaching a given $T_{72}$ (left panel) or maximal blind-triad $Z_l$ (right panel), in the Poisson Monte Carlo (solid black), random longitudes (dashed red), and redshift permutations (dot-dashed blue), with the observed values (dotted green).
        \label{fig:Nulls}
    }
\end{figure*}

In each calibration of \S\ref{subsec:Significance}, $p$ is the fraction of null realizations that exceed the observed statistic.
The Poisson Monte Carlo, random longitudes, and redshift permutations give similar null distributions of $T_{72}$ and of the maximal blind-triad $Z_l$ (Fig.~\ref{fig:Nulls}).
In redshift permutations, which keep the cluster photons, $T_{72}$ averages $0.32$, partly from surviving signal, which conservatively lowers their $Z_g$.

The analytic \smash{\v{S}id\'{a}k} estimate $Z_g^{\smash[t]{(a)}}$ follows from $1-p_g\simeq(1-p_l)^{N_t}$, with global $p_g$ and local $p_l$ and an effective number of trials $N_t$ equal to the logarithmic scan range divided by the exclusion width $2f$ (half the square of this ratio for the two-mass nonads).
The Gaussian noise of $Z_g^{\smash[t]{(n)}}$ is drawn on the detrended native bins, bypassing detrending.
Both estimates assume a unit Gaussian $Z_l$ rather than the narrower detrended null, so are conservative,
falling \eg $\sim1$ below the Monte Carlo $Z_g$ for the \SAll\ nonad, and by less for its triads; the nominal calibrations are free of this bias, as they pass every null realization through the same detrending as the data.

Channel-masked detrending (\S\ref{subsec:Detrending}) gives zero weight to the channels near each tested line energy.
As the GPR keeps the nominal amplitude $A$, each mask defines a fixed matrix, applied alike to the data and to the null stacking realizations.
The Poisson Monte Carlo here restores the smooth excess of the data detrended with the nine lines of the model masked, so no putative line enters the null.
Masking moves the strongest blind features by $\le0.4\GeV$, within their nominal $\Delta\textrm{TS}=1$ bounds.
The masked $T_{72}$ is insensitive to the mask half-width: its Gaussian significance in \SAll, computed from its exact null mean and standard deviation, is $2.7$, $2.8$, and $2.8$ for half-widths of $0.035$, $0.05$ (nominal), and $0.07$.

Errors from the observed rather than the nominal field-expected counts lower $T_{72}$ by only $0.3$.
Variants that add or remove photons change $T_{72}$ as expected from photon noise and signal distribution.
Using synthetic apertures instead of IRF filters, asymmetric exclusion windows, and a lower HEALPix order recover the same features, if less sensitively.

\section{Calibration of the catalog masses}
\label{app:MassCal}

\begin{figure}[h!]
    \centering
    \includegraphics[width=\columnwidth]{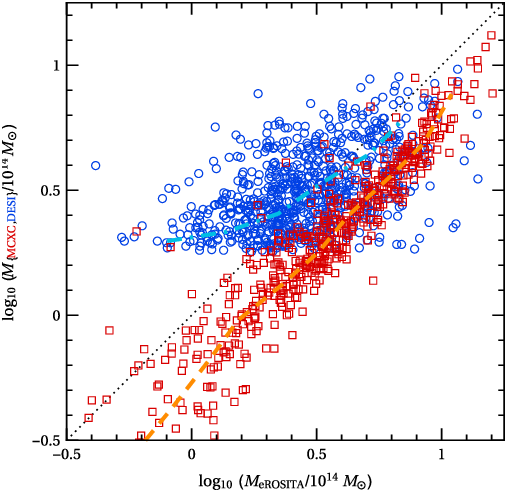}\\
    \includegraphics[width=\columnwidth]{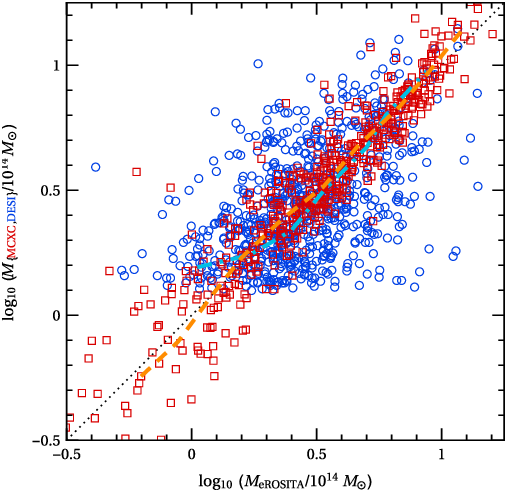}
    \caption{
    Masses $M_{500}$ of clusters listed both in eROSITA (abscissa) and in either (ordinate) MCXC (red squares) or DESI (blue circles), as cataloged (top) and Eq.~\eqref{eq:MassCal}-calibrated (bottom).
    After calibration, the median profiles of MCXC (orange dashed) and DESI (cyan dot-dashed) follow the diagonal (dotted).
    \label{fig:MassCal}}
\end{figure}

The $\mathcal{J}$-factor of a cluster in Eq.~\eqref{eq:Jc} depends on its catalog $M_{500}$, which MCXC infers from the X-ray luminosity, eROSITA from the X-ray count rate, and DESI from the richness.
Here and below, we consider the catalogs after our cuts (in \S\ref{subsec:Samples}) but before removing catalog overlaps.
The MCXC and DESI masses are then rescaled to match, on average, the eROSITA masses of their eROSITA-overlapping counterparts.

To do so, each MCXC or DESI cluster that has an eROSITA counterpart of valid $M_{500}$ within a projected $0.5\Mpc$, at a redshift $z_{\textrm{e}}$ satisfying $|z-z_{\textrm{e}}|/(1+z)<0.05$, is paired with the counterpart closest in redshift.
Of the resulting 512 MCXC and 721 DESI pairs, only $\sim1\%$ are chance coincidences.

The MCXC pairs show a constant offset, with a median $\log_{10}(M_{\textrm{eROSITA}}/M_{\textrm{MCXC}})=0.238\pm0.005$.
The DESI pairs show a mass-dependent offset, with $M_{\textrm{eROSITA}}\propto M_{\textrm{DESI}}^{1.44\pm0.05}$.
Hence, we adopt calibrated masses
\begin{equation}\label{eq:MassCal}
  \frac{\hat{M}_{\textrm{MCXC}}}{M_{\textrm{MCXC}}} = 1.73
  \mbox{ ;}\quad
  \frac{\hat{M}_{\textrm{DESI}}}{10^{14}M_\odot}=0.54\left(\frac{M_{\textrm{DESI}}}{10^{14}M_\odot}\right)^{1.44} \coma
\end{equation}
as illustrated in Fig.~\ref{fig:MassCal}.
At low eROSITA masses, the DESI masses level off, as seen in the figure, as only clusters scattered above the $\lambda_{500}>40$ cut enter DESI there.

Table~\ref{tab:MassCalResults} repeats relevant tests using the calibrated masses (eROSITA masses used for paired clusters).
The calibration mainly shifts the share of $\tilde{\mathcal{J}}$ from $\Delta$DESI to MCXC (the catalog of lowest $\Sigma$; Table~\ref{tab:Fixed}), so the cross sections of the cluster subsets agree less and the mean $\Sigma$ of the $\mathcal{N}_j$ falls, whereas the $L_{43}$
cross section of Virgo and Fornax agrees with the $\Sigma_{\gamma Z}$ of \SAll.

\begin{table}[t]
    \centering
    \setlength{\tabcolsep}{4pt}
    \caption{Catalog vs.\ calibrated $M_{500}$.\label{tab:MassCalResults}}
    \begin{tabular}{@{}l c c@{}}
        \toprule
        Quantity & Catalog & Calibrated \\
        \midrule
        Median $M_{500}/(10^{14}M_\odot)$ of MCXC & $1.7$ & $2.9$ \\
        Median $M_{500}/(10^{14}M_\odot)$ of $\Delta$DESI & $2.3$ & $1.8$ \\
        Share of $\tilde{\mathcal{J}}$ in MCXC & $22\%$ & $43\%$ \\
        Share of $\tilde{\mathcal{J}}$ in $\Delta$DESI & $45\%$ & $30\%$ \\
        $\chi^2$ of the five $\mathcal{N}_j$ (4 dof) & $3.0$ & $9.5$ \\
        $\chi^2$ of the mass tertiles (2 dof) & $4.7$ & $8.8$ \\
        $\chi^2$ of the three catalogs (2 dof) & $2.8$ & $6.1$ \\
        Mean $\Sigma/(10^{-19}\cm^3\se^{-1})$ of $\mathcal{N}_j$ & $0.96\pm0.30$ & $0.33\pm0.14$ \\
        $\Sigma/(10^{-19}\cm^3\se^{-1})$ of \SAll & $1.7\pm0.6$ & $1.4\pm0.5$ \\
        Expected $T_{72}$$^\dagger$ & $\sim2.9$ & $\sim1.2$ \\
        $L_{43}$ above the $\gamma Z$ of \SAll$^\ddagger$ & $1.5$ & $0.7$ \\
        \bottomrule
    \end{tabular}
    \TblNote{
    Results shown for all (including 160 nearby) \SAll\ clusters.\\
    $^\dagger$Noise-free $T_{72}$ of a triad of equal channel brightness whose measured $\Sigma$ equals the mean of $\mathcal{N}_j$ (observed: $T_{72}\simeq2.2$).\\
    $^\ddagger$The $L_{43}$ cross section of Virgo and Fornax for uniform emission (Appendix~\ref{app:Line43}), in combined standard deviations.}
\end{table}

\section{Subsample spectra}
\label{app:SubSpectra}

\begin{figure*}[htbp]
    \centering
    \includegraphics[width=\textwidth]{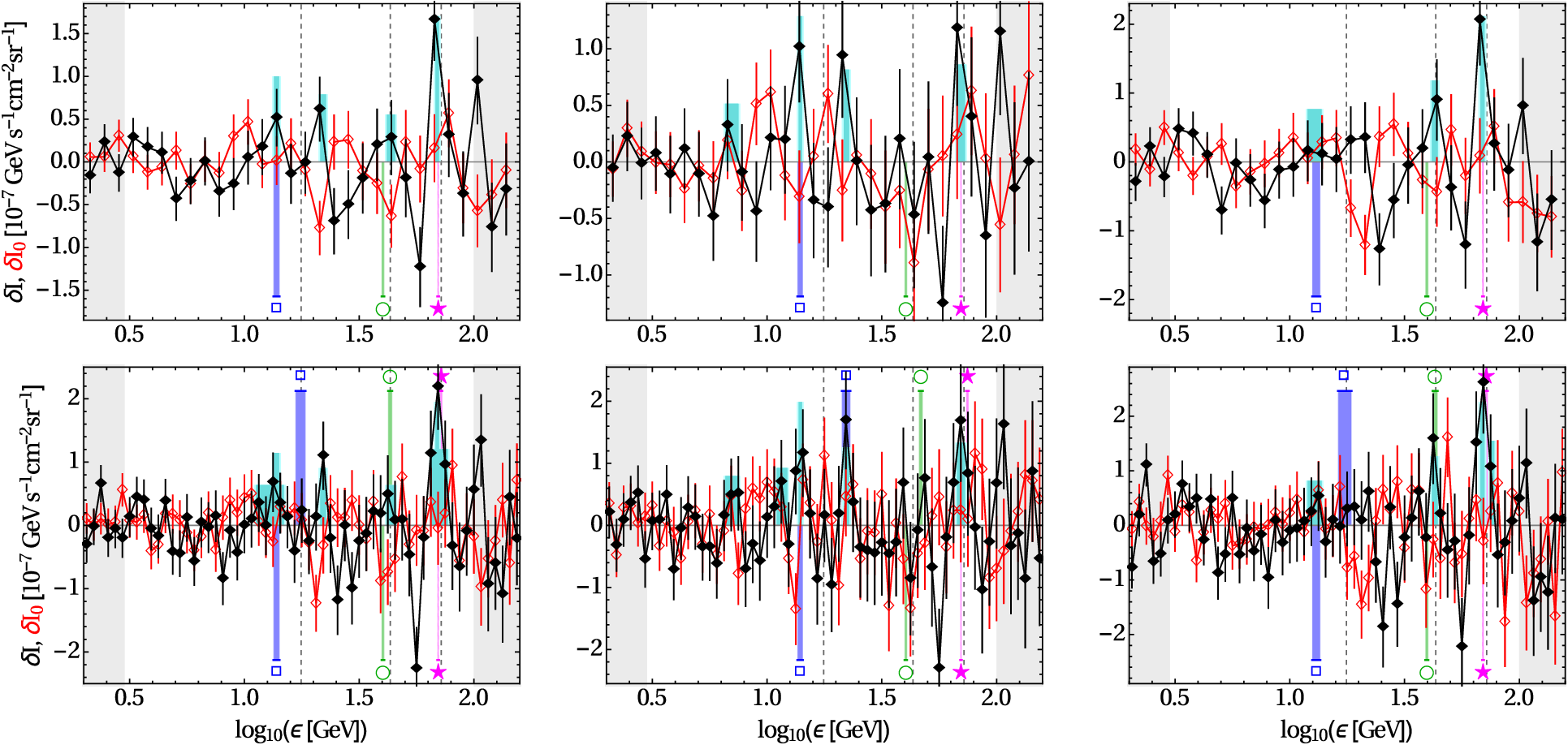}
    \caption{
        Same as the top two rows of Fig.~\ref{fig:Spectra}, for \SAll\ (left), \SE\ (middle), and \SW\ (right), but the bottom row marks the channels of the strongest (negative bars) and second-strongest (positive bars) blind triads instead of the nonad.
        \label{fig:SpectraEW}
    }
\end{figure*}

\begin{figure*}[tbp]
    \centering
    \includegraphics[width=0.38\textwidth]{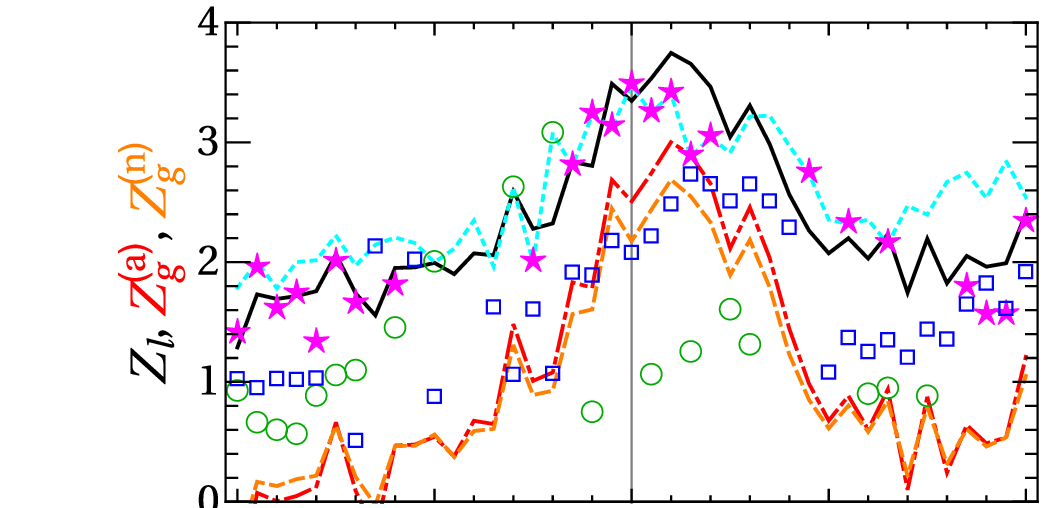}\hfill
    \includegraphics[width=0.30\textwidth]{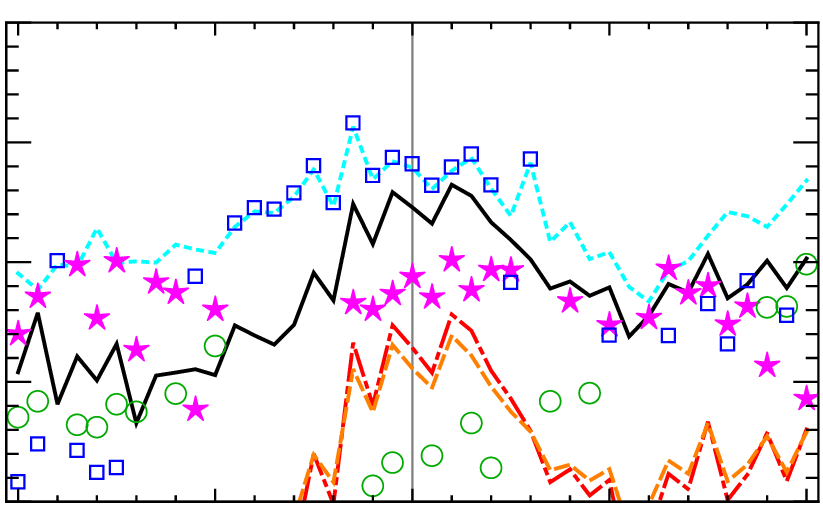}\hfill
    \includegraphics[width=0.30\textwidth]{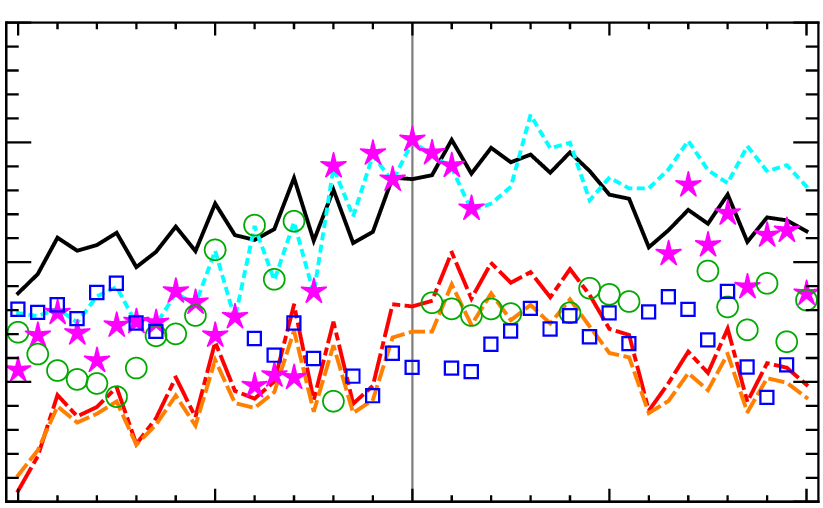}\\
    \includegraphics[width=0.38\textwidth]{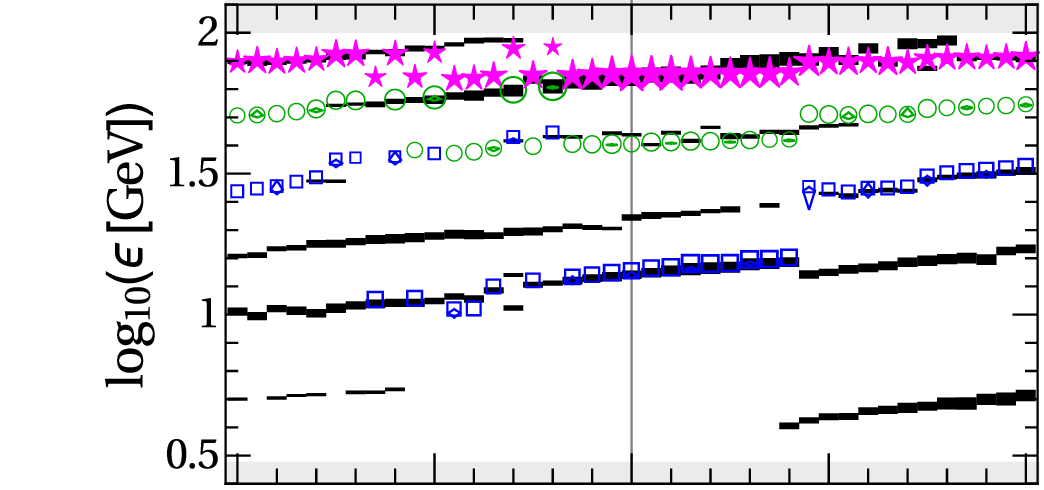}\hfill
    \includegraphics[width=0.30\textwidth]{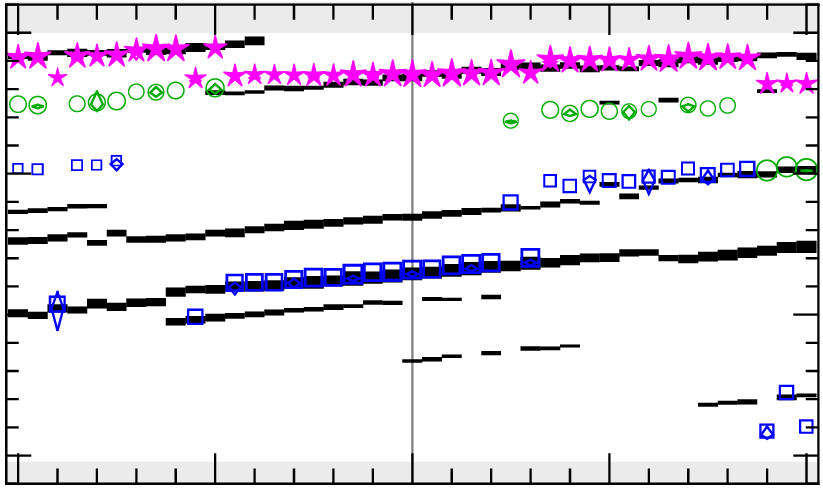}\hfill
    \includegraphics[width=0.30\textwidth]{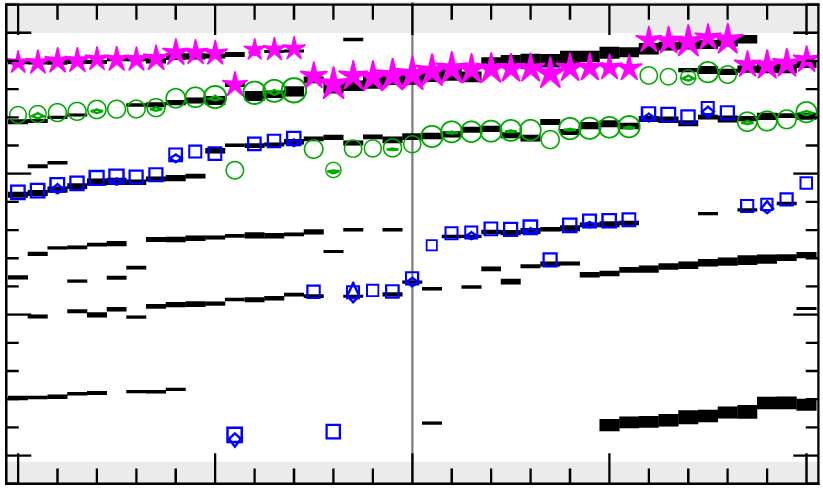}\\
    \includegraphics[width=0.38\textwidth]{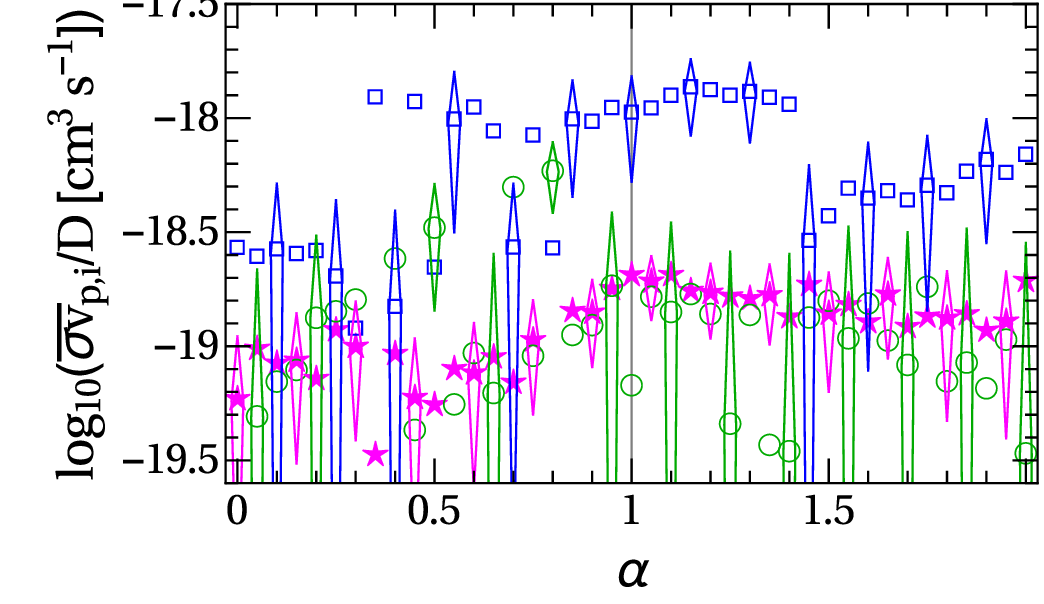}\hfill
    \includegraphics[width=0.30\textwidth]{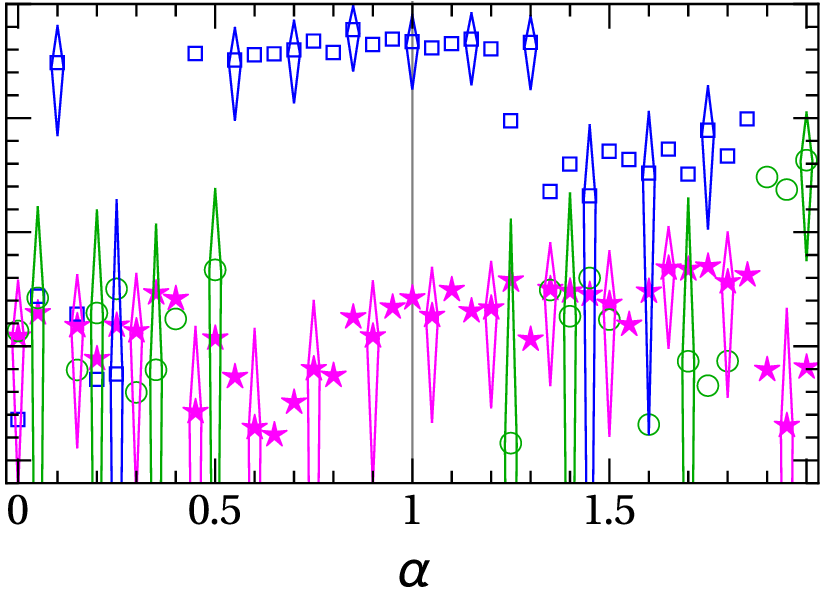}\hfill
    \includegraphics[width=0.30\textwidth]{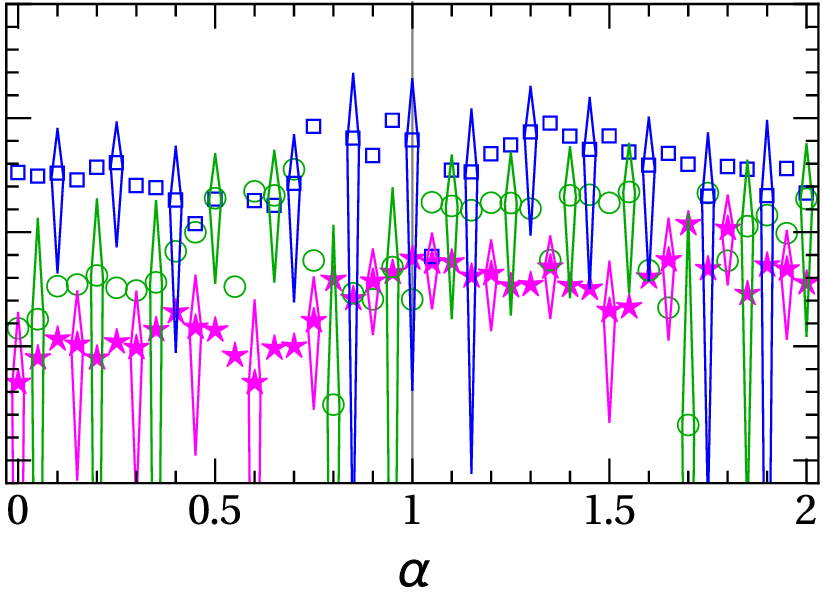}
    \caption{
        Same as the left column of Fig.~\ref{fig:Alpha}, for \SAll\ (left), \SE\ (middle), and \SW\ (right).
        \label{fig:AlphaEW}
    }
\end{figure*}

Both east and west hemispheres show $\sim70\GeV$ cluster-frame features in their stacked spectra (Fig.~\ref{fig:SpectraEW}) and in their $\alpha$ scans (Fig.~\ref{fig:AlphaEW}); \SE\ lies outside the \SCW\ sky region, whereas \SW\ contains it.
In \SW, $T_{72}$ peaks at $2.7$ for $\alpha\simeq1.05$, and the strongest triad peaks at $Z_l\simeq3.0$ for $\alpha\simeq1.1$.
In \SE, $T_{72}$ remains $\lesssim1$ at all $\alpha$, but the $69.9\GeV$ triad peaks at $\alpha\simeq1$--$1.1$ ($Z_l\simeq2.3$--$2.6$).

Tables~\ref{tab:BlindTriads} and \ref{tab:Fixed} list, for \SAll\ and each subsample, the strongest blind triad and the a priori $T_{72}$, $C_{72}$, and $L_{43}$.
The strongest blind line of every subsample but MCXC (the smallest catalog) is one of the nine lines of the two-species model (\S\ref{subsec:ModelTests}), and none of these strongest lines exceeds $2.1\sigma$.
In every sample, $T_{72}$ and its $\gamma\gamma$ channel are positive, and in \SE, so are all three triads of the model.

The four Galactic quadrants of \SAll\ give values of $T_{72}$ consistent with a uniform signal, i.e., with the $T_{72}$ of \SAll\ scaled by the square root of the fraction of clusters in each quadrant ($\chi^2\simeq3.5$ for 3 degrees of freedom).
Of the three readings of $T_{72}$ within a nonad (\S\ref{subsec:FixedTests}), the intermediate reading $N_{72}$ gives the largest $Z_g$ in \SE, \SW, \SO, and $\Delta$DESI (Table~\ref{tab:FixedNonads}), as in \SAll, whereas the high reading prevails in $\Delta$eROSITA ($2.7\sigma$, against $2.3\sigma$ for $N_{72}$).

\begin{table*}[t]
    \centering
    \caption{Strongest blind triad in each sample.\label{tab:BlindTriads}}
    \begin{tabular}{l c c c c}
        \toprule
        Sample & $\overline{m}_\chi c^2$ & $Z_l$ ($Z_g$) & $\{Z_{\gamma\gamma}, Z_{\gamma Z}, Z_{\gamma h}\}$ & $\{\overline{\sigma v}_{p,\gamma\gamma},\overline{\sigma v}_{p,\gamma Z},\overline{\sigma v}_{p,\gamma h}\}/D$ \\
         & (GeV) & & & ($10^{-19}\cm^3\se^{-1}$) \\
        \midrule
        {\bfseries\boldmath \SAll} & {\bfseries\boldmath $69.9_{-0.2}^{+0.2}$} & {\bfseries\boldmath 3.3 (2.8)} & {\bfseries\boldmath $\{3.4, 0.3, 2.0\}$} & {\bfseries\boldmath $\{2.0\pm0.6, 0.7\pm2.1, 10.2\pm5.0\}$} \\
        \SE & $69.9_{-0.2}^{+0.2}$ & 2.5 (1.5) & $\{1.8, -0.4, 2.8\}$ & $\{1.6\pm0.9, -1.1\pm3.1, 20.8\pm7.5\}$ \\
        \SW & $69.5_{-0.2}^{+0.3}$ & 2.7 (1.8) & $\{3.0, 0.6, 1.1\}$ & $\{2.4\pm0.8, 1.6\pm2.8, 7.7\pm7.1\}$ \\
        \SO & $74.5_{-0.3}^{+0.4}$ & 2.4 (1.5) & $\{1.4, 0.6, 2.3\}$ & $\{1.0\pm0.8, 1.5\pm2.4, 9.7\pm4.3\}$ \\
        MCXC & $73.5_{-0.8}^{+0.4}$ & 1.5 ($-0.3$) & $\{0.9, -0.9, 2.7\}$ & $\{0.6\pm0.7, -2.2\pm2.3, 10.8\pm4.0\}$ \\
        $\Delta$eROSITA & $87.9_{-0.5}^{+0.5}$ & 3.0 (2.3) & $\{0.8, 1.4, 3.1\}$ & $\{1.0\pm1.3, 5.3\pm3.9, 14.9\pm4.8\}$ \\
        $\Delta$DESI & $69.9_{-0.1}^{+0.2}$ & 3.5 (2.9) & $\{3.0, -0.4, 3.4\}$ & $\{2.9\pm1.0, -1.4\pm3.5, 29.1\pm8.6\}$ \\
        \bottomrule
    \end{tabular}
    \TblNote{Mean masses with $\Delta\textrm{TS}=1$ bounds, and channel cross sections before loss correction (\S\ref{subsec:Losses}).}
\end{table*}

\begin{table*}[t]
    \centering
    \setlength{\tabcolsep}{3pt}
    \caption{The triad $T_{72}$, with $C_{72}$ and $L_{43}$, in each sample.\label{tab:Fixed}}
    \begin{tabular}{l r c c c c c c}
        \toprule
        Sample & $N_c$ & $T_{72}$ & $\{Z_{\gamma\gamma}, Z_{\gamma Z}, Z_{\gamma h}\}$ & $\{Z_g, Z_g^{(r)}, Z_g^{(z)}\}$ & $C_{72}$ ($Z_g$) & $L_{43}$ ($Z_g$) & $\{\overline{\sigma v}_{p,\gamma\gamma},\overline{\sigma v}_{p,\gamma Z},\overline{\sigma v}_{p,\gamma h}\}/D$ \\
         & & & & & & & ($10^{-19}\cm^3\se^{-1}$) \\
        \midrule
        {\bfseries\boldmath \SAll} & {\bfseries\boldmath 23343} & {\bfseries\boldmath 2.2} & {\bfseries\boldmath $\{2.8, 0.9, 0.1\}$} & {\bfseries\boldmath $\{2.6, 2.8, 2.7\}$} & {\bfseries\boldmath 2.1 (2.2)} & {\bfseries\boldmath 1.1 (1.1)} & {\bfseries\boldmath $\{1.7\pm0.6, 1.8\pm2.0, 0.6\pm4.1\}$} \\
        \SE & 10274 & 0.6 & $\{1.7, -0.2, -0.3\}$ & $\{0.8, 0.8, 1.7\}$ & 1.0 (1.1) & $-0.2$ ($-0.2$) & $\{1.5\pm0.9, -0.8\pm3.1, -1.8\pm6.2\}$ \\
        \SW & 13069 & 2.4 & $\{2.3, 1.4, 0.5\}$ & $\{2.8, \textrm{---}, 2.3\}$ & 1.9 (2.0) & 1.6 ($1.6$) & $\{1.9\pm0.8, 3.9\pm2.7, 2.5\pm5.4\}$ \\
        \SO & 14992 & 1.4 & $\{1.5, 0.2, 0.8\}$ & $\{1.7, 1.7, 2.0\}$ & 1.6 (1.8) & 0.3 ($0.3$) & $\{1.1\pm0.7, 0.5\pm2.5, 3.8\pm4.9\}$ \\
        MCXC & 1628 & 0.7 & $\{0.7, -1.1, 1.6\}$ & $\{0.8, 0.8, 0.7\}$ & 1.6 (1.7) & $-1.1$ ($-1.1$) & $\{0.5\pm0.7, -2.6\pm2.3, 7.0\pm4.4\}$ \\
        $\Delta$eROSITA & 7833 & 1.9 & $\{1.2, 2.7, -0.6\}$ & $\{2.2, \textrm{---}, 2.0\}$ & 0.4 (0.4) & 3.2 ($3.1$) & $\{1.3\pm1.1, 9.7\pm3.6, -4.5\pm7.3\}$ \\
        $\Delta$DESI & 13882 & 1.3 & $\{2.5, -0.5, 0.2\}$ & $\{1.5, 1.6, 1.1\}$ & 1.9 (2.0) & $-0.6$ ($-0.6$) & $\{2.6\pm1.0, -1.8\pm3.4, 1.1\pm7.0\}$ \\
        \bottomrule
    \end{tabular}
    \par
    \TblNote{The $Z_g$ of $T_{72}$ calibrated by Monte Carlo, random longitudes ($Z_g^{(r)}$), and redshift permutations ($Z_g^{(z)}$), and those of $C_{72}$ and $L_{43}$ by Monte Carlo. Cross sections as in Table~\ref{tab:BlindTriads}.
    }
\end{table*}

\section{Redshift broadening}
\label{app:Dz}

In the \SCW\ analysis, broadened spectral apertures tested for a range of redshift contributions along the line of sight.
In stacking, each cluster is already shifted to its own frame, so broadening the apertures tests for residual, \eg photometric, redshift errors.
Each line template is averaged over line energies redshifted by $z'$ distributed uniformly between $0$ and $\Delta z$. Thus, a  positive (negative) $\Delta z$ tests for redshifts scattered above (below) the catalog values.

\begin{figure}[h]
    \centering
    \includegraphics[width=\columnwidth]{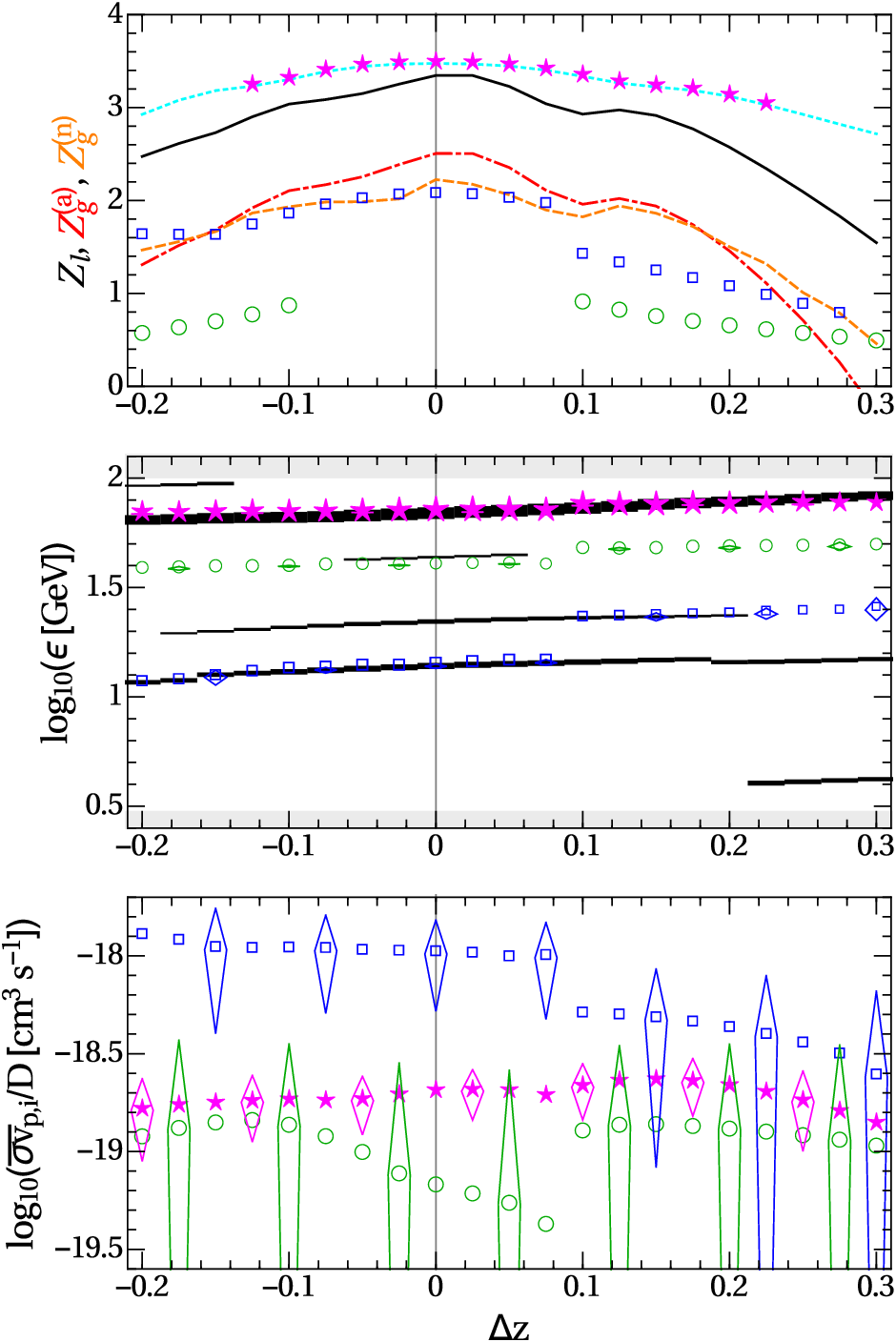}
    \caption{
        Same as the left column of Fig.~\ref{fig:Alpha}, but for apertures broadened (not shifted) uniformly in redshift between $\min(\Delta z,0)$ and $\max(0,\Delta z)$ (no broadening at $\Delta z=0$: vertical line).
        \label{fig:Dz}
    }
\end{figure}

As Fig.~\ref{fig:Dz} shows, the strongest (nominal, $69.9\GeV$) triad peaks at $Z_l\simeq3.3$ for $\Delta z\simeq0$--$0.025$ and weakens steadily with broadening in either direction, to $Z_l\simeq2.5$ at $\Delta z=-0.2$ and $1.5$ at $\Delta z=0.3$.
No broadening is thus preferred, as expected for narrow lines emitted in the cluster frame, and no
directional scatter is identified with respect to catalog redshift.
For $\Delta z>0$, a broadened template lies below its cluster-frame energies, so the mass of the strongest triad rises by $\sim\Delta z/2$;
the jump seen in the figure at $\Delta z\simeq0.1$ marks a switch from the nominal triad to $T_{72}$.

\section{The eROSITA map and \SCW}
\label{app:eROSITA}

Following \SCW\cite{Keshet26Corr}, we use the first eROSITA all-sky survey (eRASS1) $0.2$--$2.3\keV$ map of the western Galactic hemisphere\cite{MerloniEtAl24eRASS1} on an order-10 HEALPix grid, with point sources
masked.
The X-ray-correlated LAT brightness, whose detrended spectrum is denoted $\delta I_\theta$, is the exposure-weighted regression coefficient of the LAT brightness on the zero-mean X-ray contrast of this map on scales below $2\dgr$, so each LAT photon is weighted by the contrast of its pixel.
The ambient brightness of the same pixels has the detrended spectrum $\delta\llangle\boldsymbol{I}\rrangle$.

To construct \SMW\ (\S\ref{sec:Joint}), every analyzed pixel within $0\dgrdot1$ of any of the 23,343 stacked clusters is masked.
The contrast is then shifted to zero mean over the remaining pixels, and the X-ray-correlated spectrum, its errors, its detrending, and its shuffled-contrast skies are recomputed.

While the LSS contributing to \SCW\ lies at low redshift, $\gamma\gamma$ line shifts are comparable to the triad separation.
Redshift-corrected, the $71.9\GeV$ \SCW\ triad corresponds to a rest-frame mass of $72.2\GeV$ (its channel energies scaled by $1.03$, or an aperture broadened uniformly over $0\le z\le0.05$) to $72.6\GeV$ (the $\sim0.9\%$ triad offset of the \SCW\ mocks; \S\ref{subsec:FixedMass}).
The mass shifts by a smaller fraction than the line energies because, by Eq.~\eqref{eq:kin}, the $\gamma Z$ ($\gamma h$) line energy near $72\GeV$ changes with mass $\sim2.3$ ($\sim7$) times faster than the $\gamma\gamma$ one, in fractional terms.

Whereas stacking normalizes $D=1$ for a WIMP density $\Omega_\chi/\Omega_m$ times the NFW total-mass profile of each cluster, with its Jeans $\beta^2$ (Eqs.~\eqref{eq:LineFlux} and \eqref{eq:SigmaStack}), the corresponding factor $D_{\textrm{X}}$ of \SCW\ is unity for a density $\Omega_\chi/\Omega_b$ times that of
the X-ray-emitting gas, integrated along the line of sight, and an approximate $\beta^2=10^{-5}$.
The ratio $K\equiv D/D_{\textrm{X}}$ depends on the signal distribution among and within the clusters, and is estimated as follows.

Each \SAll\ cluster is assigned two sets of $T_{72}$ fluxes.
One is the synthetic Eq.~\eqref{eq:LineFlux} signal, fixed by the catalog mass and redshift.
The other is an X-ray-scaled signal based on $L_{500}$ from the catalog (MCXC, eROSITA) or the luminosity–mass relation (DESI), converted to $T_{72}$ fluxes by the \SCW\ line-to-X-ray brightness ratio\cite{Keshet26Corr} in each channel (here for templates at $z=0.03$).
The two signals are spread on the sky like the NFW emission of each cluster (as for $\psi_\Myc$), stacked, and compared before detrending, giving $K\simeq\{4.3, 4.3, 4.4\}$ in the $\{\gamma\gamma, \gamma Z, \gamma h\}$ channels ($K\simeq 5.5$ with the calibrated Appendix~\ref{app:MassCal} masses).

The $\overline{\sigma v}_{p,\gamma\gamma}/D_{\textrm{X}}$ loss-corrected \SCW\ cross section, when divided by $K$, now gives $\Sigma_0=(3.4\pm2.2)\times10^{-19}\cm^3\se^{-1}$ (or $0.29$ times this value if the modeled redshift distribution is replaced by one uniform over $z<0.05$) for templates at $z=0.03$.
These values assume that bremsstrahlung produces all of the X-ray-contrast variance ($\mathcal{F}_b=1$); the $\mathcal{F}_b<1$ bound \cite{Keshet26Corr} would raise $\Sigma_0$ by factors $\gtrsim1.6$.
Detrended in \SAll, the X-ray-scaled signal retains fractions $\mathcal{F}^{\textrm{X}}_i\simeq\{0.73, 0.50, 0.58\}$ of its line brightness (varying by $\lesssim0.03$ among the samples), so \SCW\ implies that stacking would measure $\Sigma_{\textrm{X}}\equiv\mathcal{F}^{\textrm{X}}_{\gamma\gamma}\Sigma_0\simeq(2.5\pm1.6)\times10^{-19}\cm^3\se^{-1}$ (or $0.7\times10^{-19}\cm^3\se^{-1}$ for the uniform distribution; Fig.~\ref{fig:Scaling}).

\section{The $43.2\GeV$ line}
\label{app:Line43}

As $L_{43}$ is a $\gamma Z$ line in our model, samples should show it and its $72.1\GeV$ ($\gamma\gamma$) and $17.7\GeV$ ($\gamma h$) companions $C_{72}$ both overall and in the Virgo and Fornax of \SCT\ (Ophiuchus lies below the latitude cut).

Among the six subsamples of Table~\ref{tab:Fixed}, $L_{43}$ is highest in $\Delta$eROSITA ($3.1\sigma$; \S\ref{subsec:FixedMass}).
Virgo and Fornax are MCXC clusters; removing the few $\Delta$eROSITA clusters within their $\theta_{500}$ leaves $L_{43}$ unchanged.
Removing its five strongest contributors still leaves $L_{43}\simeq2.3$.
Its $C_{72}$ companions are weak in $\Delta$eROSITA ($Z_l\simeq0.4$; Table~\ref{tab:Fixed}).

In Virgo and Fornax, we count photons within $\pm7\%$ of each line energy, subtract the background scaled from an outer annulus, and subtract the continuum interpolated from the adjacent energy bands.
The point-source mask removes the core of Virgo around M87 and with it nearly all $L_{43}$ photons ($1\pm5$ remain).

Unmasked, $L_{43}$ shows $11\pm7$ photons, consistent with the $\sim10$--$16$ fitted in \SCTb, whereas $-3\pm4$ ($6\pm11$) photons are found at $72.1\GeV$ ($17.7\GeV$).
Scaling the $L_{43}$ photons by the exposure and by the channel ratios of the $T_{72}$ of \SCW\ (as observed) predicts $\sim7$ ($\sim10$) photons, with wide uncertainties from the ratios.
The companion counts thus fall short by $\sim1.8$ ($\sim0.4$) standard deviations, or by $\sim2.0$ ($\sim1.2$) for the common ratios of \S\ref{subsec:ModelTests}.
With such small photon numbers, these counts neither confirm nor exclude the $\gamma Z$ reading of $L_{43}$.

Converted by Eq.~\eqref{eq:LineFlux} as for the stacked clusters and multiplied by the fraction $\mathcal{F}_{\gamma Z}\simeq0.61$ that stacking retains, the $L_{43}$ fluxes fitted to Virgo and Fornax in \SCTb\ agree with the $\gamma Z$ cross section of \SAll\ (Table~\ref{tab:Fixed}) for an NFW annihilation profile (to within $0.01$ combined standard deviations), and lie $1.5$ combined standard deviations above it for uniform emission.

\section{Statistics of \SJ}
\label{app:JointStats}

\begin{table*}[t]
    \centering
    \caption{Strongest blind line, triad, and nonad of \SJ\ in each treatment of the \SCW\ redshift.\label{tab:JointBlind}}
    \setlength{\tabcolsep}{3.5pt}
    \begin{tabular}{l c c c c c c c c c}
        \toprule
        Treatment & $\epsilon_0$ (GeV) & $Z_l$ & $Z_g$ ($Z_g^{(n)}$) & $\overline{m}_\chi c^2$ (GeV) & $Z_l$ & $Z_g$ ($Z_g^{(n)}$) & $\{m_1, m_2\}c^2$ (GeV) & $Z_l$ & $Z_g$ ($Z_g^{(n)}$) \\
        \midrule
        {\bfseries\boldmath $z=0.03$} & {\bfseries\boldmath $70.1_{-1.0}^{+1.4}$} & {\bfseries\boldmath 3.7} & {\bfseries\boldmath 2.4 (2.4)} & {\bfseries\boldmath $72.2_{-0.2}^{+0.2}$} & {\bfseries\boldmath 3.7} & {\bfseries\boldmath 2.9 (2.9)} & {\bfseries\boldmath $\{69.1, 74.9\}$} & {\bfseries\boldmath 3.9} & {\bfseries\boldmath 3.8 (3.6)} \\
        $\Delta z=0$ & $69.6_{-1.0}^{+0.8}$ & 3.9 & 2.7 (2.7) & $72.0_{-0.2}^{+0.1}$ & 3.6 & 2.7 (2.7) & $\{66.1, 71.9\}$ & 4.0 & 3.9 (3.7) \\
        $\Delta z=0.05$ & $70.1_{-1.1}^{+1.0}$ & 3.7 & 2.4 (2.5) & $72.1_{-0.2}^{+0.2}$ & 3.6 & 2.9 (2.9) & $\{69.1, 74.9\}$ & 3.9 & 3.9 (3.7) \\
        $\Delta z=0.08$ & $70.1_{-1.3}^{+1.4}$ & 3.6 & 2.2 (2.3) & $72.3_{-0.3}^{+0.2}$ & 3.5 & 2.7 (2.8) & $\{69.5, 74.5\}$ & 4.0 & 3.9 (3.7) \\
        Shifted & $71.0_{-1.4}^{+1.4}$ & 3.4 & 2.0 (2.0) & $69.7_{-0.1}^{+0.2}$ & 3.8 & 3.1 (3.1) & $\{68.1, 72.3\}$ & 4.0 & 4.0 (3.7) \\
        Folded & $70.1_{-1.3}^{+1.6}$ & 3.4 & 2.0 (2.1) & $69.7_{-0.3}^{+0.2}$ & 3.4 & 2.6 (2.6) & $\{69.8, 74.5\}$ & 3.8 & 3.9 (3.7) \\
        \bottomrule
    \end{tabular}
    \TblNote{Nominal treatment ($z=0.03$) in bold. Bounds are $\Delta\textrm{TS}=1$, for the line ($f=0.1$) and the triad. The $Z_g$ are from $4\times10^5$ null pairs with the shuffled-contrast \SMW\ skies ($Z_g^{\smash[t]{(n)}}$: with Gaussian \SMW\ realizations).}
\end{table*}

In every treatment of the \SCW\ redshift (\S\ref{subsec:FixedMass}), the calibrations by the shuffled-contrast skies and by Gaussian realizations agree to within $0.15\sigma$, except for the blind nonads; see Tables~\ref{tab:JointBlind} and \ref{tab:JointFixed}.
Detrending, which the skies share with the data but the Gaussian realizations lack, suppresses the null maxima of the extended nonad filters the most, so the skies calibrate the blind nonads $0.2$--$0.3\sigma$ higher, and their Gaussian $Z_g^{\smash[t]{(n)}}$ are deliberately low.
We find $\{N_{72}, T_{72}, C_{72}\}$ null means $\{0.68, 0.56, 0.42\}$ and dispersions $\{0.72, 0.82, 0.89\}$ for \SMW\ over its skies.
A smooth version of the shuffled-contrast \SMW\ skies, consisting of $10^5$ multivariate normal realizations of their $Z_l$ with the same mean and covariance, reproduces their calibration to within $0.1\sigma$.

\begin{table}[h]
    \centering
    \caption{A priori tests of \SJ.\label{tab:JointFixed}}
    \begin{tabular}{l c c}
        \toprule
        Treatment & $Z_l$ (\SMW, \SAll) & $Z_g$ ($Z_g^{(n)}$) \\
        \midrule
        \multicolumn{3}{l}{$L_{43}$} \\
        {\bfseries\boldmath $z=0.03$} & {\bfseries\boldmath 1.8 (1.4, 1.1)} & {\bfseries\boldmath 1.6 (1.8)} \\
        $\Delta z=0$ & 2.0 (1.8, 1.1) & 1.9 (2.0) \\
        $\Delta z=0.05$ & 1.9 (1.6, 1.1) & 1.7 (1.9) \\
        $\Delta z=0.08$ & 1.9 (1.5, 1.1) & 1.7 (1.9) \\
        Shifted & 1.7 (1.2, 1.1) & 1.5 (1.7) \\
        Folded & 1.9 (1.5, 1.1) & 1.7 (1.9) \\
        \midrule
        \multicolumn{3}{l}{$C_{72}$} \\
        {\bfseries\boldmath $z=0.03$} & {\bfseries\boldmath 3.2 (2.5, 2.1)} & {\bfseries\boldmath 3.2 (3.3)} \\
        \midrule
        \multicolumn{3}{l}{$T_{72}$} \\
        {\bfseries\boldmath $z=0.03$} & {\bfseries\boldmath 3.6 (2.9, 2.2)} & {\bfseries\boldmath 3.8 (3.8)} \\
        $\Delta z=0$ & 3.5 (2.8, 2.2) & 3.7 (3.8) \\
        $\Delta z=0.05$ & 3.6 (2.9, 2.2) & 3.8 (3.9) \\
        $\Delta z=0.08$ & 3.4 (2.6, 2.2) & 3.6 (3.7) \\
        Shifted & 3.1 (2.2, 2.2) & 3.2 (3.4) \\
        Folded & 3.1 (2.1, 2.2) & 3.2 (3.3) \\
        \midrule
        \multicolumn{3}{l}{$N_{72}$ ($\{69.7, 74.4\}\GeV$)} \\
        {\bfseries\boldmath $z=0.03$} & {\bfseries\boldmath 3.9 (2.4, 3.1)} & {\bfseries\boldmath 4.5 (4.2)} \\
        \bottomrule
    \end{tabular}
    \TblNote{Nominal treatment ($z=0.03$) in bold. The $Z_g$ are from all null pairs with the $20000$ shuffled-contrast \SMW\ skies ($Z_g^{\smash[t]{(n)}}$: with Gaussian \SMW\ realizations).}
\end{table}

Let $t$ and $\mathbf{t}_o$ be the $Z_l$, minus their null means, for a triad and its two companions within a nonad.
The conditional significance
\begin{equation}\label{eq:Zc}
Z_c=\left(V-\mathbf{v}^{\textrm{T}}V_o^{-1}\mathbf{v}\right)^{-1/2}\left(t-\mathbf{v}^{\textrm{T}}V_o^{-1}\mathbf{t}_o\right)
\end{equation}
quantifies the part of $t$ not predicted by $\mathbf{t}_o$ through null correlations.
Here, $V$ is the null variance of $t$, $V_o$ the null covariance of $\mathbf{t}_o$, and $\mathbf{v}$ their null cross-covariance.
The measured $Z_c$ of the three triads in \SAll\ and in \SMW\ are compared with those expected for
a model pattern of the nine lines, scaled separately in each dataset, through a $\chi^2$ calibrated by null pairs fitted likewise.

Each \SMW\ channel, after subtracting its mean amplitude over the shuffled-contrast skies, is converted to a cross section as \SCW\ is in Appendix~\ref{app:eROSITA}.
The values $\overline{\sigma v}_d\pm\Delta_d$ of the two datasets are combined as $\sum_d(\overline{\sigma v}_d/\Delta_d)/\sum_d\Delta_d^{-1}\pm\smash{\sqrt{2}}/\sum_d\Delta_d^{-1}$, which gives the combination the significance of their co-added $Z_l$.

\section{Branching ratios}
\label{app:BR}

To crudely estimate the $\gamma\gamma:\gamma Z:\gamma h$ branching ratios (\S\ref{subsec:ModelTests}), we take them equal for the three triads of the two-species model (its nine lines; Table~\ref{tab:NineLines}) and fit them jointly to two datasets: \SAll\ with \SMW, or \SO\ with the unmasked \SCW.
In dataset $d$, the loss-corrected cross section of channel $i$ of triad $j$ is modeled as $u_{dj} r_i$, with a free amplitude $u_{dj}$ and the common ratios $r_i$ ($r_{\gamma\gamma}=1$).

The $\chi^2$ fit accounts for the overlap of adjacent filters through the response of each filter to each line.
In stacking, this response comes from injected, detrended lines, so it includes the losses, and the errors include the channel correlations of the null spectra.
In \SMW\ and the unmasked \SCW, where the conversion to cross sections (Appendix~\ref{app:eROSITA}) corrects the losses, it comes from the overlaps of the filter templates.

The \SO+\SCW\ and \SAll+\SMW\ fits agree, but they share most photons, so the agreement shows only that assigning these shared photons to \SAll\ or to \SCW\ gives similar results.
The common $\gamma h/\gamma Z\simeq4.7$ ratio is dominated by the lower limits imposed by the outer triads; the intermediate triad gives $0.3$--$2.8$.
A separate $\gamma h/\gamma Z$ ratio for each triad, which tests the assumption that triads share the same branching ratios,
lowers $\chi^2$ by only $1.1$ for two degrees of freedom ($p\simeq0.57$).

\end{document}

%% file: Definitions.tex
\newcommand{\ie}{\emph{i.e.} }
\newcommand{\eg}{\emph{e.g.,} }

\newcommand{\be}{\begin{equation}}
\newcommand{\ee}{\end{equation}}
\newcommand{\bea}{\begin{equation*}}
\newcommand{\eea}{\end{equation*}}
\newcommand{\beqr}{\begin{eqnarray} \nonumber}
\newcommand{\eeqr}{\end{eqnarray}}
\newcommand{\beqrb}{\begin{eqnarray}}
\newcommand{\eeqrb}{\nonumber \end{eqnarray}}
\newcommand{\fin}{\mbox{ .}}
\newcommand{\coma}{\mbox{ ,}}

\newcommand{\cm}{\mbox{ cm}}
\newcommand{\sr}{\mbox{ sr}}
\newcommand{\se}{\mbox{ s}}

\newcommand{\Mpc}{\mbox{ Mpc}}

\newcommand{\keV}{\mbox{ keV}}

\newcommand{\GeV}{\mbox{ GeV}}

\newcommand{\gama}{$\gamma$}

\newcommand{\mynewcommand}[2]{\ifdefined #1 \else \newcommand{#1}{#2} \fi}

\mynewcommand{\apj}{ApJ}     
\mynewcommand{\apjl}{ApJL}     
\mynewcommand{\apjs}{ApJS}    
\mynewcommand{\aap}{A\&A}    
\mynewcommand{\nat}{Nature}  

\newcommand{\dgr}{^{\circ}}

\newcommand{\dgrdot}{{\overset{^\circ}{.}}}
\newcommand{\mindot}{{\overset{{'}}{.}}}